\documentclass[sn-mathphys,Numbered]{sn-jnl}

\usepackage{graphicx}%
\usepackage{multirow}%
\usepackage{amsmath,amssymb,amsfonts}%
\usepackage{amsthm}%
\usepackage{mathrsfs}%
\usepackage[title]{appendix}%
\usepackage{xcolor}%
\usepackage{textcomp}%
\usepackage{manyfoot}%
\usepackage{booktabs}%
\usepackage{algorithm}%
\usepackage{algorithmicx}%
\usepackage{algpseudocode}%
\usepackage{listings}%
\usepackage{longtable, lscape}
\usepackage{rotating}
\usepackage{comment}
\usepackage{float}
\usepackage[para]{threeparttablex}
\usepackage{subcaption} 
\usepackage{afterpage}
\usepackage{wrapfig}
\usepackage{multirow}
\newcommand{\myitem}{\item[]}

\begin{document}

\title[Discovering Motifs to Fingerprint Multi-Layer Networks: a Case Study on the Connectome of \textit{C.~elegans}]{Discovering Motifs to Fingerprint Multi-Layer Networks: a Case Study on the Connectome of \textit{C.~Elegans}}


\author*[1]{\fnm{Deepak} \sur{Sharma}}\email{des@informatik.uni-kiel.de}
\author[1]{\fnm{Matthias} \sur{Renz}}\email{mr@informatik.uni-kiel.de}


\author[2]{\fnm{Philipp} \sur{H{\"o}vel}}\email{philipp.hoevel@uni-saarland.de}

\affil*[1]{\orgdiv{Department of Computer Science}, \orgname{Christian-Albrechts Universit\"at zu Kiel}, \orgaddress{\street{Christian-Albrechts-Platz 4}, \city{Kiel}, \postcode{24118}, \country{Germany}}}

\affil[2]{\orgdiv{Theoretical Physics and Center for Biophysics}, \orgname{Saarland University}, \orgaddress{\street{Campus E2 6}, \city{Saarbr\"ucken}, \postcode{66123}, \country{Germany}}}

\abstract{Motif discovery is a powerful and insightful method to quantify network structures and explore their function. As a case study, we present a comprehensive analysis of regulatory motifs in the connectome of the model organism \textit{Caenorhabditis elegans} (\textit{C.~elegans}). Leveraging the Efficient Subgraph Counting Algorithmic PackagE (ESCAPE) algorithm, we identify network motifs in the multi-layer nervous system of \textit{C.~elegans} and link them to functional circuits. We further investigate motif enrichment within signal pathways and benchmark our findings with random networks of similar size and link density. Our findings provide valuable insights into the organization of the nerve net of this well documented organism and can be easily transferred to other species and disciplines alike.}



\maketitle


\section{Introduction}
The nematode \textit{Caenorhabditis elegans}, commonly known as \textit{C.~elegans}, has long been a prominent model organism in genetics and developmental biology research. Its relatively simple and well-defined anatomy, short life cycle, and fully sequenced genome make it an ideal subject for investigating fundamental biological processes \cite{chase2007biogenic,yan2017network}. Over the years, extensive studies on \textit{C.~elegans} have provided valuable insights into various aspects of cellular and molecular biology, contributing significantly to our understanding of eukaryotic biology \cite{towlson2013rich,dag2023dissecting, towlson2020synthetic,chew2017recordings}.


In this study, we take a network-science driven angle \cite{barabasi2023neuroscience} and focus on motif discovery in the nervous system of \textit{C.~elegans} to shed light on (i) the composition of the multi-layer connectome and (ii) the underlying regulatory circuits that govern the behavior of this model organism. The multi-layer nature arises from the presence of electrical (gap junction) and synaptic connections between the neurons.
To achieve this, we adopt the \textit{Efficient Subgraph Counting Algorithmic PackagE }(ESCAPE) algorithm \cite{pinar2017escape}, a powerful computational method that has proven effective in identifying regulatory motifs in complex networks extracted from empirical data. 

We focus on network motifs that emerge from sets of 4 or 5 nodes and their connecting links such as triangles and cliques. Applying the ESCAPE algorithm to individual layers, the aggregated network, and the locomotory circuit, we compare the distribution of these small-scale, recurrent network structures to random networks of similar size and link density. This allows identification of a fingerprint of the nervous system structure. 
We hypothesize that the ESCAPE algorithm will uncover novel and functionally relevant motifs in \textit{C.~elegans} that play crucial roles for behaviors such as locomotion. 
By pursuing these objectives and testing our central hypothesis, we aim to provide valuable insights into the anatomical and functional landscape of the \textit{C.~elegans} connectome, exemplifying a universal approach to characterize network data in biology and beyond. This study holds the potential to advance our knowledge of networks in general and has implications in biological contexts in particular.
Our approach is universally applicable and can be extended to network data (including directed links) of other sources and invites benchmarking against any network model. Insights into the motif distribution can also be used for artificial network generation for realisitic \textit{in-silico} studies.

The rest of the manuscript is organized as follows: In Section~\ref{sec:methods}, we elaborate on the workflow and data set considered. Section~\ref{sec:results} presents our findings and highlights exemplary cases of significant and insignificant motif occurrences. Finally, we conclude in Section~\ref{sec:conclusions}. We stress that we focus on selected examples of motifs involving 3, 4, and 5 nodes in the main text and provide the full set of results in Appendices~\ref{app:4non-induced}--\ref{app:5induced_random}.


\clearpage
\section{Methods and Data}
\label{sec:methods}
This section discusses the methodology applied to the \textit{C.~elegans} data for the motif discovery. We start with a summary of the multi-layer network and then, briefly review the ESCAPE framework \cite{pinar2017escape}.

\subsection {Multi-Layer Network of \textit{C.~elegans}}

The complexity of connectome structure in \textit{C.~elegans} extends beyond simple regulatory interactions, necessitating a comprehensive approach to capture the intricate relationships between anatomy and function. To achieve a more holistic understanding of the regulatory landscape, we consider a multi-layer network representation of \textit{C.~elegans}. This multi-layer network incorporates different layers of biological information, providing a powerful framework to investigate the interplay between neurons (network nodes), connections (network links), and their regulatory motifs (small, recurrent network structures). As discussed in the following, the multi-layer structure consists of (i) electrical connections and (ii) synapses that involve different neurotransmitters, monoamines, and peptides (see Fig.~\ref{fig:connectome_layers} for the largest four layers in a circular layout):

\begin{figure}[ht!]
\begin{subfigure}[h]{0.45\textwidth}
\includegraphics[width=\textwidth]{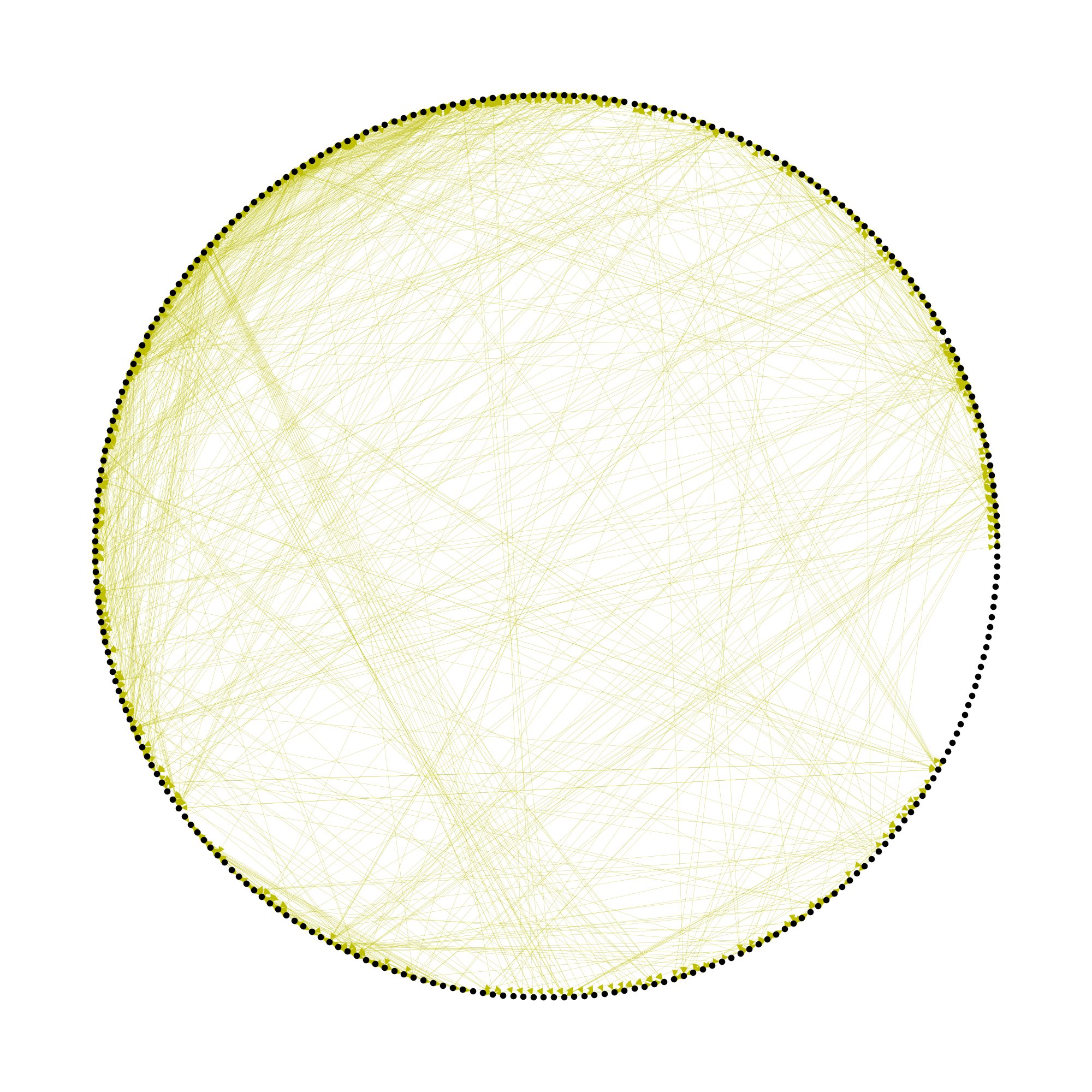}
\caption {Acetylcholine layer (33.4\%)}
\end{subfigure}
\begin{subfigure}[h]{0.45\textwidth}
\includegraphics[width=\textwidth]{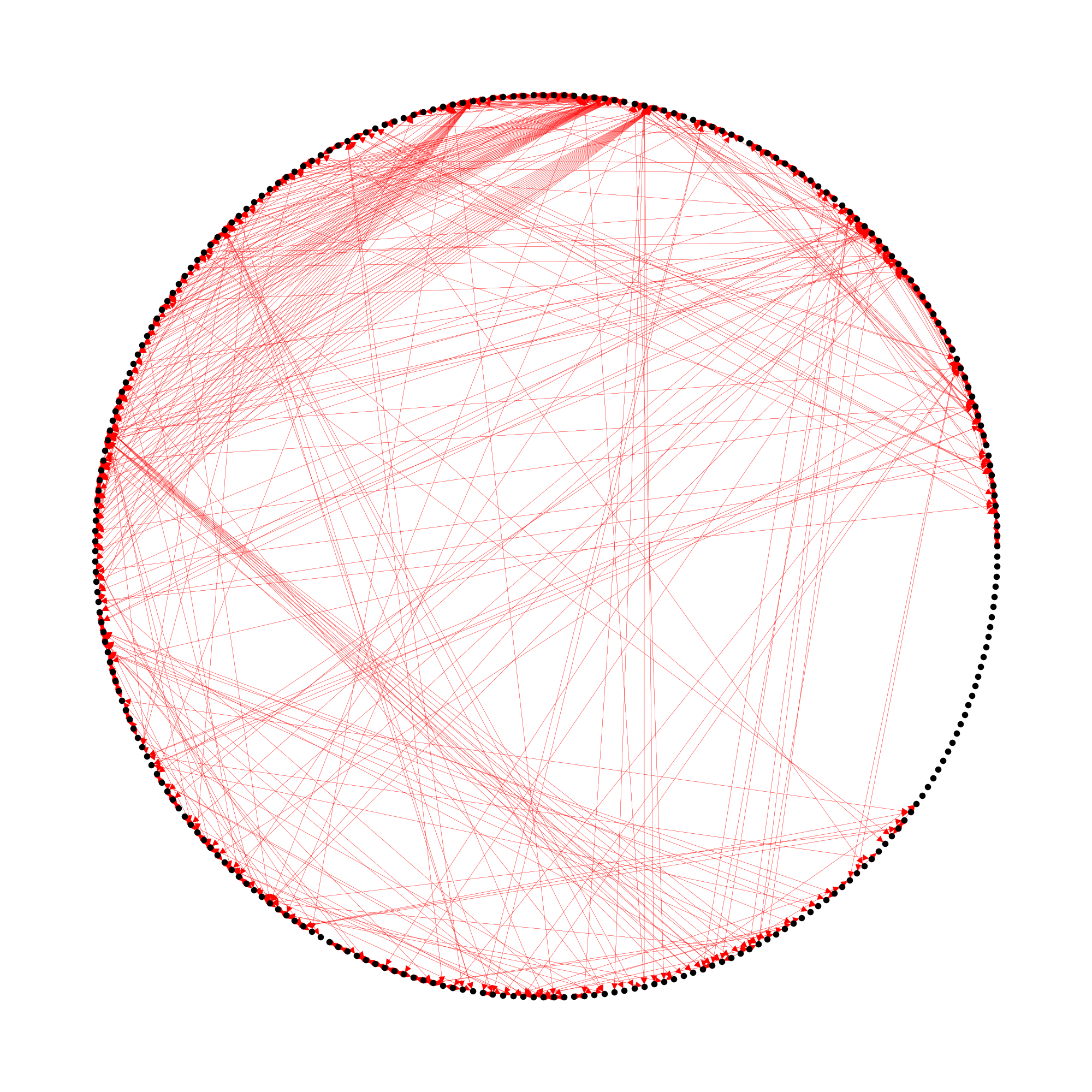}
\caption{Electrical layer (29.1\%)}
\end{subfigure}
\begin{subfigure}[h]{0.45\textwidth}
\includegraphics[width=\textwidth]{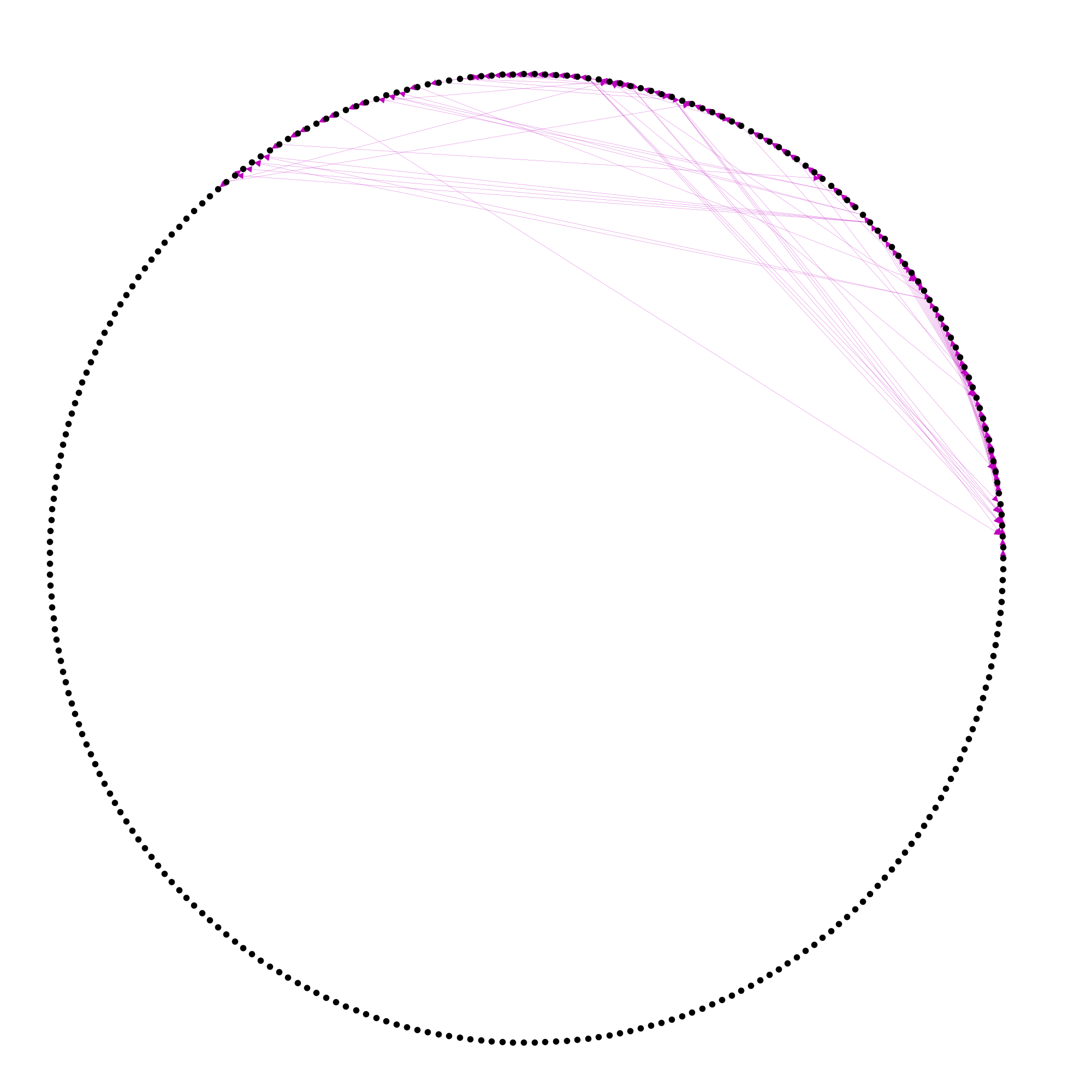}
\caption {gamma-Aminobutyric acid layer (3.8\%)}
\end{subfigure}
\begin{subfigure}[h]{0.45\textwidth}
\includegraphics[width=\textwidth]{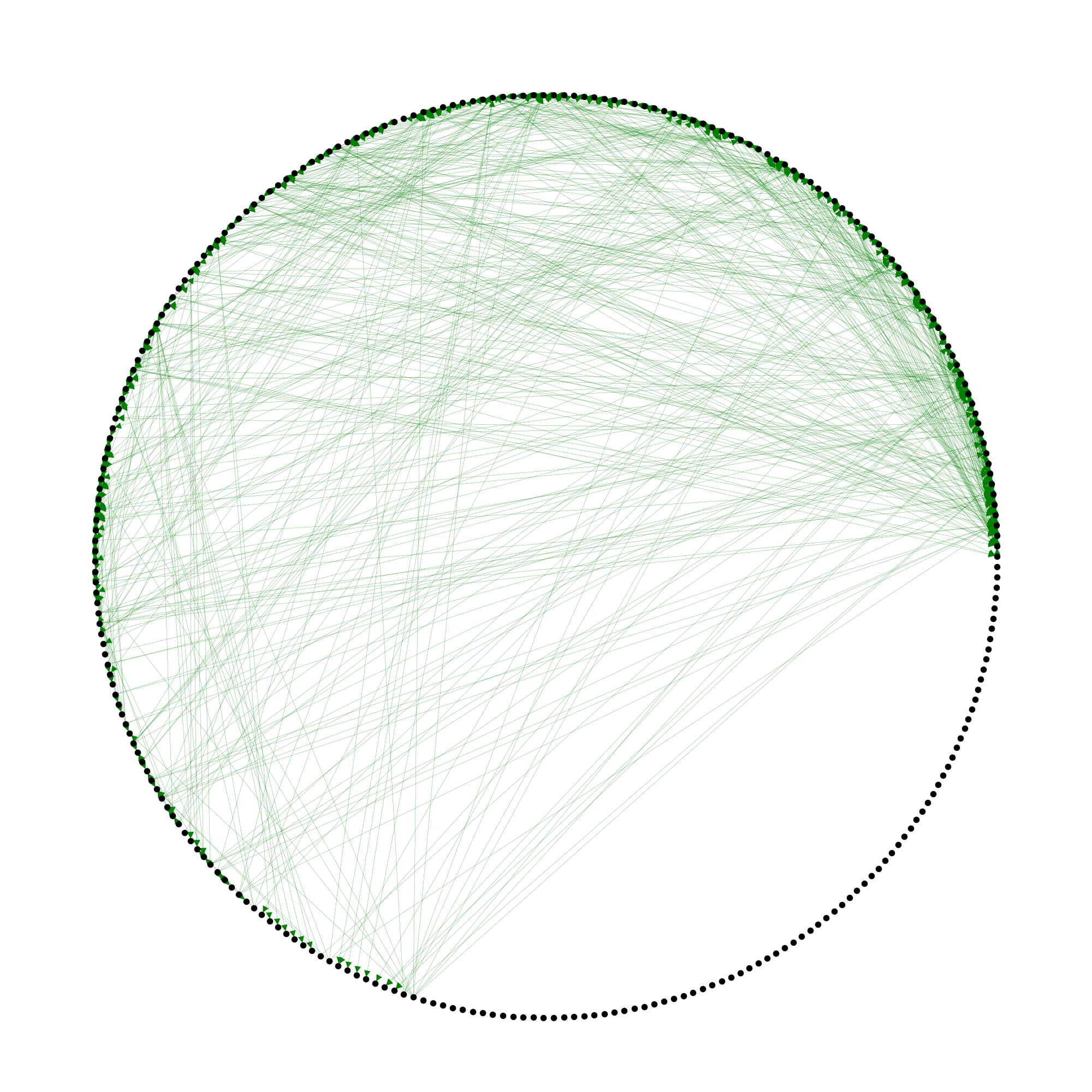}
\caption {Glutamate layer (20.5\%)}
\end{subfigure}
\centering
    \caption{The \textit{C.~elegans} multi-layer network consists of nodes (circles) representing 279 neurons in a circular layout. In the aggregated network, there are 3,538 distinct connections (lines) among the nodes. Panels (a)-(d) show the links of 4 largest layers in terms of number of links: Acetylcholine (yellow), electrical (red), gamma-Aminobutyric acid (magenta), and Glutamate (green) layers, respectively.}
\label{fig:connectome_layers}
\end{figure}

\textbf{Acetylcholine Layer (ACh):} The Acetylcholine (ACh) layer, the largest component of the network, encompasses approximately 33.4\% of the total 3,538 connections. It is distributed extensively across all neurons without showing a specific preference for any particular neuron type. In humans, ACh functions at skeletal neuromuscular junctions, interfaces between the vagus nerve and cardiac muscle fibers, as well as various locations within the central nervous system. Although ACh's role at neuromuscular junctions is well-documented, its precise function within the central nervous system remains a subject of incomplete understanding \cite{purves2019neurosciences}. In the context of \textit{C.~elegans}, ACh is intricately involved in a multitude of behaviors, including but not limited to locomotion \cite{winnier1999unc,hallam2000c}, egg-laying \cite{bany2003genetic}, feeding \cite{raizen1995interacting}, and defecation \cite{thomas1990genetic}.

\textbf{Electrical Layer:} Electrical transmission, comprising 29.1\% of network connections, is the second-largest layer after ACh. Electrical synapses are universally present in nervous systems, allowing direct bidirectional flow of electrical current between neurons. This mode of transmission is exceptionally fast, virtually instantaneous, enabling communication without delay, a departure from typical chemical synapses. Electrical synapses are located where rapid coordination and synchronization of network activity are required. For example, certain brainstem neurons synchronize via electrical synapses for rhythmic breathing, and similar synchronization occurs in interneuron populations in the cerebral cortex, thalamus, and cerebellum \cite{purves2019neurosciences}. In \textit{C.~elegans}, electrical transmission plays a significant role in locomotion behavior and development \cite{hall2017gap,simonsen2014gap}.

\textbf{Gamma-Aminobutyric acid Layer (GABA):} The gamma-Aminobutyric acid (GABA) layer stands as the smallest component within the network, accounting for approximately 3.8\% of the connections. Most of these connections are established by interneurons and motor neurons. GABA serves as the neurotransmitter for approximately one third of the synapses in the brain and is predominantly found in interneurons within local circuits. Diverging from the excitatory nature of ACh and Glu, GABA exerts an inhibitory effect \cite{purves2019neurosciences}. Notably, in \textit{C.~elegans}, GABA can also function as an excitatory transmitter, depending on the specific neuroreceptor involved. In its inhibitory role, GABA regulates head movements during foraging \cite{white1986structure} and relaxes muscle cells during locomotion \cite{mclntire1993genes}.

\textbf{Glutamate Layer (Glu):} The Glutamate (Glu) layer constitutes the third-largest component of the network, utilizing approximately 20.5\% of its connections. In contrast to the ACh layer, this layer involves significantly fewer motor neurons. In the human brain, Glu plays a pivotal role as the primary neurotransmitter for normal brain function, with an estimated release from more than half of all brain synapses \cite{purves2019neurosciences}. In the context of \textit{C.~elegans}, Glu contributes to various behavioral aspects, including foraging behavior \cite{hills2004dopamine}, long-term memory processes \cite{rose2003glr}, and spontaneous transitions from forward to backward movement, commonly referred to as reversals \cite{zheng1999neuronal,brockie2001c}.

\textbf{Monoamine Layer (MA):} MA transmitters constitute approximately 5.8\% of network connections, with no specific neuron-type preference, although most motor neurons primarily receive postsynaptic signals. Among MA transmitters, dopamine accounts for about two-thirds. In humans, MA transmitters play a crucial role in regulating various brain functions, extending from central homeostatic processes to cognitive phenomena like attention. Dysfunctions in MA-mediated signal processing can lead to psychiatric disorders, exemplified by Parkinson's disease resulting from dopaminergic neuron degeneration \cite{purves2019neurosciences}. In \textit{C.~elegans}, MA transmitters influence a range of behaviors, including egg-laying, pharyngeal pumping, locomotion, and learning, as detailed in \cite{chase2007biogenic}. For instance, dopamine modulates locomotion behavior and learning, facilitating the worms' responsiveness to environmental changes \cite{sawin2000c} and efficient exploration of new food sources \cite{hills2004dopamine}. Learning allows worms to adapt their behavior based on prior experiences, such as responding to non-localized mechanical stimuli like plate tapping by either moving backward or increasing forward locomotion, with repeated tapping leading to habituation and reduced reversal frequency \cite{rose2001analyses}.

\textbf{Peptide Layer:}  Peptides account for approximately 11.8\% of network connections, predominantly established by interneurons and sensory neurons, with most originating from the FMRFamide-like peptide (FLP) and peptide deformylase (PDF) families. Roughly 31\% of these peptides act as neurotransmitters, while about 69\% function as co-transmitters. In humans, peptides play roles in pain perception, emotional modulation, and complex responses to stress. Peptide co-transmission can modulate synaptic activity, impacting functions such as food intake, metabolism, social behavior, learning, and memory \cite{russo2017verview,purves2019neurosciences}. In \textit{C.~elegans}, peptides influence a wide array of behaviors, including locomotion, dauer-stage formation, egg-laying, sleep, learning, social behavior, mechano- and chemosensation. Notably, some FLP family peptides are involved in feeding behavior \cite{li2018neuropeptides}, while others, like nlp-22, regulate \textit{C.~elegans}' sleep-like state (lethargus) during larval transitions\cite{nelson2013neuropeptide}. Neuropeptides, constituting the third layer of information flow in neuronal communication alongside chemical and electrical transmission, are still not fully understood in \textit{C.~elegans}, with over 300 neuropeptides identified \cite{ringstad2017neuromodulation,chew2018neuropeptides}. Similar to monoamines, neuropeptides form a small wired network but a vast wireless network\cite{bentley2016multilayer}, suggesting their potential involvement in all \textit{C.~elegans} behaviors.

In the subsequent sections, we present our findings based on this multi-layer network analysis, highlighting the discovery of regulatory motifs. This paves the way to understand their functional relevance, and the implications for understanding mechanisms governing the \textit{C.~elegans} nervous system. We anticipate that the insights gained from this comprehensive multi-layer network analysis will contribute significantly to the broader understanding of the \textit{C.~elegans} connectome and provide a valuable resource for the scientific community in further investigations that also invites the transfer of methodology to other model organisms and contexts.

\subsection{\textbf{\textit{E}}fficient \textbf{\textit{S}}ubgraph \textbf{\textit{C}}ounting \textbf{\textit{A}}lgorithmic Packag\textbf{\textit{E}}: \textit{ESCAPE}}
In a nutshell, the applied algorithmic framework ESCAPE revolves around breaking down complex patterns into smaller subpatterns and leveraging counts of these smaller patterns to obtain larger pattern counts. By employing this divide-and-conquer strategy, we can efficiently handle the enumeration and computation of subgraphs, even for large and complex networks. This approach allows us to avoid the excessive computational overhead associated with naively counting subgraphs.
A key feature of the algorithm \cite{pinar2017escape} is the exploitation of degree orientations within the network. By judiciously utilizing information about the degree distributions, we can further optimize runtime, resulting in faster computations. This optimization ensures that our algorithm can scale efficiently even for massive-sized networks.

\begin{figure*}[]
    \centering
    \includegraphics[height= 8cm, width=12cm]{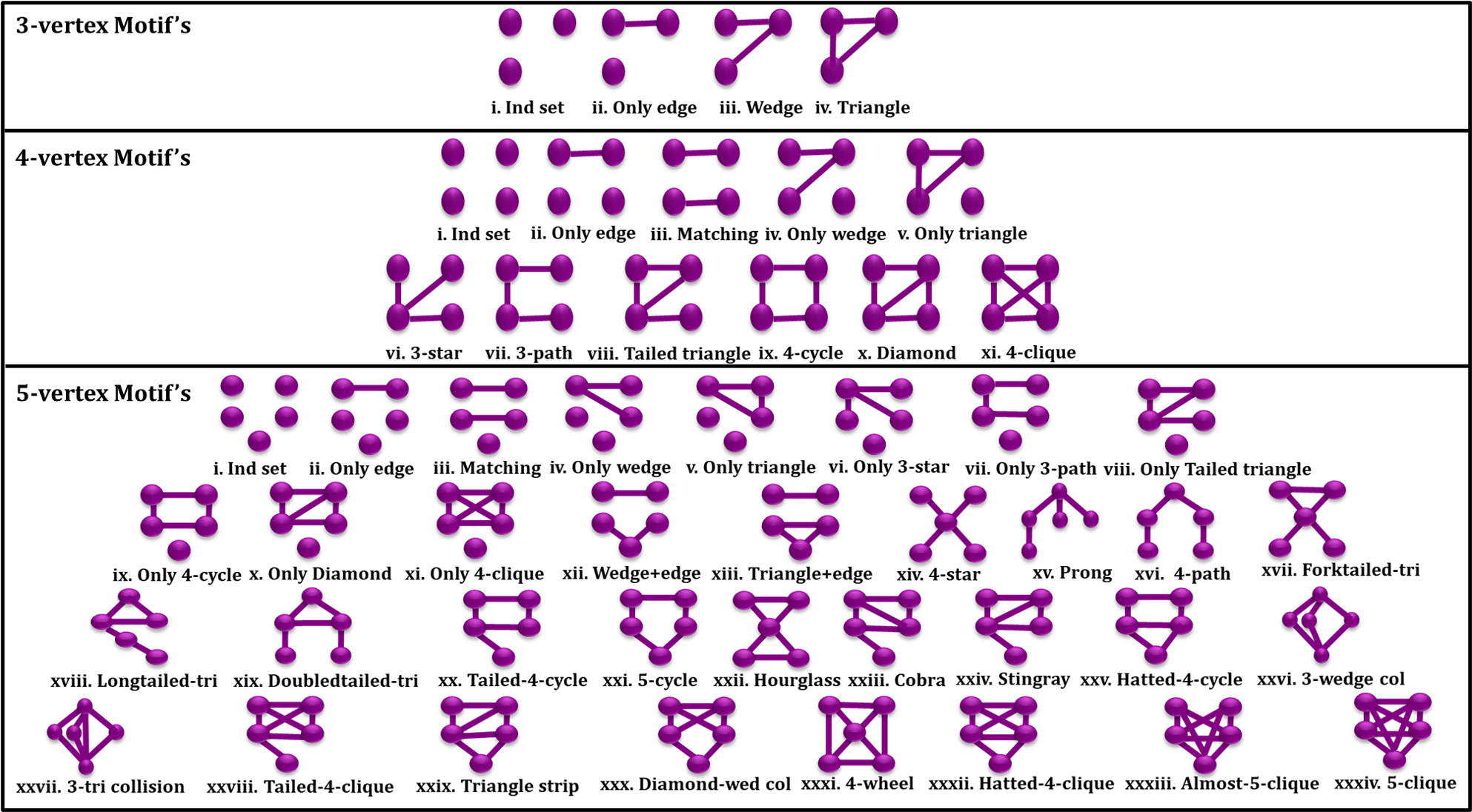}
    \caption{Complete list 3-, 4- and 5-node motifs including names for later identification.}
    \label{fig:allmotifs} 
\end{figure*}


\subsection{Motif Discovery Algorithm}
The enumeration of motifs, also called graphlets, particularly in the context of large-scale networks with millions of nodes and links, presents a formidable challenge due to the exponential growth in computational complexity. In the past, it was widely believed that subgraphs beyond three nodes were exceptionally difficult to enumerate, and conventional enumeration algorithms, which necessitate traversing each subgraph, were deemed impractical in terms of termination within a reasonable timeframe.

In response to the combinatorial explosion that hampers typical enumeration algorithms, the ESCAPE method was introduced by Pinar et al. \cite{pinar2017escape}. This approach adopts a divide-and-conquer strategy, effectively identifying substructures within each counting subgraph and partitioning them into smaller patterns. 
In other words, calculating all 5-node motifs counts requires enumerating only a small set of patterns. The corresponding formal framework \textit{cuts} a pattern into smaller subpatterns, which is the main reason for its practical feasibility. A degree ordering applied to this subsets of fundamental patterns significantly cuts down the combinatorial expansion and results in a practical technique that is sufficient to count all 5-vertex patterns.

Despite its broad applicability, the method allows for tailored decomposition choices that facilitate the derivation of a set of formulas to compute the frequency of each subgraph efficiently. While the original paper delineates the resulting formulas for subgraphs up to size 5, diligent efforts can extend the application to larger sizes. For further details, we refer to Ref.~\cite{pinar2017escape}.

To the best of our knowledge, ESCAPE stands as one of the most efficient algorithms available for counting undirected subgraphs and orbits up to size 5. Its strength lies in the ability to mitigate the exponential growth in complexity, enabling the enumeration of motifs even in sizable networks with millions of nodes and links. Through its divide-and-conquer paradigm and strategic substructure identification, ESCAPE opens new avenues for motif discovery, offering unprecedented opportunities to unravel recurring patterns within complex networks. Its scalability and performance make it a compelling choice for motif analysis in diverse domains, such as social networks, biological networks, and recommendation systems.

\subsection{Proposed Methodology}
\begin{figure*}[ht!]
    \centering
    \includegraphics[width=\linewidth]{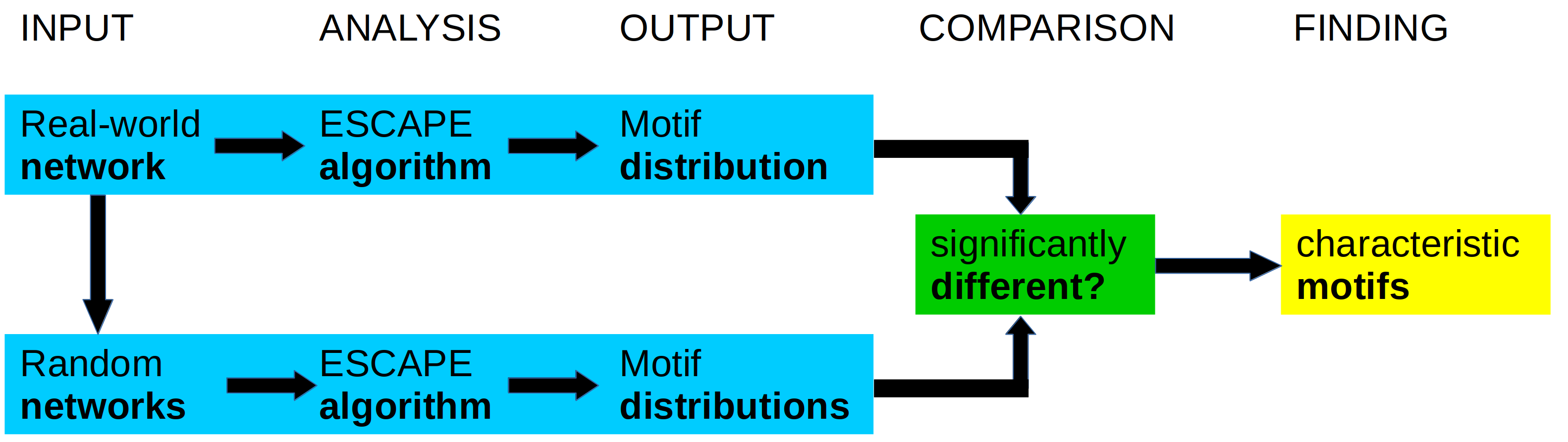}
    \caption{Pipeline for the motif discovery.}
    \label{fig:pipeline} 
\end{figure*}

Next, we describes the stepwise methodology of the proposed work for \textit{C.~elegans} motif discovery. As schematically depicted in Fig.~\ref{fig:pipeline}, the pipeline includes the following steps:
\begin{itemize}
   \myitem  \textbf{Step 1:} Input: Reading multi-layer network data and generation of random networks of the same size
    \myitem  \textbf{Step 2:} Analysis: Application of the ESCAPE algorithm
    \myitem  \textbf{Step 3:} Output: Motif distribution
    \myitem  \textbf{Step 3:} Comparison: Benchmarking empirical against random networks
    \myitem  \textbf{Step 4:} Comparison: Identification of significant motifs
\end{itemize}

The selection of random networks is one of many choices of network classes to generate artificial networks for benchmarking. Other prominent examples include regular, small-world, or scale-free networks, to name but a few. 
For the proper determination of a motif's subnetwork significance, we need comparison by an ensemble of a reference network class, e.g., random networks. Thus, the generation of this ensemble due to a given random network model is a necessary step of the algorithm. One of the popular random-network models on which we also focused is to preserve the degree sequence of the original network. Note there has been some research concerning the problem of subnetwork distribution within such networks for
directed sparse random networks \cite{maslov2002specificity,milo2003uniform}. 

We follow a randomizing approach to generate reference networks of the same number of nodes and links.  In that approach, similar to Milo's random model \cite{maslov2002specificity,milo2003uniform}, switching operations are repeatedly applied on the randomly chosen vertices of the input network and their links, until the network is well randomized. By applying these switching operations, an ensemble of random networks is generated for comparing the real network to obtain the significance of each motif.

For the significance of occurrence of each motif, statistical measures are introduced that lead us to the probable motifs in the input network:
(i) \textit{Frequency}: This is the simplest measurement for estimating the significance of a motif.
The frequency is defined as the number of occurrences of a motif $P$ in the network. 
(ii) \textit{KS test}: We employ the Kolmogorov-Smirnov (KS) test to identify significant and insignificant motif frequencies by comparing the input network's motif frequency with those from 1000 generated random networks. First, the motif $P$ frequency is counted in the input network. Then, the same motif is counted in each of the random networks, creating a distribution of motif frequencies for comparison. 
(iii) \textit{p-value}: The KS test compares the motif frequency distribution of the input network against the random networks' distributions, providing a p-value to indicate significance. A low p-value signifies that the motif frequency is significantly different from random, indicating the motif's significance, while a high p-value indicates insignificance. Therefore, p-values range from $0$ to $1$. The smaller the p-value, the more significant the motif. Here, we consider a p-value of $0.05$, below which we declare the occurrence of a motif as significant, that is, characteristic.


\clearpage
\section{Results and Performance Evaluation}
\label{sec:results}
 This section provides the details of dataset and performance indicators used for the experiment analysis.

\subsection{Data}
The dataset \cite{maertens2021multilayer} used in this study encompasses a comprehensive network representation of a biological system, capturing the interactions between various types of molecular entities. The data consists of nodes and links, representing different neuronal units and their connections within the biological system. The dataset contains a total of 279 nodes and 3,538 links. These nodes refer to neurons, each contributing to the overall functionality of the nervous system. The links in the network represent the connections or interactions between the neurons. A total of 3,538 links have been identified in the dataset across several layers (cf. Fig.~\ref{fig:connectome_layers}), signifying the complex web of interactions within the biological system.

\begin{table}[!ht]
\caption{Data description: Number of nodes and links for the different layers, the aggregated network (All), and the locomotory circuit (LC).}
\label{tab1}
\begin{tabular}{@{}lllllllll@{}}
\toprule
& \textbf{ACh}& \textbf{Electrical}& \textbf{GABA} & \textbf{Glu}& \textbf{MA}& \textbf{Peptides}& \textbf{All}& \textbf{LC}\\
\midrule
Nodes&258&253&102&197&131&162&279&83 \\ 
Links &1145&1028&133&727&212&455&3700&435 \\
Undirected links &1080&514&130&708&203&401&2287&435 \\
\bottomrule
\end{tabular}
\end{table}

Table~\ref{tab1} provides an overview of the data, showcasing the different layers and the number of neurons involved as nodes in the network and the number of links that connect them. Each network layer corresponds to a specific neurotransmitter or signaling molecule involved in the biological system under investigation. The last two columns represent the total number of nodes in the aggregated network and the locomotory circuit, which is added as a functionally relevant subnetwork. In total, there are 3700 aggregated links, which amounts to 3,538 distinct connections between the neurons. Note that the electrical layer is -- for reasons of neuro-physiology -- bi-directional. All other layers are uni-directional. For the purpose of this study, we ignore the directionality and treat all connections as  undirected. The last row of Tab.~\ref{tab1} shows the number of links in the considered undirected network.

\subsection{Network motifs}

Next, we present the results of our analysis on the multi-layer network of \textit{C.~elegans} using the ESCAPE algorithm \cite{pinar2017escape} for counting small subnetwork patterns. Specifically, we focus on motifs consisting of 3, 4, and 5 nodes and their prevalence across the different layers within the network. Additionally, we discuss the findings and implications of our motif counting analysis.

\subsubsection{3-node motifs}

Table~\ref{3node_nonind_motif} summarizes the number of occurrence of all possible motifs involving 3 nodes, from the \textit{independent set} (3 nodes with no links) to the \textit{triangle} (3-clique or all-to-all connected set of 3 nodes). Obviously, the former is given be the binomial coefficient ${N\choose 3}$, where $N$ denotes the number of nodes in the layer (cf. Tab.~\ref{tab1}). Evaluating the occurrences, it is important to note that a brute-force calculation of the motifs inevitably leads to double and multiple countings. For instance, a triangle automatically includes 3 motifs of \textit{wedge} type. Here, network science/graph theory comes to our aid and provides the notion of non-induced and induced subnetworks.

\begin{table}[!ht]
\caption{3-node non-induced motifs in \textit{C.~elegans} multi-layer network}
\label{3node_nonind_motif}
\begin{tabular}{@{}lccccccccc@{}}
\toprule
& \textbf{Motif}& \textbf{ACh}& \textbf{Electrical}& \textbf{GABA} & \textbf{Glu}& \textbf{MA}& \textbf{Peptides}& \textbf{All}& \textbf{LC}\\
\midrule
Ind set& \raisebox{-.5\totalheight}{\includegraphics[height=1cm,width=1cm]{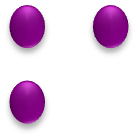}}&2829056&2667126&171700&1254890&366145&695520&3580779&91881\\ \hline
Only link&\raisebox{-.5\totalheight}{\includegraphics[height=1cm,width=1cm]{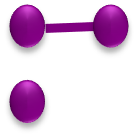}}&276480&129014&13000&138060&26187&64160&633499&35235\\ \hline
Wedge&\raisebox{-.5\totalheight}{\includegraphics[height=1cm,width=1cm]{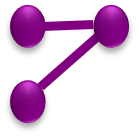}}&16848&3972&678&8035&1565&3028&56984&7056\\ \hline
Triangle&\raisebox{-.5\totalheight}{\includegraphics[height=1cm,width=1cm]{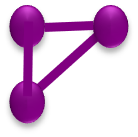}}&1083&170&18&376&28&125&4055&680\\ \hline
\end{tabular}
\end{table}

\begin{figure}[ht!]
\includegraphics[width=\textwidth]{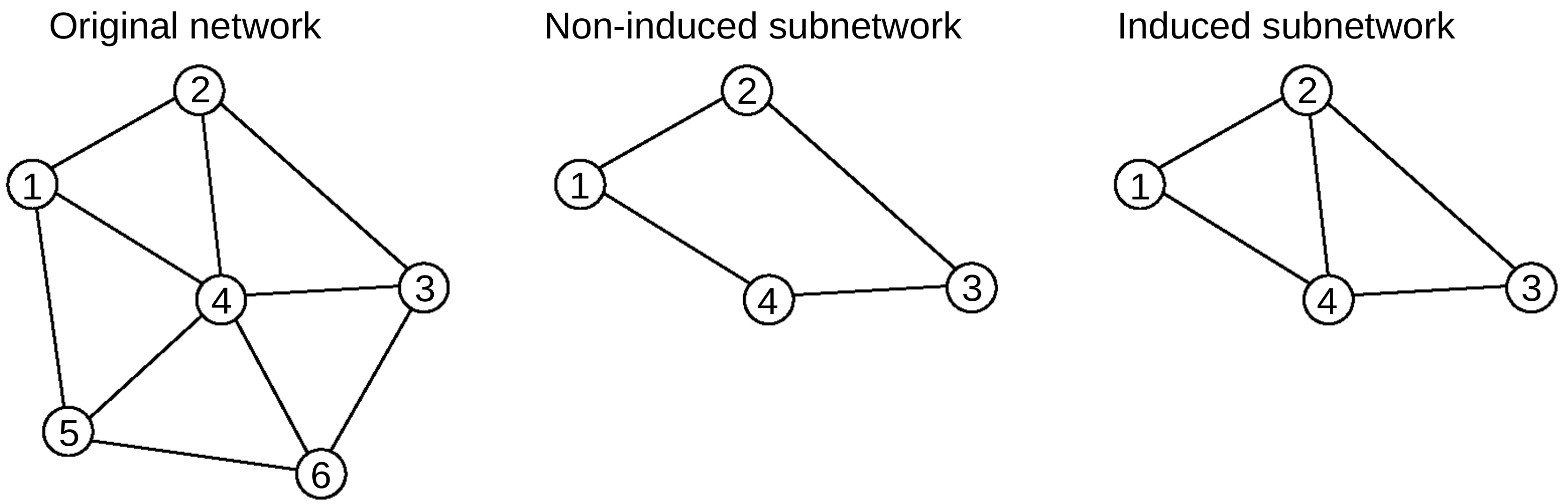}
\caption {Schematics of a 6-node network, a non-induced 4-node subnetwork, and an induced 4-node subnetwork. The latter consists of the maximum number of nodes and links taken from the original network.}
\label{fig:induced-vs-noninduced}
\end{figure}

As illustrated in Fig.~\ref{fig:induced-vs-noninduced}, a \textit{non-induced} subnetwork consists of a set of nodes, but not all possible edges that are present in the original network. In contrast, a motif is called \textit{induced} if it contains the maximum set of links connecting its nodes in the original network. In the schematic example, the link between nodes 2 and 4 is missing in the non-induced subnetwork of nodes $1, 2, 3$, and $4$, while it is accounted for in the induced one. One can express the induced as a linear combination of the non-induced count and vice versa, that is, by multiplying the respective count vectors with a transformation matrix (or its inverse) that provides the combinatorics of lower-motif inclusions. For 3- and 4-node motifs, it is given by
\begin{align}
    A_{3} =
    \left(
    \begin{array}{cccc}
        1 & 1 & 1 & 1\\
        0 & 1 & 2 & 3\\
        0 & 0 & 1 & 3\\
        0 & 0 & 0 & 1
    \end{array}
    \right),
    \qquad
    A_{4} =
    \left(
    \begin{array}{ccccccccccc}
        1&1&1&1&1&1&1&1&1&1&1\\
        0&1&2&2&3&3&3&4&4&5&6 \\
        0&0&1&0&0&0&1&1&2&2&3 \\
        0&0&0&1&3&3&2&5&4&8&12 \\
        0&0&0&0&1&0&0&1&0&2&4\\
        0&0&0&0&0&1&0&1&0&2&4\\
        0&0&0&0&0&0&1&2&4&6&12\\
        0&0&0&0&0&0&0&1&0&4&12\\
        0&0&0&0&0&0&0&0&1&1&3\\
        0&0&0&0&0&0&0&0&0&1&6 \\
        0&0&0&0&0&0&0&0&0&0&1
    \end{array}
    \right),
\end{align}
where the ordering of motifs corresponds to Fig.~\ref{fig:allmotifs}. For example, considering the last column of $A_{4}$, a \textit{4-clique} gives rise to the following non-induced motifs: 6 \textit{diamonds}, three \textit{4-cycles}, 12 \textit{tailed triangles} etc. For the case of connected 5-node motifs, that is, 21 out of the 34 motifs in Fig.~\ref{fig:allmotifs}, see the the 21$\times$21 transformation matrix $A_5$ given in Appendix~B of Ref.~\cite{pinar2017escape}.

\begin{table}[!ht]
\caption{3-node induced motifs in \textit{C.~elegans} multi-layer network}
\label{3node_ind_motif}
\begin{tabular}{@{}llcccccccc@{}}
\toprule
\textbf{Motif}& 
\textbf{Motif}& \textbf{ACh}& \textbf{Electrical}& \textbf{GABA} & \textbf{Glu}& \textbf{MA}& \textbf{Peptides}& \textbf{All}& \textbf{LC}\\
\midrule
Ind set& \raisebox{-.5\totalheight}{\includegraphics[height=1cm,width=1cm]{3_IndSet.png}}&2568341&2541914&159360&1124489&341495&634263&3000209&63022\\ \hline
Only link& \raisebox{-.5\totalheight}{\includegraphics[height=1cm,width=1cm]{3_Oneedge.png}}&246033&121580&11698&123118&23141&58479&531696&23163\\ \hline
Wedge&\raisebox{-.5\totalheight}{\includegraphics[height=1cm,width=1cm]{3_Wedge.png}}&13599&3462&624&6907&1481&2653&44819&5016\\ \hline
Triangle&\raisebox{-.5\totalheight}{\includegraphics[height=1cm,width=1cm]{3_Triangle.png}}
&1083&170&18&376&28&125&4055&680\\ \hline
\end{tabular}
\end{table}

Therefore, it is helpful to focus on the occurrences of induced motifs, which is given in Tab.~\ref{3node_ind_motif}. That way, the count reflects subnetworks that are not part of a motif with the same set of nodes connected with more links. One can quantify the different via the ratio of induced to non-induced motifs. As shown in Tab.~\ref{3node_nonind_motif_ratio} that ratio is rather high, that is close to unity. For instance, considering the ACh layer, we find that 13599 \textit{wedges} do not stem from \textit{triangles} with one link left out. In fact, one can easily verify that the difference between the number of \textit{wedges} of the induced to the non-induced case equals 3 times the number of triangles. In the following, we will consider only induced motifs.  For 4- and 5-node motifs, which will be discussed in the subsequent sections, the tables for the non-induced case for both the original and reference networks can be found in Appendices~\ref{app:4non-induced}, \ref{app:4non-induced_reference}, \ref{app:5non-induced}, and \ref{app:5non-induced_reference}.

\begin{table}[!ht]
\caption{Ratio of 3-node induced vs non-induced motifs in \textit{C.~elegans} multi-layer network}
\label{3node_nonind_motif_ratio}
\begin{tabular}{@{}lccccccccc@{}}
\toprule
&\textbf{Motif}& \textbf{ACh}& \textbf{Electrical}& \textbf{GABA} & \textbf{Glu}& \textbf{MA}& \textbf{Peptides}& \textbf{All}& \textbf{LC}\\
\midrule
Ind set&\raisebox{-.5\totalheight}{\includegraphics[height=1cm,width=1cm]{3_IndSet.png}}&0.91&0.95&0.93&0.9&0.93&0.91&0.84&0.69 \\ \hline
Only link&\raisebox{-.5\totalheight}{\includegraphics[height=1cm,width=1cm]{3_Oneedge.png}}&0.89&0.94&0.9&0.89&0.88&0.91&0.84&0.66 \\ \hline
Wedge&\raisebox{-.5\totalheight}{\includegraphics[height=1cm,width=1cm]{3_Wedge.png}}&0.81&0.87&0.92&0.86&0.95&0.88&0.79&0.71 \\ \hline
Triangle&\raisebox{-.5\totalheight}{\includegraphics[height=1cm,width=1cm]{3_Triangle.png}}&1.00&1.00&1.00&1.00&1.00&1.00&1.00&1.00 \\ \hline
\end{tabular}
\end{table}

\subsubsection{4-node motifs}
%
%

\begin{figure*}[ht!]
    \centering
    \includegraphics[width=\linewidth]{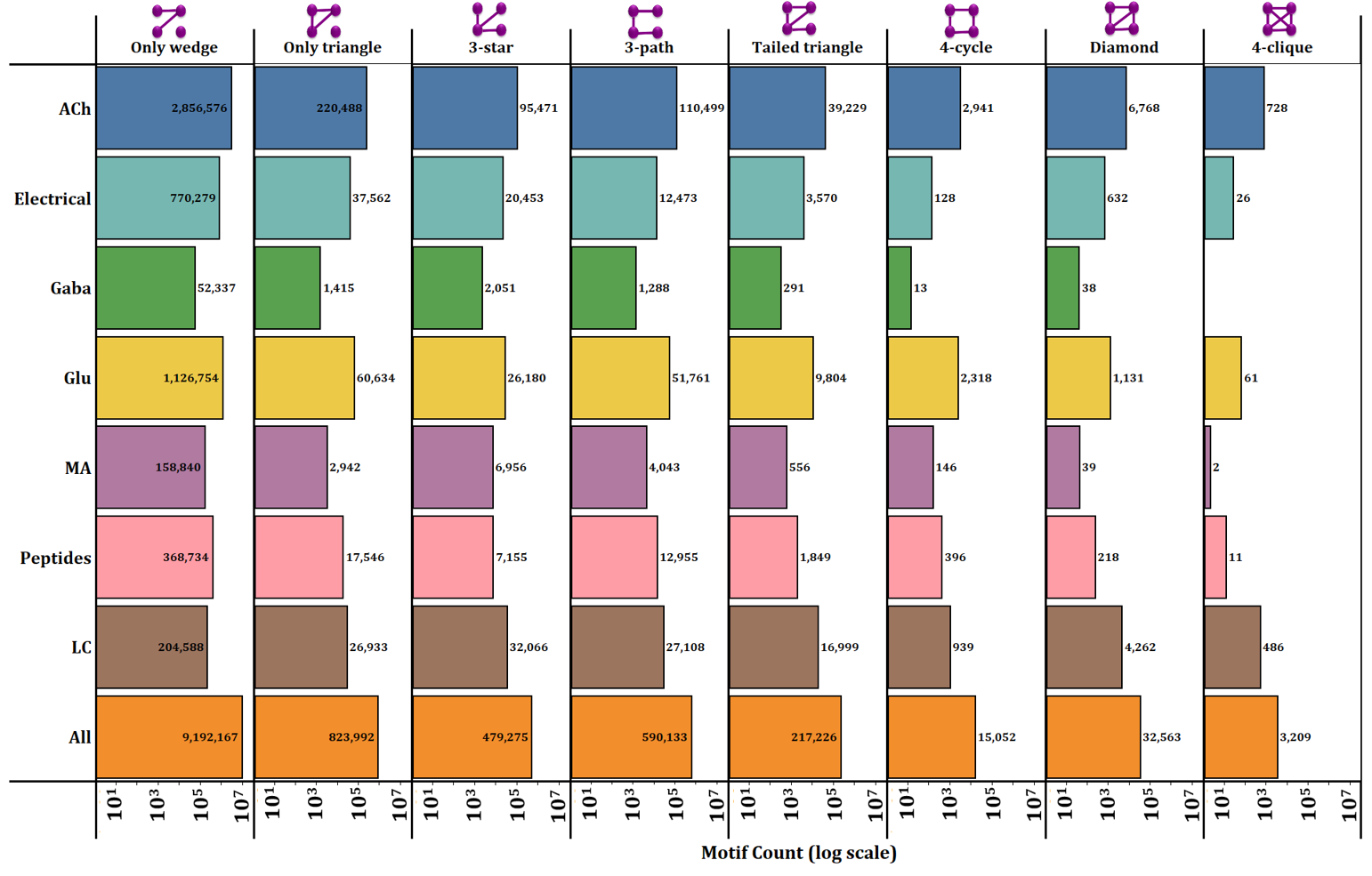}
    \caption{Histogram of induced 4-node motifs in each layer. See Tab.~\ref{Tab:4node_indvsnonind_motif} for a comparison to the non-induced counts given in Tab.~\ref{Tab:4node_nonind_motif}.}
    \label{fig:4induced}
\end{figure*}

Figure~\ref{fig:4induced} showcases the histogram of the counts of various 4-node motifs identified within the multi-layer network, classified by different molecular entity types, including ACh, Electrical, GABA, Glu, Monoamines, Peptides, and All (representing the total, aggregated network). Each motif type is represented by a corresponding graphical depiction. Note that the result of the independent set, only edge, and matching motifs are not shown. The respective numbers can be found in Appendix~\ref{app:4induced}. 

To highlight some numbers: The motif \textit{only edge} representing a wedge (an unconnected pair of nodes sharing a common neighbor) plus an independent node  is prevalent across all molecular entity types, with counts ranging from 2856576 in the ACh layer to 52337 in the GABA layer. In the aggregated network, we observe a total count of more than 9 million \textit{only wedge} motifs. The motifs \textit{only triangle} (three nodes interconnected in a closed loop plus a single independent node) are also frequently observed across the network layers, with counts varying from 220488 in the ACh layer to 1415 in the GABA layer and an overall count of 823992 in the combined-layer network. 

As the motif structure becomes more elaborate, the number of occurrences decreases. The \textit{diamond} (two pairs of nodes connected by a central node) and \textit{4-clique} (four fully interconnected nodes) motifs feature the least among all 4-node motifs. In fact, there is not a single occurrence of a \textit{4-clique} in the GABA layer.

\begin{table}[!ht]
    \centering
    \caption{Ratios of 4-node induced vs non-induced motifs in \textit{C.~elegans} multi-layer network. The values in bold indicate that the discovered induced motif count is not significant.} 
    \label{Tab:4node_indvsnonind_motif}
\begin{tabular}{@{}lccccccccc@{}}
\toprule
& \textbf{Motif} & \textbf{ACh}& \textbf{Electrical}& \textbf{GABA} & \textbf{Glu}& \textbf{MA}& \textbf{Peptides}& \textbf{All}& \textbf{LC}\\
\midrule
Ind set&\raisebox{-.5\totalheight}{\includegraphics[height=1cm,width=1cm]{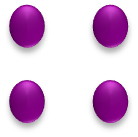}}&0.83&0.91&0.86&0.81&0.87&0.83&0.71&0.50 \\ \hline
Only link&\raisebox{-.5\totalheight}{\includegraphics[height=1cm,width=1cm]{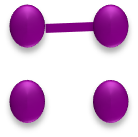}}&0.77&0.88&0.79&0.77&0.77&0.81&0.67&0.41 \\ \hline
Matching&\raisebox{-.5\totalheight}{\includegraphics[height=1cm,width=1cm]{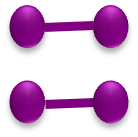}}&0.70&0.86&0.78&0.72&0.74&0.79&0.64&0.36 \\ \hline
Only wedge & \raisebox{-.5\totalheight}{\includegraphics[height=1cm,width=1cm]{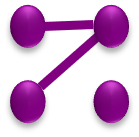}}&0.66&0.78&0.78&0.72&0.79&0.77&0.58&0.36 \\ \hline
Only triangle&\raisebox{-.5\totalheight}{\includegraphics[height=1cm,width=1cm]{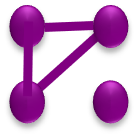}}&0.80&0.88&0.79&0.83&0.82&0.88&0.74&0.50 \\ \hline
3-star&\raisebox{-.5\totalheight}{\includegraphics[height=1cm,width=1cm]{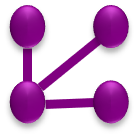}}&0.63&0.81&0.85&0.68&0.92&0.75&0.62&0.54 \\ \hline
3-path&\raisebox{-.5\totalheight}{\includegraphics[height=1cm,width=1cm]{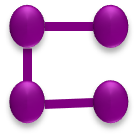}}&0.44&\textbf{0.51}&\textbf{0.60}&0.59&0.67&0.66&0.45&0.28 \\ \hline
Tailed triangle&\raisebox{-.5\totalheight}{\includegraphics[height=1cm,width=1cm]{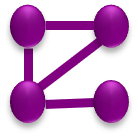}}&0.52&0.56&0.66&0.65&0.76&0.65&0.56&0.43 \\ \hline
4-cycle&\raisebox{-.5\totalheight}{\includegraphics[height=1cm,width=1cm]{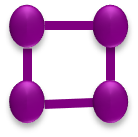}}&0.25&0.15&\textbf{0.25}&0.64&0.76&0.61&0.26&0.14 \\ \hline
Diamond&\raisebox{-.5\totalheight}{\includegraphics[height=1cm,width=1cm]{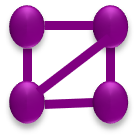}}&0.61&0.80&1.00&0.76&0.76&0.77&0.63&0.59 \\ \hline
4-clique&\raisebox{-.5\totalheight}{\includegraphics[height=1cm,width=1cm]{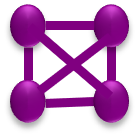}}&1.00&1.00&\textbf{-}&1.00&1.00&1.00&1.00&1.00 \\ \hline
\end{tabular}
\end{table}

The ratios of the induced and non-induced 4-node motifs are given in Table~\ref{Tab:4node_indvsnonind_motif} (See Appendix~\ref{app:4non-induced} for the absolute values of the non-induced motifs). We find that for sparser layers such as Glu and MA, motifs with more links do not tend to be part of higher-order motifs, that is, they are induced. See, for instance, ratios of 0.92 (0.63) or 0.76 (0.25) for \textit{3-star} and \textit{4-cycle} motifs in the MA (ACh) layer.

\begin{table}[!ht]
    \centering
    \caption{Percentage of 4-node induced motifs in \textit{C.~elegans} multi-layer network relative to ALL (aggregated network). The values in bold indicate that the discovered motif count is not significant.} 
    \label{Tab:Percentage_4node_motif}
\begin{tabular}{lccccccc}
\toprule
& \textbf{Motif}& \textbf{ACh}& \textbf{Electrical}& \textbf{GABA} & \textbf{Glu}& \textbf{MA}& \textbf{Peptides}\\
\midrule
Ind set&\raisebox{-.5\totalheight}{\includegraphics[height=1cm,width=1cm]{4_IndSet.png}}&85.29&86.57&2.1&28.03&5.85&13.17\\ \hline
Only link&\raisebox{-.5\totalheight}{\includegraphics[height=1cm,width=1cm]{4_Oneedge.png}}&46.24&23.98&0.87&17.51&2.18&6.98\\ \hline
Matching&\raisebox{-.5\totalheight}{\includegraphics[height=1cm,width=1cm]{4_Matching.png}}&23.98&6.7&0.37&10.55&0.85&3.72\\ \hline
Only wedge &\raisebox{-.5\totalheight}{\includegraphics[height=1cm,width=1cm]{4_OnlyWedge.png}}&31.08&8.38&0.57&12.26&1.73&4.01\\ \hline
Only triangle&\raisebox{-.5\totalheight}{\includegraphics[height=1cm,width=1cm]{4_OnlyTriangle.png}}&26.76&4.56&0.17&7.36&0.36&2.13\\ \hline
3-star&\raisebox{-.5\totalheight}{\includegraphics[height=1cm,width=1cm]{4_3-Star.png}}&19.92&4.27&0.43&5.46&1.45&1.49\\ \hline
3-path&\raisebox{-.5\totalheight}{\includegraphics[height=1cm,width=1cm]{4_3-Star.png}}&18.72&\textbf{2.11}&\textbf{0.22}&8.77&0.69&2.20\\ \hline
Tailed triangle&\raisebox{-.5\totalheight}{\includegraphics[height=1cm,width=1cm]{4_TailedTriangle.png}}&18.06&1.64&0.13&4.51&0.26&0.85\\ \hline
4-cycle&\raisebox{-.5\totalheight}{\includegraphics[height=1cm,width=1cm]{4_4-Cycle.png}}&19.54&0.85&\textbf{0.09}&15.4&0.97&2.63\\ \hline
Diamond&\raisebox{-.5\totalheight}{\includegraphics[height=1cm,width=1cm]{4_Diamond.png}}&20.78&1.94&0.12&3.47&0.12&0.67\\ \hline
4-clique&\raisebox{-.5\totalheight}{\includegraphics[height=1cm,width=1cm]{4_4-Clique.png}}&22.69&0.81&\textbf{0.00}&1.9&0.06&0.34\\ 
\botrule
\end{tabular}
\end{table}

The results of the occurrences of motifs in each layered compared to the aggregated network are summarized in Tab.~\ref{Tab:Percentage_4node_motif}. Here, one can see the difference between edges formed by electrical gap junctions and chemical synapses. Although the electrical layer form the second-largest layer, in terms of both number of nodes and edges, its contribution to the motif count is rather small. In other works, most of the motifs are found in the chemical layers. Furthermore, the ACh layer, while represented at almost 50\% of the connections in the aggregated network (cf. Tab.~\ref{tab1}) do not feature that prominently in the motif count. This means that many chemical links involve more than just a single neurotransmitter.

\begin{figure}[htbp]
    \begin{center}
    \begin{subfigure}[h]{0.48\textwidth}
    \includegraphics[width=\textwidth]{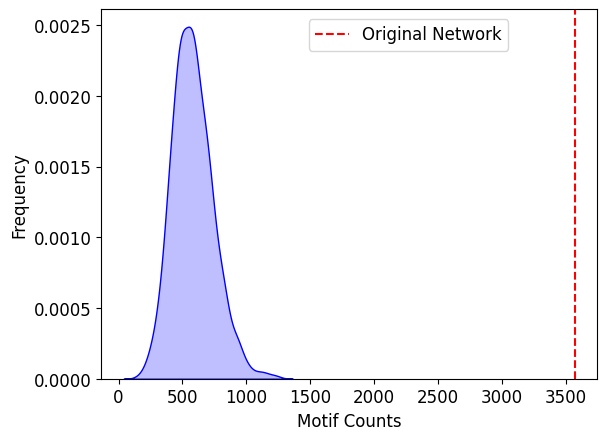}
    \caption{Significance of tailed-triangle motif in the electrical layer as shown in Fig.~\ref{fig:4induced} and Table~\ref{Tab:4nodemotif}.}
    \end{subfigure}
    \hfill
    \begin{subfigure}[h]{0.48\textwidth}
    \includegraphics[width=\textwidth]{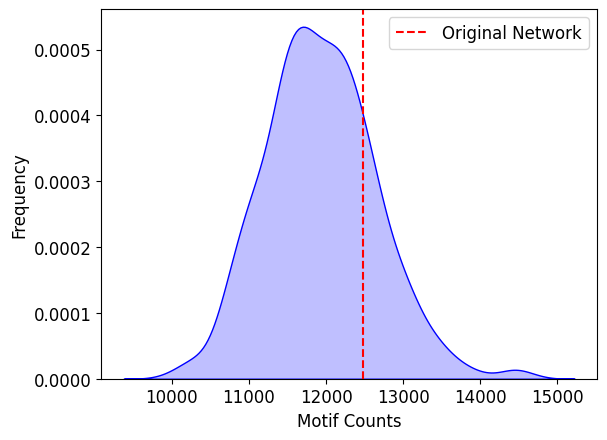}
    \caption{Insignificance of 3-path motif in electrical layer as shown in  Fig.~\ref{fig:4induced} and Table~\ref{Tab:4nodemotif}.}
    \end{subfigure} 
    \end{center}
\caption{Examples for a significant and an insignificant 4-node induced motif in the electrical layer of original network (cf. Fig.~\ref{fig:4induced}) as compared to the random reference networks. }
\label{fig:4node-examples}
\end{figure}

In order to identify the occurrences of significant motifs, we compare the motif distributions of the original network to 1000 random networks with the same number of nodes and edges and repeat the motif analysis. The resulting average counts of the different induced and non-induced 4-node motifs are given in 
Appendices~\ref{app:4non-induced_reference} and \ref{app:4induced_reference}, respectively. As an exemplary illustration, Fig.~\ref{fig:4node-examples} depicts the case of a significant motif (\textit{tailed triangle}) in panel (a) and an insignificant motif (\textit{3-path}) in panel (b). One can clearly see that the number of \textit{tailed triangle} motifs (3570) lies outside the distribution calculated for the random networks, while the the number of \textit{3-path} motifs (12473) can also be found in a substantial amount of random reference networks. Interestingly, we find only a few cases of insignificant induced motifs. These are indicated by bold values in Tabs.~\ref{Tab:4node_indvsnonind_motif}, \ref{Tab:Percentage_4node_motif}, and \ref{Tab:4nodemotif}. Moreover, in almost all cases of significant motifs, the counts extracted from the original network is far away from those of the random reference networks.

\clearpage
\subsubsection{5-node motifs}
\label{subsec:5-node_motifs´}
Next, we turn our attention to motifs that involve 3 nodes, and focus on those subnetwork structures that cannot be part of a 4-node motif with 1 additional disconnected node. This includes 23 out of 34 possible 5-node motifs (cf. Fig.~\ref{fig:allmotifs}). Note that we still include the \textit{wedge+edge} and \textit{triangle+edge} motif.

Table~\ref{Tab:Percentage_5node_induced_motif} summarizes the findings in terms of the percentage of 5-node induced motifs relative the aggregated network case. Again, we find that almost all motifs are significant in their occurrence and cannot be expected from random reference networks. It is noteworthy that except the \textit{wedge+link} motif, the overwhelming majority of insignificant motif counts appear for very small fractions below 1 percent. In those cases, their existence is unlikely in random networks as well. The absolute values and numbers for non-induced 5-node motifs can be found in Appendices~\ref{app:5non-induced} to \ref{app:5induced_random} for the sake of completeness, but are not displayed here to improve readability. 

\small{
\begin{longtable}{lccccccc}
    \caption{Percentage of 5-node induced motifs in \textit{C.~elegans} multi-layer network relative to ALL (aggregated network). The values in bold indicate that the discovered motif count is not significant.} \\
\toprule
& \textbf{Motif}& \textbf{ACh}& \textbf{Electrical}& \textbf{GABA} & \textbf{Glu}& \textbf{MA}& \textbf{Peptides}\\
\midrule
\endhead
\hline
\multicolumn{8}{l}{Continued}\\   \bottomrule
\endfoot
\bottomrule
\endlastfoot
\label{Tab:Percentage_5node_induced_motif} 
Wedge+link&\raisebox{-.5\totalheight}{\includegraphics[height=1cm,width=1cm]{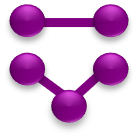}}&16.15&2.67&0.10&\textbf{5.80}&0.35&1.49\\ \hline
Triangle+link&\raisebox{-.5\totalheight}{\includegraphics[height=1cm,width=1cm]{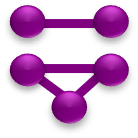}}&13.08&1.43&0.03&3.33&0.08&0.78\\ \hline
4-star&\raisebox{-.5\totalheight}{\includegraphics[height=1cm,width=1cm]{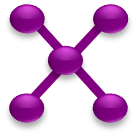}}&12.38&2.33&0.1&1.41&0.50&0.28\\ \hline
Prong&\raisebox{-.5\totalheight}{\includegraphics[height=1cm,width=1cm]{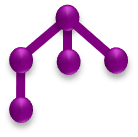}}&12.55&1.01&0.08&3.60&0.31&0.66\\ \hline
4-path&\raisebox{-.5\totalheight}{\includegraphics[height=1cm,width=1cm]{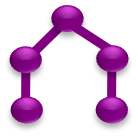}}&11.41&\textbf{0.80}&\textbf{0.04}&5.33&0.21&0.93\\ \hline
Forktailed-tri&\raisebox{-.5\totalheight}{\includegraphics[height=1cm,width=1cm]{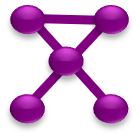}}&11.25&0.91&0.04&1.42&0.11&0.16\\ \hline
Longtailed-tri&\raisebox{-.5\totalheight}{\includegraphics[height=1cm,width=1cm]{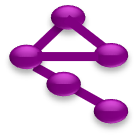}}&10.15&0.48&0.02&2.86&0.05&0.39\\ \hline
Doubletailed-tri&\raisebox{-.5\totalheight}{\includegraphics[height=1cm,width=1cm]{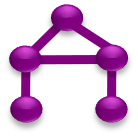}}&12.77&0.46&0.04&2.51&0.12&0.27\\ \hline
Tailed-4-cycle&\raisebox{-.5\totalheight}{\includegraphics[height=1cm,width=1cm]{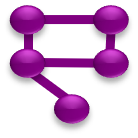}}&11.72&0.4&\textbf{0.02}&6.95&0.35&\textbf{0.82}\\ \hline
5-cycle&\raisebox{-.5\totalheight}{\includegraphics[height=1cm,width=1cm]{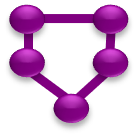}}&8.76&\textbf{0.33}&\textbf{0.01}&4.11&0.14&0.63\\ \hline
Hourglass&\raisebox{-.5\totalheight}{\includegraphics[height=1cm,width=1cm]{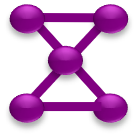}}&7.30&0.31&0.01&1.58&0.02&0.15\\ \hline
Cobra&\raisebox{-.5\totalheight}{\includegraphics[height=1cm,width=1cm]{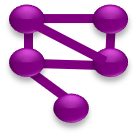}}&13.39&0.41&0.02&2.22&\textbf{0.01}&0.24\\ \hline
Stingray&\raisebox{-.5\totalheight}{\includegraphics[height=1cm,width=1cm]{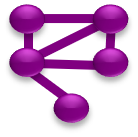}}&13.56&0.80&0.03&1.29&0.06&0.12\\ \hline
Hatted-4-cycle&\raisebox{-.5\totalheight}{\includegraphics[height=1cm,width=1cm]{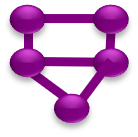}}&9.55&0.16&\textbf{0.01}&3.72&0.08&\textbf{0.39}\\ \hline
3-wedge-col&\raisebox{-.5\totalheight}{\includegraphics[height=1cm,width=1cm]{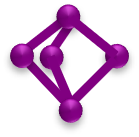}}&18.5&\textbf{0.16}&\textbf{0.02}&20.50&0.97&\textbf{1.40}\\ \hline
3-tri-collision&\raisebox{-.5\totalheight}{\includegraphics[height=1cm,width=1cm]{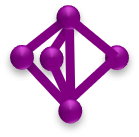}}&15.83&1.96&0.04&0.72&0.02&0.09\\ \hline
Tailed-4-clique&\raisebox{-.5\totalheight}{\includegraphics[height=1cm,width=1cm]{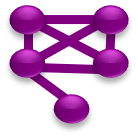}}&15.66&0.28&\textbf{0.00}&0.95&\textbf{0.00}&0.07\\ \hline
Triangle-strip&\raisebox{-.5\totalheight}{\includegraphics[height=1cm,width=1cm]{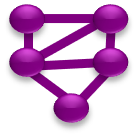}}&12.06&0.31&0.01&1.31&\textbf{0.00}&0.13\\ \hline
Diamond-wed-col&\raisebox{-.5\totalheight}{\includegraphics[height=1cm,width=1cm]{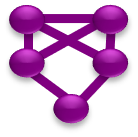}}&13.78&0.15&0.01&5.88&0.04&0.51\\ \hline
4-wheel&\raisebox{-.5\totalheight}{\includegraphics[height=1cm,width=1cm]{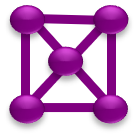}}&16.19&0.21&0.01&3.11&\textbf{0.00}&0.22\\ \hline
Hatted-4-clique&\raisebox{-.5\totalheight}{\includegraphics[height=1cm,width=1cm]{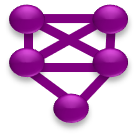}}&15.89&0.46&\textbf{0.00}&0.56&0.00&0.08\\ \hline
Almost-5-clique&\raisebox{-.5\totalheight}{\includegraphics[height=1cm,width=1cm]{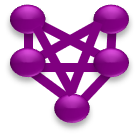}}&19.34&0.13&\textbf{0.00}&0.45&0.01&0.01\\ \hline
5-clique&\raisebox{-.5\totalheight}{\includegraphics[height=1cm,width=1cm]{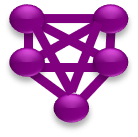}}&23.11&\textbf{0.00}&\textbf{0.00}&0.06&\textbf{0.00}&\textbf{0.00}\\ 
\end{longtable}
}

\begin{figure*}[ht!]
    \centering
    \includegraphics[width=12cm, height=18cm]{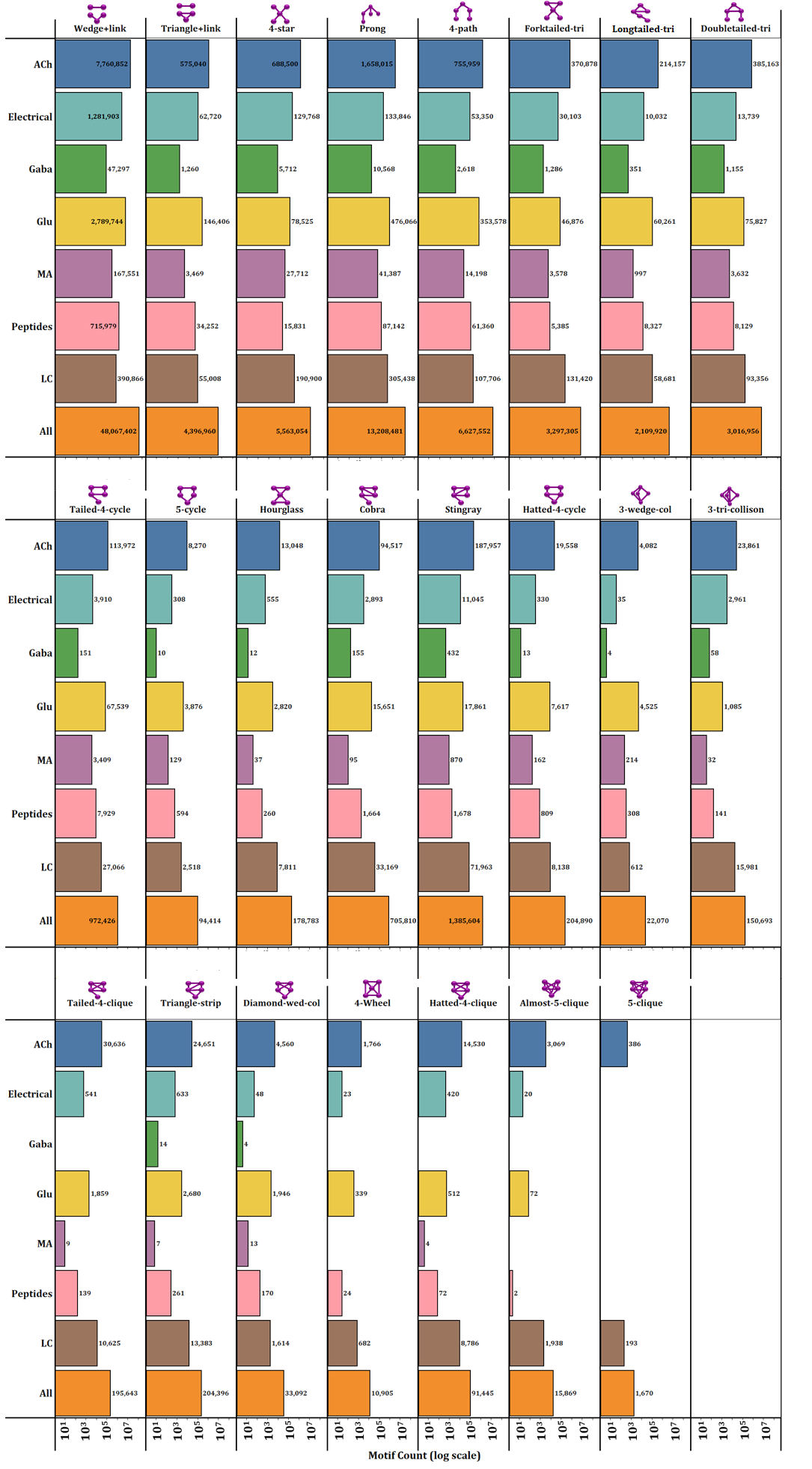}
    \caption{Histogram of induced 5-node motifs in each layer. See Tab.~\ref{Tab:5node_indVsnonind_motif} for a comparison to the non-induced counts given in Tab.~\ref{Tab:5node_nonind_motif}.}
    \label{fig:5induced}
\end{figure*}

The exact numbers are also included in Fig.~\ref{fig:5induced} in addition to the histograms. We find that in general, more complex motifs are present in lower numbers and are completely absent in small layers. At the same time, layers with more nodes and links have the tendency to allow for these motifs, e.g., \textit{4-wheel},  \textit{hatted-4-clique}, \textit{almost-5-clique}, or \textit{5-clique} (see bottom panel of Fig.~\ref{fig:5induced}), and the corresponding motifs remain significant to characterize the layer.

\clearpage
\section{Conclusions}
\label{sec:conclusions}
We have presented the application of a motif-discovery algorithm for network analysis. As a case study, the algorithm has been applied to the multi-layer connectome of \textit{C.~elegans}. We have considered motifs formed by 3, 4, or 5 nodes. Our analysis of the motif occurrences has yielded several important findings. First, the abundance of wedge motifs indicates the prevalence of shared regulators across different layers corresponding to different molecular entities (neurotransmitters), reflecting the existence of conserved regulatory mechanisms. Additionally, the presence of motifs with closed paths such as triangles, 4- or 5-cycles and their various variants with extensions to the core cycle suggests the occurrence of regulatory feedback. Similarly, stars of spokes-and-hub motifs point to central regulators in the network.

To identify significant motifs, we have compared the original network to an ensemble of 1000 random networks of the same number of nodes and links, that is, same link density. For almost all layers and motif types, we find that the occurrence of motifs cannot be explained by a random network model.  Notably, the 3-path and 4-path motifs stand out as an exception of insignificant motifs in the electrical and GABA layer, which are not suited to characterize the network at hand. Interestingly, the presence of 4-cycles and diamonds across multiple layers implies recurrent regulatory patterns and interconnected information pathways. Conversely,  4- and 5-clique motifs feature less prominently, which suggests that fully interconnected regulatory clusters are less prevalent in the network.

The proposed approach can be easily extended in several directions: (i) It invites the characterization of significant and insignificant motifs in other empirical networks from various backgrounds. (ii) We have chosen random networks are reference case. The pipeline, however, allows to consider other classes of networks as well. That way, a network extracted from an empirical dataset can be fingerprinted and uniquely characterized by its composition of small network substructures. (iii) Information on the motif composition can also inform generative models to generate ensembles of realstic networks that can be used in simulations. (iv) We have focused on the occurrences of motifs in different layers. The general concept of these small network structures, however, can be extended to configuration where links are part of diffent layers. That way, information pathways across layers become accessible.

\section*{Acknowledgment}
We are grateful to Karlheinz Ochs, Sebastian Jenderny, and Christian Beth for stimulating discussions. This work was funded by the Deutsche Forschungsgemeinschaft (DFG--German Research Foundation) under Project ID: 434434223-SFB 1461.



\clearpage
\appendix
\begin{appendices}

\section{Results for non-induced 4-node motifs in the original network}
\label{app:4non-induced}
\begin{sidewaystable}[!ht]
    \centering
    \caption{4-node non-induced motifs in \textit{C.~elegans} multi-layer network. The values in bold indicate that the discovered motif count is not significant.} 
    \label{Tab:4node_nonind_motif}
\begin{tabular}{@{}lccccccccc@{}}
\toprule
& \textbf{Motif} & \textbf{ACh}& \textbf{Electrical}& \textbf{GABA} & \textbf{Glu}& \textbf{MA}& \textbf{Peptides}& \textbf{All}& \textbf{LC}\\
\midrule
Ind set & \raisebox{-.5\totalheight}{\includegraphics[height=1cm,width=1cm]{4_IndSet.png}} & 180352320&166695375&4249575&60862165&11716640&27646920&247073751&1837620\\ \hline
Only link & \raisebox{-.5\totalheight}{\includegraphics[height=1cm,width=1cm]{4_Oneedge.png}} & 35251200&16126750&643500&13391820&1675968&5100720&87422862&1409400\\ \hline
Matching & \raisebox{-.5\totalheight}{\includegraphics[height=1cm,width=1cm]{4_Matching.png}}& 565812&127869&7707&242243&18938&77172&2557057&87339\\ \hline
Only wedge & \raisebox{-.5\totalheight}{\includegraphics[height=1cm,width=1cm]{4_OnlyWedge.png}}&4296240&993000&67122&1558790&200320&481452&15727584&564480\\ \hline
Only triangle&\raisebox{-.5\totalheight}{\includegraphics[height=1cm,width=1cm]{4_OnlyTriangle.png}}&276165&42500&1782&72944&3584&19875&1119180&54400\\ \hline
3-star&\raisebox{-.5\totalheight}{\includegraphics[height=1cm,width=1cm]{4_3-Star.png}}&151148&25391&2418&38490&7598&9484&774463&59533\\ \hline
3-path&\raisebox{-.5\totalheight}{\includegraphics[height=1cm,width=1cm]{4_3-Path.png}}&250065&24229&2150&88159&5997&19677&1318679&96266\\ \hline
Tailed triangle&\raisebox{-.5\totalheight}{\includegraphics[height=1cm,width=1cm]{4_TailedTriangle.png}}&75037&6410&443&15060&736&2853&385986&39879\\ \hline
4-cycle&\raisebox{-.5\totalheight}{\includegraphics[height=1cm,width=1cm]{4_4-Cycle.png}}&11893&838&51&3632&191&647&57242&6659\\ \hline
Diamond&\raisebox{-.5\totalheight}{\includegraphics[height=1cm,width=1cm]{4_Diamond.png}}&11136&788&38&1497&51&284&51817&7178\\ \hline
4-clique&\raisebox{-.5\totalheight}{\includegraphics[height=1cm,width=1cm]{4_4-Clique.png}}&728&26&\textbf{0$^{\star \dagger}$}&61&2&11&3209&486\\ \hline
\multicolumn{10}{l}{\textbf{$\star$ indicates link-swapping random network generation method}} \\
\multicolumn{10}{l}{\textbf{$\dagger$ indicates gnm random network generation method}}
\end{tabular}
\end{sidewaystable}

\begin{table}[!ht]
    \centering
    \caption{Percentage of 4-node non-induced motifs in the \textit{C.~elegans} multi-layer network relative to ALL.} 
    \label{Tab:Percentage_4node_noninduced_motif}
\begin{tabular}{lccccccc}
\toprule
\textbf{Motif}& \textbf{Motif} &\textbf{ACh}& \textbf{Electrical}& \textbf{GABA} & \textbf{Glu}& \textbf{MA}& \textbf{Peptides}\\
\midrule
Ind set&\raisebox{-.5\totalheight}{\includegraphics[height=1cm,width=1cm]{4_IndSet.png}}&73.00&67.47&1.72&24.63&4.74&11.19\\ \hline
Only link&\raisebox{-.5\totalheight}{\includegraphics[height=1cm,width=1cm]{4_Oneedge.png}}&40.32&18.45&0.74&15.32&1.92&5.83\\ \hline
Matching&\raisebox{-.5\totalheight}{\includegraphics[height=1cm,width=1cm]{4_Matching.png}}&22.13&5.00&0.30&9.47&0.74&3.02\\ \hline
Only wedge& \raisebox{-.5\totalheight}{\includegraphics[height=1cm,width=1cm]{4_OnlyWedge.png}}&27.32&6.31&0.43&9.91&1.27&3.06\\ \hline
Only triangle&\raisebox{-.5\totalheight}{\includegraphics[height=1cm,width=1cm]{4_OnlyTriangle.png}}&24.68&3.80&0.16&6.52&0.32&1.78\\ \hline
3-star&\raisebox{-.5\totalheight}{\includegraphics[height=1cm,width=1cm]{4_3-Star.png}}&19.52&3.28&0.31&4.97&0.98&1.22\\ \hline
3-path&\raisebox{-.5\totalheight}{\includegraphics[height=1cm,width=1cm]{4_3-Path.png}}&18.96&1.84&0.16&6.69&0.45&1.49\\ \hline
Tailed triangle&\raisebox{-.5\totalheight}{\includegraphics[height=1cm,width=1cm]{4_TailedTriangle.png}}&19.44&1.66&0.11&3.9&0.19&0.74\\ \hline
4-cycle&\raisebox{-.5\totalheight}{\includegraphics[height=1cm,width=1cm]{4_4-Cycle.png}}&20.78&1.46&0.09&6.34&0.33&1.13\\ \hline
Diamond&\raisebox{-.5\totalheight}{\includegraphics[height=1cm,width=1cm]{4_Diamond.png}}&21.49&1.52&0.07&2.89&0.10&0.55\\ \hline
4-clique&\raisebox{-.5\totalheight}{\includegraphics[height=1cm,width=1cm]{4_4-Clique.png}}&22.69&0.81&0.00&1.9&0.06&0.34\\ 

\botrule
\end{tabular}
\end{table}

\clearpage
\section{Results for non-induced 4-node motifs in the random reference networks}
\label{app:4non-induced_reference}
\small{
\begin{sidewaystable}[!ht]
    \centering
    \caption{Average of 4-node non-induced motifs in the \textit{C.~elegans} multi-layer network using link-swapping random network generation method.} 
    \label{Tab:4nodemotif_avg_es_nonind}
\begin{tabular}{@{}lccccccccc@{}}
\toprule
 & \textbf{Motif} & \textbf{ACh}& \textbf{Electrical}& \textbf{GABA} & \textbf{Glu}& \textbf{MA}& \textbf{Peptides}& \textbf{All}& \textbf{LC}\\
\midrule
Ind set & \raisebox{-.5\totalheight}{\includegraphics[height=1cm,width=1cm]{4_IndSet.png}}
&\shortstack{180352320\\$\pm$0.00}
&\shortstack{166695375\\$\pm$0.00}
&\shortstack{4249575\\$\pm$0.00}
&\shortstack{60862165\\$\pm$0.00}
&\shortstack{11716640\\$\pm$0.00}
&\shortstack{27646920\\$\pm$0.00}
&\shortstack{247073751\\$\pm$0.00}
&\shortstack{1837620\\$\pm$0.00}
\\ \hline

Only link & \raisebox{-.5\totalheight}{\includegraphics[height=1cm,width=1cm]{4_Oneedge.png}}
&\shortstack{35251200\\$\pm$0.00}
&\shortstack{16126750\\$\pm$0.00}
&\shortstack{643500\\$\pm$0.00}
&\shortstack{13391820\\$\pm$0.00}
&\shortstack{1675968\\$\pm$0.00}
&\shortstack{5100720\\$\pm$0.00}
&\shortstack{87422862\\$\pm$0.00}
&\shortstack{1409400\\$\pm$0.00}
\\ \hline

Matching &\raisebox{-.5\totalheight}{\includegraphics[height=1cm,width=1cm]{4_Matching.png}} 
&\shortstack{571090.02\\$\pm$233.64}
&\shortstack{129108.29\\$\pm$105.09}
&\shortstack{7942.18\\$\pm$32.88}
&\shortstack{244268.16\\$\pm$116.50}
&\shortstack{19560.14\\$\pm$56.45}
&\shortstack{77882.15\\$\pm$64.37}
&\shortstack{2570562.06\\$\pm$470.75}
&\shortstack{89234.01\\$\pm$118.40}
\\ \hline

Only wedge & \raisebox{-.5\totalheight}{\includegraphics[height=1cm,width=1cm]{4_OnlyWedge.png}}
&\shortstack{2950345.67\\$\pm$59576.94}
&\shortstack{683178.75\\$\pm$26272.43}
&\shortstack{43839.48\\$\pm$3255.48}
&\shortstack{1165909.15\\$\pm$22601.23}
&\shortstack{120686.59\\$\pm$7225.18}
&\shortstack{368538.47\\$\pm$10235.52}
&\shortstack{12000186.61\\$\pm$129926.40}
&\shortstack{412879.12\\$\pm$9471.76} 
\\ \hline

Only triangle&\raisebox{-.5\totalheight}{\includegraphics[height=1cm,width=1cm]{4_OnlyTriangle.png}}
&\shortstack{55786.86\\$\pm$5167.57}
&\shortstack{6630.75\\$\pm$1455.54}
&\shortstack{542.82\\$\pm$237.64}
&\shortstack{20756.64\\$\pm$2276.22}
&\shortstack{1428.10\\$\pm$469.40}
&\shortstack{5347.33\\$\pm$974.25}
&\shortstack{325675.03\\$\pm$12114.61}
&\shortstack{21905.92\\$\pm$1619.39}
\\ \hline

3-star&\raisebox{-.5\totalheight}{\includegraphics[height=1cm,width=1cm]{4_3-Star.png}}
&\shortstack{57968.20\\$\pm$4106.35}
&\shortstack{8549.89\\$\pm$1218.69}
&\shortstack{806.81\\$\pm$184.07}
&\shortstack{19700.29\\$\pm$1090.82}
&\shortstack{2429.84\\$\pm$404.29}
&\shortstack{5117.26\\$\pm$405.77}
&\shortstack{348002.15\\$\pm$14307.91}
&\shortstack{24723.02\\$\pm$1944.71}
\\ \hline

3-path&\raisebox{-.5\totalheight}{\includegraphics[height=1cm,width=1cm]{4_3-Path.png}}
&\shortstack{120828.99\\$\pm$4828.86}
&\shortstack{13480.70\\$\pm$954.50}
&\shortstack{1294.32\\$\pm$186.02}
&\shortstack{50174.57\\$\pm$1963.08}
&\shortstack{3519.88\\$\pm$402.83}
&\shortstack{12890.50\\$\pm$711.18}
&\shortstack{808285.87\\$\pm$16326.50}
&\shortstack{58105.20\\$\pm$2330.79}
\\ \hline

Tailed triangle&\raisebox{-.5\totalheight}{\includegraphics[height=1cm,width=1cm]{4_TailedTriangle.png}}
&\shortstack{9503.82\\$\pm$1301.51}
&\shortstack{638.43\\$\pm$197.70}
&\shortstack{76.56\\$\pm$41.38}
&\shortstack{3116.37\\$\pm$425.14}
&\shortstack{203.12\\$\pm$84.34}
&\shortstack{636.73\\$\pm$137.01}
&\shortstack{80973.83\\$\pm$4902.58}
&\shortstack{10970.35\\$\pm$1243.83}
\\ \hline

4-cycle&\raisebox{-.5\totalheight}{\includegraphics[height=1cm,width=1cm]{4_4-Cycle.png}}&\shortstack{1636.25\\$\pm$151.31}
&\shortstack{90.42\\$\pm$17.88}
&\shortstack{11.56\\$\pm$4.85}
&\shortstack{660.71\\$\pm$60.88}
&\shortstack{34.10\\$\pm$9.57}
&\shortstack{133.00\\$\pm$18.32}
&\shortstack{15690.92\\$\pm$640.10}
&\shortstack{2256.32\\$\pm$176.21}
\\ \hline

Diamond&\raisebox{-.5\totalheight}{\includegraphics[height=1cm,width=1cm]{4_Diamond.png}}
&\shortstack{378.59\\$\pm$105.57}
&\shortstack{15.05\\$\pm$11.25}
&\shortstack{1.90\\$\pm$2.50}
&\shortstack{99.69\\$\pm$29.13}
&\shortstack{4.71\\$\pm$4.19}
&\shortstack{15.89\\$\pm$7.81}
&\shortstack{3889.99\\$\pm$432.92}
&\shortstack{952.85\\$\pm$182.31}
\\ \hline

4-clique&\raisebox{-.5\totalheight}{\includegraphics[height=1cm,width=1cm]{4_4-Clique.png}}
&\shortstack{7.29\\$\pm$4.82}
&\shortstack{0.16\\$\pm$0.42}
&\shortstack{0.02\\$\pm$0.14}
&\shortstack{1.34\\$\pm$1.41}
&\shortstack{0.03\\$\pm$0.19}
&\shortstack{0.16\\$\pm$0.41}
&\shortstack{76.74\\$\pm$17.40}
&\shortstack{30.31\\$\pm$10.68} \\ 
\botrule
\end{tabular}
\end{sidewaystable}
}

\small{
\begin{landscape}
\begin{ThreePartTable}
\begin{longtable}{llllllllll}
    \caption{Average of 4-node non-induced motifs in the \textit{C.~elegans} multi-layer network using the gnm random network generation method.} \\
\toprule
 & \textbf{Motif} & \textbf{ACh}& \textbf{Electrical}& \textbf{GABA} & \textbf{Glu}& \textbf{MA}& \textbf{Peptides}& \textbf{All}& \textbf{LC}\\
\midrule
\endhead \hline
\multicolumn{10}{l}{Continued}\\   \bottomrule
\endfoot
\bottomrule
\endlastfoot
\label{Tab:4nodemotif_avg_gnm_nonind} 
Ind set & \raisebox{-.5\totalheight}{\includegraphics[height=1cm,width=1cm]{4_IndSet.png}} &
\shortstack{180352320\\$\pm$0.00}
&\shortstack{166695375\\$\pm$0.00}
&\shortstack{4249575\\$\pm$0.00}
&\shortstack{60862165\\$\pm$0.00}
&\shortstack{11716640\\$\pm$0.00}
&\shortstack{27646920\\$\pm$0.00}
&\shortstack{247073751\\$\pm$0.00}
&\shortstack{1837620\\$\pm$0.00}\\ \hline

Only link &\raisebox{-.5\totalheight}{\includegraphics[height=1cm,width=1cm]{4_Oneedge.png}} &
\shortstack{35251200\\$\pm$0.00}
&\shortstack{16126750\\$\pm$0.00}
&\shortstack{643500\\$\pm$0.00}
&\shortstack{13391820\\$\pm$0.00}
&\shortstack{1675968\\$\pm$0.00}
&\shortstack{5100720\\$\pm$0.00}
&\shortstack{87422862\\$\pm$0.00}
&\shortstack{1409400\\$\pm$0.00}\\ \hline

Matching & \raisebox{-.5\totalheight}{\includegraphics[height=1cm,width=1cm]{4_Matching.png}} &
\shortstack{573662.21\\$\pm$89.64}
&\shortstack{129764.87\\$\pm$43.98}
&\shortstack{8059.33\\$\pm$16.58}
&\shortstack{245221.81\\$\pm$68.17}
&\shortstack{19882.79\\$\pm$23.08}
&\shortstack{78230.90\\$\pm$41.15}
&\shortstack{2576705.72\\$\pm$168.58}
&\shortstack{89900.10\\$\pm$58.82}\\ \hline

Only wedge & \raisebox{-.5\totalheight}{\includegraphics[height=1cm,width=1cm]{4_OnlyWedge.png}}
&\shortstack{2294435.43\\$\pm$22859.18}
&\shortstack{519031.75\\$\pm$10995.09}
&\shortstack{32241.83\\$\pm$1640.99}
&\shortstack{980901.25\\$\pm$13224.56}
&\shortstack{79387.39\\$\pm$2954.83}
&\shortstack{313086.74\\$\pm$6543.56}
&\shortstack{10304537.28\\$\pm$46527.26}
&\shortstack{359591.92\\$\pm$4705.35}
\\ \hline

Only triangle&\raisebox{-.5\totalheight}{\includegraphics[height=1cm,width=1cm]{4_OnlyTriangle.png}}
&\shortstack{24740.87\\$\pm$2491.19}
&\shortstack{2779.00\\$\pm$827.21}
&\shortstack{266.90\\$\pm$161.03}
&\shortstack{11962.43\\$\pm$1550.42}
&\shortstack{636.42\\$\pm$273.73}
&\shortstack{3178.89\\$\pm$723.04}
&\shortstack{202457.32\\$\pm$7504.90}
&\shortstack{15230.8\\$\pm$1101.68}
\\ \hline

3-star&\raisebox{-.5\totalheight}{\includegraphics[height=1cm,width=1cm]{4_3-Star.png}}
&\shortstack{24865.06\\$\pm$764.47}
&\shortstack{2778.34\\$\pm$183.01}
&\shortstack{266.77\\$\pm$43.41}
&\shortstack{11960.71\\$\pm$492.54}
&\shortstack{622.66\\$\pm$73.81}
&\shortstack{3194.55\\$\pm$210.40}
&\shortstack{202300.62\\$\pm$2769.87}
&\shortstack{15259.81\\$\pm$604.97}
\\ \hline

3-path&\raisebox{-.5\totalheight}{\includegraphics[height=1cm,width=1cm]{4_3-Path.png}}
&\shortstack{74597.74\\$\pm$1498.80}
&\shortstack{8342.30\\$\pm$363.81}
&\shortstack{801.35\\$\pm$84.02}
&\shortstack{35872.17\\$\pm$967.87}
&\shortstack{1870.06\\$\pm$144.04}
&\shortstack{9584.51\\$\pm$408.64}
&\shortstack{607050.52\\$\pm$5479.48}
&\shortstack{45778.27\\$\pm$1178.04}
\\ \hline

Tailed triangle&\raisebox{-.5\totalheight}{\includegraphics[height=1cm,width=1cm]{4_TailedTriangle.png}}
&\shortstack{2408.34\\$\pm$265.23}
&\shortstack{134.05\\$\pm$42.38}
&\shortstack{19.57\\$\pm$13.02}
&\shortstack{1308.95\\$\pm$189.15}
&\shortstack{44.64\\$\pm$21.32}
&\shortstack{292.01\\$\pm$74.46}
&\shortstack{35757.42\\$\pm$1492.20}
&\shortstack{5803.37\\$\pm$524.11}
\\ \hline

4-cycle&\raisebox{-.5\totalheight}{\includegraphics[height=1cm,width=1cm]{4_4-Cycle.png}}
&\shortstack{605.81\\$\pm$33.94}
&\shortstack{33.49\\$\pm$6.42}
&\shortstack{5.09\\$\pm$2.39}
&\shortstack{327.86\\$\pm$23.72}
&\shortstack{10.99\\$\pm$3.66}
&\shortstack{73.12\\$\pm$10.61}
&\shortstack{8935.31\\$\pm$183.55}
&\shortstack{1450.8\\$\pm$82.48}
\\ \hline

Diamond&\raisebox{-.5\totalheight}{\includegraphics[height=1cm,width=1cm]{4_Diamond.png}}
&\shortstack{39.05\\$\pm$10.21}
&\shortstack{1.07\\$\pm$1.22}
&\shortstack{0.25\\$\pm$0.56}
&\shortstack{23.91\\$\pm$8.64}
&\shortstack{0.48\\$\pm$0.83}
&\shortstack{4.47\\$\pm$3.14}
&\shortstack{1052.04\\$\pm$89.69}
&\shortstack{366.25\\$\pm$63.17}
\\ \hline

4-clique&\raisebox{-.5\totalheight}{\includegraphics[height=1cm,width=1cm]{4_4-Clique.png}}
&\shortstack{0.21\\$\pm$0.45}
&\shortstack{0.00\\$\pm$0.04}
&\shortstack{0.00\\$\pm$0.00}
&\shortstack{0.15\\$\pm$0.41}
&\shortstack{0.00\\$\pm$0.03}
&\shortstack{0.03\\$\pm$0.18}
&\shortstack{10.32\\$\pm$3.42}
&\shortstack{7.70\\$\pm$3.48} \\ 
\end{longtable}
\end{ThreePartTable}
\end{landscape}
}

\clearpage
\section{Results for induced 4-node motifs in the original network}
\label{app:4induced}
\begin{sidewaystable}[!ht]
    \centering
    \caption{4-node induced motifs in \textit{C.~elegans} multi-layer network. The values in bold indicate that the discovered motif count is not significant.} 
    \label{Tab:4nodemotif}
\begin{tabular}{@{}lccccccccc@{}}
\toprule
& \textbf{Motif} & \textbf{ACh}& \textbf{Electrical}& \textbf{GABA} & \textbf{Glu}& \textbf{MA}& \textbf{Peptides}& \textbf{All}& \textbf{LC}\\
\midrule
Ind set &\raisebox{-.5\totalheight}{\includegraphics[height=1cm,width=1cm]{4_IndSet.png}} & 149362316&151603860&3675010&49089041&10243629&23059015&175117828&909686\\ \hline
Only link & \raisebox{-.5\totalheight}{\includegraphics[height=1cm,width=1cm]{4_Oneedge.png}}& 27262822&14136164&511106&10320884&1285524&4117934&58957465&583181\\ \hline
Matching & \raisebox{-.5\totalheight}{\includegraphics[height=1cm,width=1cm]{4_Matching.png}}& 394482&110228&6026&173597&13963&61107&1644841&31372\\ \hline

Only wedge & \raisebox{-.5\totalheight}{\includegraphics[height=1cm,width=1cm]{4_OnlyWedge.png}}&2856576&770279&52337&1126754&158840&368734&9192167&204588\\ \hline
Only triangle&\raisebox{-.5\totalheight}{\includegraphics[height=1cm,width=1cm]{4_OnlyTriangle.png}} &220488&37562&1415&60634&2942&17546&823992&26933\\ \hline
3-star&\raisebox{-.5\totalheight}{\includegraphics[height=1cm,width=1cm]{4_3-Star.png}} &95471&20453&2051&26180&6956&7155&479275&32066\\ \hline
3-path&\raisebox{-.5\totalheight}{\includegraphics[height=1cm,width=1cm]{4_3-Path.png}}&110499&\textbf{12473$^{\star}$}&\textbf{1288$^{\star}$}&51761&4043&12955&590133&27108\\ \hline
Tailed triangle&\raisebox{-.5\totalheight}{\includegraphics[height=1cm,width=1cm]{4_TailedTriangle.png}}&39229&3570&291&9804&556&1849&217226&16999\\ \hline
4-cycle&\raisebox{-.5\totalheight}{\includegraphics[height=1cm,width=1cm]{4_4-Cycle.png}}&2941&128&\textbf{13$^{\star}$}&2318&146&396&15052&939\\ \hline
Diamond&\raisebox{-.5\totalheight}{\includegraphics[height=1cm,width=1cm]{4_Diamond.png}}&6768&632&38&1131&39&218&32563&4262\\ \hline
4-clique&\raisebox{-.5\totalheight}{\includegraphics[height=1cm,width=1cm]{4_4-Clique.png}}&728&26&\textbf{0$^{\star \dagger}$}&61&2&11&3209&486\\
\bottomrule
\multicolumn{10}{l}{\textbf{$\star$ indicates link-swapping random network generation method}} \\
\multicolumn{10}{l}{\textbf{$\dagger$ indicates gnm random network generation method}}
\end{tabular}
\end{sidewaystable}

\clearpage
\section{Results for induced 4-node motifs in the random reference networks}
\label{app:4induced_reference}
\small{
\begin{sidewaystable}[!ht]
    \centering
    \caption{Average of 4-node induced motifs in the \textit{C.~elegans} multi-layer network using link-swapping random network generation method.} 
    \label{Tab:4nodemotif_avg_es_ind}
\begin{tabular}{@{}lccccccccc@{}}
\toprule
& \textbf{Motif} & \textbf{ACh}& \textbf{Electrical}& \textbf{GABA} & \textbf{Glu}& \textbf{MA}& \textbf{Peptides}& \textbf{All}& \textbf{LC}\\
\midrule

Ind set & \raisebox{-.5\totalheight}{\includegraphics[height=1cm,width=1cm]{4_IndSet.png}}
&\shortstack{148398740.4\\$\pm$48937.91}
&\shortstack{151352964.65\\$\pm$23787.07}
&\shortstack{3655298.94\\$\pm$2826.60}
&\shortstack{48793569.55\\$\pm$18748.68}
&\shortstack{10173773.45\\$\pm$6306.41}
&\shortstack{22970019.54\\$\pm$8817.63}
&\shortstack{172832526.13\\$\pm$96589.37}
&\shortstack{837903.13\\$\pm$5211.09}
\\ \hline

Only link & \raisebox{-.5\totalheight}{\includegraphics[height=1cm,width=1cm]{4_Oneedge.png}}
&\shortstack{28869369.70\\$\pm$88678.47}
&\shortstack{14585318.88\\$\pm$45356.39}
&\shortstack{547525.49\\$\pm$5292.90}
&\shortstack{10828741.94\\$\pm$34193.21}
&\shortstack{1416682.48\\$\pm$11824.25}
&\shortstack{4274943.55\\$\pm$16426.30}
&\shortstack{62359584.30\\$\pm$164862.59}
&\shortstack{671051.85\\$\pm$7349.14}
\\ \hline

Matching &\raisebox{-.5\totalheight}{\includegraphics[height=1cm,width=1cm]{4_Matching.png}} 
&\shortstack{462302.05\\$\pm$3888.41}
&\shortstack{116417.22\\$\pm$919.21}
&\shortstack{6743.77\\$\pm$186.90}
&\shortstack{198336.03\\$\pm$1718.71}
&\shortstack{16302.25\\$\pm$387.14}
&\shortstack{65863.11\\$\pm$673.64}
&\shortstack{1867082.11\\$\pm$12089.58}
&\shortstack{44797.03\\$\pm$1325.12}
\\ \hline

Only wedge & \raisebox{-.5\totalheight}{\includegraphics[height=1cm,width=1cm]{4_OnlyWedge.png}}
&\shortstack{22418545.40\\$\pm$35651.79}
&\shortstack{614110.72\\$\pm$20417.14}
&\shortstack{37615.95\\$\pm$2319.37}
&\shortstack{961632.53\\$\pm$14366.94}
&\shortstack{103187.73\\$\pm$5161.09}
&\shortstack{314954.25\\$\pm$7249.85}
&\shortstack{8800017.15\\$\pm$56788.75}
&\shortstack{213399.83\\$\pm$1752.16}
\\ \hline

Only triangle&\raisebox{-.5\totalheight}{\includegraphics[height=1cm,width=1cm]{4_OnlyTriangle.png}}
&\shortstack{47011.05\\$\pm$4139.10}
&\shortstack{6021.80\\$\pm$1296.22}
&\shortstack{470.00\\$\pm$203.38}
&\shortstack{17834.27\\$\pm$1921.69}
&\shortstack{1234.26\\$\pm$398.85}
&\shortstack{4741.71\\$\pm$855.93}
&\shortstack{252174.21\\$\pm$8434.97}
&\shortstack{12720.05\\$\pm$806.12}
\\ \hline
3-star&\raisebox{-.5\totalheight}{\includegraphics[height=1cm,width=1cm]{4_3-Star.png}}
&\shortstack{49192.39\\$\pm$3381.56}
&\shortstack{7940.94\\$\pm$1139.68}
&\shortstack{733.99\\$\pm$167.96}
&\shortstack{16777.92\\$\pm$887.06}
&\shortstack{2236.01\\$\pm$369.50}
&\shortstack{4511.65\\$\pm$351.11}
&\shortstack{274501.33\\$\pm$11171.21}
&\shortstack{15537.14\\$\pm$1249.31}
\\ \hline
3-path&\raisebox{-.5\totalheight}{\includegraphics[height=1cm,width=1cm]{4_3-Path.png}}
&\shortstack{97460.37\\$\pm$2779.32}
&\shortstack{11930.59\\$\pm$725.25}
&\shortstack{1106.20\\$\pm$136.94}
&\shortstack{41881.00\\$\pm$1333.25}
&\shortstack{3005.12\\$\pm$285.43}
&\shortstack{11178.37\\$\pm$537.60}
&\shortstack{605993.56\\$\pm$8033.02}
&\shortstack{32492.64\\$\pm$654.38}
\\ \hline
Tailed triangle&\raisebox{-.5\totalheight}{\includegraphics[height=1cm,width=1cm]{4_TailedTriangle.png}}
&\shortstack{8076.97\\$\pm$950.64}
&\shortstack{580.11\\$\pm$161.09}
&\shortstack{69.15\\$\pm$34.03}
&\shortstack{2733.74\\$\pm$330.61}
&\shortstack{184.67\\$\pm$71.09}
&\shortstack{575.16\\$\pm$113.08}
&\shortstack{66334.76\\$\pm$3441.67}
&\shortstack{7522.62\\$\pm$655.38}
\\ \hline
4-cycle&\raisebox{-.5\totalheight}{\includegraphics[height=1cm,width=1cm]{4_4-Cycle.png}}
&\shortstack{1279.54\\$\pm$91.48}
&\shortstack{75.84\\$\pm$14.80}
&\shortstack{9.70\\$\pm$4.14}
&\shortstack{565.05\\$\pm$48.52}
&\shortstack{29.48\\$\pm$8.17}
&\shortstack{117.61\\$\pm$16.08}
&\shortstack{12031.15\\$\pm$389.79}
&\shortstack{1394.39\\$\pm$92.55}
\\ \hline
Diamond&\raisebox{-.5\totalheight}{\includegraphics[height=1cm,width=1cm]{4_Diamond.png}}
&\shortstack{334.83\\$\pm$82.04}
&\shortstack{14.11\\$\pm$10.24}
&\shortstack{1.80\\$\pm$2.25}
&\shortstack{91.63\\$\pm$23.90}
&\shortstack{4.52\\$\pm$3.92}
&\shortstack{14.90\\$\pm$6.82}
&\shortstack{3429.55\\$\pm$344.29}
&\shortstack{771.02\\$\pm$126.93}
\\ \hline
4-clique&\raisebox{-.5\totalheight}{\includegraphics[height=1cm,width=1cm]{4_4-Clique.png}}
&\shortstack{7.29\\$\pm$4.82}
&\shortstack{0.16\\$\pm$0.42}
&\shortstack{0.02\\$\pm$0.14}
&\shortstack{1.34\\$\pm$1.41}
&\shortstack{0.03\\$\pm$0.19}
&\shortstack{0.16\\$\pm$0.41}
&\shortstack{76.74\\$\pm$17.40}
&\shortstack{30.31\\$\pm$10.68}
\\
\botrule
\end{tabular}
\end{sidewaystable}
}

\small{
\begin{sidewaystable}[!ht]
    \centering
    \caption{Average of 4-node induced motifs in the \textit{C.~elegans} multi-layer network using the gnm random network generation method.} 
    \label{Tab:4nodemotif_avg_gnm_ind}
\begin{tabular}{@{}lccccccccc@{}}
\toprule
& \textbf{Motif} & \textbf{ACh}& \textbf{Electrical}& \textbf{GABA} & \textbf{Glu}& \textbf{MA}& \textbf{Peptides}& \textbf{All}& \textbf{LC}\\
\midrule
Ind set & \raisebox{-.5\totalheight}{\includegraphics[height=1cm,width=1cm]{4_IndSet.png}}
&\shortstack{147847989.29\\$\pm$20017.98}
&\shortstack{151203688.45\\$\pm$10336.32}
&\shortstack{3645065.54\\$\pm$1479.88}
&\shortstack{48638285.79\\$\pm$11392.78}
&\shortstack{10136868.19\\$\pm$2674.92}
&\shortstack{22921920.4\\$\pm$5751.07}
&\shortstack{171563974.56\\$\pm$36841.32}
&\shortstack{808338.75\\$\pm$2749.56}
\\ \hline

Only link & \raisebox{-.5\totalheight}{\includegraphics[height=1cm,width=1cm]{4_Oneedge.png}}
&\shortstack{29875753.09\\$\pm$37600.77}
&\shortstack{14870190.88\\$\pm$20116.82}
&\shortstack{566805.37\\$\pm$2831.83}
&\shortstack{11112531.25\\$\pm$21240}
&\shortstack{1486594.95\\$\pm$5120.39}
&\shortstack{4364520.16\\$\pm$10841.47}
&\shortstack{64522228.72\\$\pm$65758.43}
&\shortstack{711991.01\\$\pm$4098.07}
\\ \hline

Matching & \raisebox{-.5\totalheight}{\includegraphics[height=1cm,width=1cm]{4_Matching.png}}
&\shortstack{502606.98\\$\pm$1435.32}
&\shortstack{121621.46\\$\pm$390.27}
&\shortstack{7287.23\\$\pm$93.82}
&\shortstack{211266.93$\pm$926.73}
&\shortstack{18078.39\\$\pm$155.42}
&\shortstack{69075.81\\$\pm$407.95}
&\shortstack{2021210.12\\$\pm$4711.66}
&\shortstack{52117.39\\$\pm$817.93}
\\ \hline

Only wedge & \raisebox{-.5\totalheight}{\includegraphics[height=1cm,width=1cm]{4_OnlyWedge.png}}
&\shortstack{2010577.25 \\ $\pm$ 17088.42}
&\shortstack{486470.77 \\$\pm$ 9709.98}
&\shortstack{29154.33\\$\pm$1342.49}
&\shortstack{845054.16\\$\pm$9569.94}
&\shortstack{72133.35\\$\pm$2412.16}
&\shortstack{276514.61\\$\pm$4978.44}
&\shortstack{8082398.31\\$\pm$28192.74}
&\shortstack{208545.94\\$\pm$1489.13}
\\ \hline

Only triangle&\raisebox{-.5\totalheight}{\includegraphics[height=1cm,width=1cm]{4_OnlyTriangle.png}}
&\shortstack{22409.78 \\$\pm$ 2251.02}
&\shortstack{2647.09\\$\pm$788.21}
&\shortstack{247.84\\$\pm$149.64}
&\shortstack{10700.71\\$\pm$1381.60}
&\shortstack{592.73\\$\pm$254.78}
&\shortstack{2895.68\\$\pm$656.18}
&\shortstack{168762.69\\$\pm$6223.75}
&\shortstack{10129.15\\$\pm$716.04}
\\ \hline
3-star&\raisebox{-.5\totalheight}{\includegraphics[height=1cm,width=1cm]{4_3-Star.png}}&\shortstack{22533.98\\$\pm$681.52}
&\shortstack{2646.43\\$\pm$175.64}
&\shortstack{247.70\\$\pm$40.50}
&\shortstack{10699.00\\$\pm$433.63}
&\shortstack{578.98\\$\pm$68.30}
&\shortstack{2911.34\\$\pm$188.43}
&\shortstack{168605.99\\$\pm$2310.03}
&\shortstack{10158.16\\$\pm$397.68}
\\ \hline
3-path&\raisebox{-.5\totalheight}{\includegraphics[height=1cm,width=1cm]{4_3-Path.png}}
&\shortstack{67589.56 \\$\pm$1243.26}
&\shortstack{7946.66\\$\pm$336.59}
&\shortstack{743.34\\$\pm$74.25}
&\shortstack{32084.52\\$\pm$788.98}
&\shortstack{1739.69\\$\pm$125.84}
&\shortstack{8734.42\\$\pm$341.76}
&\shortstack{505982.83\\$\pm$4004.72}
&\shortstack{30473.47\\$\pm$565.05}
\\ \hline
Tailed triangle&\raisebox{-.5\totalheight}{\includegraphics[height=1cm,width=1cm]{4_TailedTriangle.png}}
&\shortstack{2254.69 \\$\pm$ 234.36}
&\shortstack{129.78\\$\pm$39.73}
&\shortstack{18.57\\$\pm$11.69}
&\shortstack{1215.06\\$\pm$162.99}
&\shortstack{42.73\\$\pm$19.62}
&\shortstack{274.53\\$\pm$66.38}
&\shortstack{31673.12\\$\pm$1182.17}
&\shortstack{4430.73\\$\pm$314.52}
\\ \hline
4-cycle&\raisebox{-.5\totalheight}{\includegraphics[height=1cm,width=1cm]{4_4-Cycle.png}}
&\shortstack{567.40 \\$\pm$32.38}
&\shortstack{32.42\\$\pm$6.30}
&\shortstack{4.84\\$\pm$2.32}
&\shortstack{304.39\\$\pm$22.03}
&\shortstack{10.51\\$\pm$3.61}
&\shortstack{68.75\\$\pm$10.08}
&\shortstack{7914.24\\$\pm$163.60}
&\shortstack{1107.64\\$\pm$62.96}
\\ \hline
Diamond&\raisebox{-.5\totalheight}{\includegraphics[height=1cm,width=1cm]{4_Diamond.png}}
&\shortstack{37.77\\$\pm$9.63}
&\shortstack{1.06 \\$\pm$1.19}
&\shortstack{0.25\\$\pm$0.56}
&\shortstack{23.04\\$\pm$7.95}
&\shortstack{0.47\\$\pm$0.79}
&\shortstack{4.27\\$\pm$2.85}
&\shortstack{990.11\\$\pm$79.34}
&\shortstack{320.07\\$\pm$48.76}
\\ \hline
4-clique&\raisebox{-.5\totalheight}{\includegraphics[height=1cm,width=1cm]{4_4-Clique.png}}
&\shortstack{0.21\\$\pm$0.45}
&\shortstack{0.00\\$\pm$0.04}
&\shortstack{0.00\\$\pm$0.00}
&\shortstack{0.15\\$\pm$0.41}
&\shortstack{0.00\\$\pm$0.03}
&\shortstack{0.03\\$\pm$0.18}
&\shortstack{10.32\\$\pm$3.42}
&\shortstack{7.70\\$\pm$3.48}
\\ 
\botrule
\end{tabular}
\end{sidewaystable}
}

\clearpage
\section{Results for non-induced 5-node motifs in the original network}
\label{app:5non-induced}
\begin{landscape}
\begin{ThreePartTable}
    \begin{longtable}{llllllllll}
    \caption{5-node non-induced motifs in the \textit{C.~elegans} multi-layer network. The values in bold indicate that the discovered motif count is not significant.} \\
\toprule
& \textbf{Motif} & \textbf{ACh}& \textbf{Electrical}& \textbf{GABA} & \textbf{Glu}& \textbf{MA}& \textbf{Peptides}& \textbf{All}& \textbf{LC}\\
\midrule
\endhead 
\hline
\multicolumn{10}{l}{Continued}\\   \bottomrule
\endfoot
\bottomrule
\endlastfoot
\label{Tab:5node_nonind_motif} 
Ind set&\raisebox{-.5\totalheight}{\includegraphics[height=1cm,width=1cm]{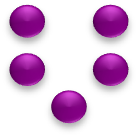}}&9161897856&8301429675&83291670&2349279569&297602656&873642672&13589056305&29034396\\ \hline
Only link&\raisebox{-.5\totalheight}{\includegraphics[height=1cm,width=1cm]{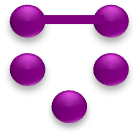}}&2984601600&1338520250&21021000&861540420&70949312&268637920&8013762350&37114200\\ \hline
Matching&\raisebox{-.5\totalheight}{\includegraphics[height=1cm,width=1cm]{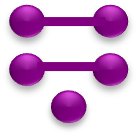}}&143716248&31839381&755286&46752899&2405126&12193176&703190675&6899781\\ \hline
Wedge+link&\raisebox{-.5\totalheight}{\includegraphics[height=1cm,width=1cm]{5_wedge+edge.png}}&17205321&1908523&75176&5379794&279693&1139991&125235528&2682077\\ \hline
Triangle+link&\raisebox{-.5\totalheight}{\includegraphics[height=1cm,width=1cm]{5_Triangle+edge.png}}&1091354&80460&1843&250020&4864&46897&8875634&253881\\ \hline
4-star&\raisebox{-.5\totalheight}{\includegraphics[height=1cm,width=1cm]{5_4-star.png}}&1405355&179490&7573&154375&32288&24010&11375923&483097\\ \hline
Prong&\raisebox{-.5\totalheight}{\includegraphics[height=1cm,width=1cm]{5_Prong.png}}&6398697&366720&19903&1233301&71213&168770&49972614&2135837\\ \hline
4-path&\raisebox{-.5\totalheight}{\includegraphics[height=1cm,width=1cm]{5_4-path.png}}&3881202&171761&7042&986172&32579&132148&31321978&1325556\\ \hline
Forktailed-tri&\raisebox{-.5\totalheight}{\includegraphics[height=1cm,width=1cm]{5_Forktailed-tri.png}}&1262834&78343&2568&114927&5679&12009&9593669&566754\\ \hline
Longtailed-tri&\raisebox{-.5\totalheight}{\includegraphics[height=1cm,width=1cm]{5_Lontailed-tri.png}}&1079519&30415&859&165198&1884&18624&8506399&480460\\ \hline
Doubletailed-tri&\raisebox{-.5\totalheight}{\includegraphics[height=1cm,width=1cm]{5_Doubledtailed-tri.png}}&1777988&73070&2788&202772&6143&20643&12624398&745673\\ \hline
Tailed-4-cycle&\raisebox{-.5\totalheight}{\includegraphics[height=1cm,width=1cm]{5_Tailed-4-cycle.png}}&1148058&47031&1240&195598&6411&19712&8027537&532415\\ \hline
5-cycle&\raisebox{-.5\totalheight}{\includegraphics[height=1cm,width=1cm]{5_5-cycle.png}}&120769&2419&\textbf{49$^{\star}$}&20889&338&2256&911648&61511\\ \hline
Hourglass&\raisebox{-.5\totalheight}{\includegraphics[height=1cm,width=1cm]{5_Hourglass.png}}&94495&2194&28&7649&58&725&708143&54653\\ \hline
Cobra&\raisebox{-.5\totalheight}{\includegraphics[height=1cm,width=1cm]{5_Cobra.png}}&472031&9902&199&39076&218&3759&3067571&228874\\ \hline
Stingray&\raisebox{-.5\totalheight}{\includegraphics[height=1cm,width=1cm]{5_Stingray.png}}&721513&36472&812&43572&1167&4327&4724609&375170\\ \hline
Hatted-4-cycle&\raisebox{-.5\totalheight}{\includegraphics[height=1cm,width=1cm]{5_Hatted-4-cycle.png}}&292288&5064&69&29689&276&2779&1906636&160352\\ \hline
3-wedge-col&\raisebox{-.5\totalheight}{\includegraphics[height=1cm,width=1cm]{5_3-wedgeCol.png}}&66701&3590&68&9044&267&747&399286&38039\\ \hline
3-tri-collision&\raisebox{-.5\totalheight}{\includegraphics[height=1cm,width=1cm]{5_3-TriCollison.png}}&51458&3441&58&1823&39&219&306445&32511\\ \hline
Tailed-4-clique&\raisebox{-.5\totalheight}{\includegraphics[height=1cm,width=1cm]{5_Tailed-4-clique.png}}&85830&1501&\textbf{0$^{\star \dagger}$}&3335&23&295&507147&43685\\ \hline
Triangle-strip&\raisebox{-.5\totalheight}{\includegraphics[height=1cm,width=1cm]{5_TriangleStrip.png}}&168237&2765&18&7440&41&681&999638&97719\\ \hline
Diamond-wed-col&\raisebox{-.5\totalheight}{\includegraphics[height=1cm,width=1cm]{5_Diamond-wedCol.png}}&65355&740&8&4492&26&356&361078&36360\\ \hline
4-wheel&\raisebox{-.5\totalheight}{\includegraphics[height=1cm,width=1cm]{5_4-Wheel.png}}&16763&83&1&570&3&30&83562&9391\\ \hline
Hatted-4-clique&\raisebox{-.5\totalheight}{\includegraphics[height=1cm,width=1cm]{5_Hatted-4-clique.png}}&44524&540&\textbf{0$^{\star \dagger}$}&974&10&84&236759&26204\\ \hline
Almost-5-clique&\raisebox{-.5\totalheight}{\includegraphics[height=1cm,width=1cm]{5_Almost-5-clique.png}}&6929&20&\textbf{0$^{\star \dagger}$}&82&1&2&32569&3868\\ \hline
5-clique&\raisebox{-.5\totalheight}{\includegraphics[height=1cm,width=1cm]{5_5-clique.png}}&386&\textbf{0$^{\star \dagger}$}&\textbf{0$^{\star \dagger}$}&1&\textbf{0$^{\star \dagger}$}&\textbf{0$^{\star \dagger}$}&1670&193\\ 
\bottomrule
\multicolumn{10}{l}{\textbf{$\star$ indicates link-swapping random network generation method}} \\
\multicolumn{10}{l}{\textbf{$\dagger$ indicates gnm random network generation method}}
\end{longtable}
\end{ThreePartTable}
\end{landscape}

\clearpage
\section{Results for non-induced 5-node motifs in the random reference networks}
\label{app:5non-induced_reference}
\small{
\begin{landscape}
\begin{ThreePartTable} 
    \label{Tab:5nodemotif_es_avg_nonind} 
\begin{longtable}{llllllllll}
    \caption{Average of 5-node non-induced motifs in \textit{C.~elegans} multi-layer network using link-swapping random network generation method.}\\
\toprule
& \textbf{Motif} & \textbf{ACh}& \textbf{Electrical}& \textbf{GABA} & \textbf{Glu}& \textbf{MA}& \textbf{Peptides}& \textbf{All}& \textbf{LC}\\
\midrule
\endhead
\hline
\multicolumn{10}{l}{Continued}\\   \bottomrule
\endfoot
\bottomrule
\endlastfoot
Ind set&\raisebox{-.5\totalheight}{\includegraphics[height=1cm,width=1cm]{5_Indset.png}}&\shortstack{9161897856\\$\pm$0.00}&\shortstack{8301429675\\$\pm$0.00}&\shortstack{83291670\\$\pm$0.00}&\shortstack{2349279569\\$\pm$0.00}&\shortstack{297602656\\$\pm$0.00}&\shortstack{873642672\\$\pm$0.00}&\shortstack{13589056305\\$\pm$0.00}&\shortstack{29034396\\$\pm$0.00}\\ \hline
Only link&\raisebox{-.5\totalheight}{\includegraphics[height=1cm,width=1cm]{5_Onlyedge.png}}&\shortstack{2984601600\\$\pm$0.00}&\shortstack{1338520250\\$\pm$0.00}&\shortstack{21021000\\$\pm$0.00}&\shortstack{861540420\\$\pm$0.00}&\shortstack{70949312\\$\pm$0.00}&\shortstack{268637920\\$\pm$0.00}&\shortstack{8013762350\\$\pm$0.00}&\shortstack{37114200\\$\pm$0.00}\\ \hline
Matching&\raisebox{-.5\totalheight}{\includegraphics[height=1cm,width=1cm]{5_Matching.png}}&\shortstack{145056864.32\\$\pm$59343.31}&\shortstack{32147962.97\\$\pm$26167.34}&\shortstack{778333.35\\$\pm$3222.59}&\shortstack{47143754.69\\$\pm$22484.73}&\shortstack{2484137.27\\$\pm$7168.73}&\shortstack{12305379.38\\$\pm$10171.15}&\shortstack{706904567.33\\$\pm$129455.66}&\shortstack{7049486.87\\$\pm$9353.36}\\ \hline
Wedge+link&\raisebox{-.5\totalheight}{\includegraphics[height=1cm,width=1cm]{5_wedge+edge.png}}&\shortstack{12056222.76\\$\pm$231072.67}&\shortstack{1346459.44\\$\pm$48631.08}&\shortstack{51655.83\\$\pm$3364.07}&\shortstack{4083176.75\\$\pm$75292.68}&\shortstack{175152.91\\$\pm$9507.48}&\shortstack{883589.27\\$\pm$23152.13}&\shortstack{96685252.89\\$\pm$1003103.60}&\shortstack{2043507.33\\$\pm$41058.72}\\ \hline
Triangle+link&\raisebox{-.5\totalheight}{\includegraphics[height=1cm,width=1cm]{5_Triangle+edge.png}}&\shortstack{226113.62\\$\pm$20613.83}&\shortstack{12914.83\\$\pm$2798.90}&\shortstack{619.78\\$\pm$267.07}&\shortstack{72313.69\\$\pm$7868.82}&\shortstack{2028.28\\$\pm$656.43}&\shortstack{12748.41\\$\pm$2307.97}&\shortstack{2614105.05\\$\pm$95735.62}&\shortstack{107321.62\\$\pm$7574.21}\\ \hline
4-star&\raisebox{-.5\totalheight}{\includegraphics[height=1cm,width=1cm]{5_4-star.png}}&\shortstack{301106.87\\$\pm$46498.56}&\shortstack{33079.52\\$\pm$9116.55}&\shortstack{1484.22\\$\pm$626.73}&\shortstack{53861.05\\$\pm$5850.24}&\shortstack{6173.14\\$\pm$1848.78}&\shortstack{9272.43\\$\pm$1467.69}&\shortstack{2768053.59\\$\pm$265697.79}&\shortstack{109192\\$\pm$18122.95}\\ \hline
Prong&\raisebox{-.5\totalheight}{\includegraphics[height=1cm,width=1cm]{5_Prong.png}}&\shortstack{1760219.93\\$\pm$155356.56}&\shortstack{113109.56\\$\pm$18421.72}&\shortstack{6235.86\\$\pm$1796.77}&\shortstack{485613.14\\$\pm$36040.32}&\shortstack{23162.02\\$\pm$5091.33}&\shortstack{82459.19\\$\pm$8623.10}&\shortstack{18841230.95\\$\pm$897712.1}&\shortstack{787686.15\\$\pm$69551.88}\\ \hline
4-path&\raisebox{-.5\totalheight}{\includegraphics[height=1cm,width=1cm]{5_4-path.png}}&\shortstack{1256407.62\\$\pm$75951.64}&\shortstack{66762.97\\$\pm$7091.59}&\shortstack{3725.12\\$\pm$783.62}&\shortstack{416025.01\\$\pm$24649.83}&\shortstack{13623.97\\$\pm$2246.02}&\shortstack{71207.59\\$\pm$5869.56}&\shortstack{14955798.09\\$\pm$444045.61}&\shortstack{642239.82\\$\pm$36449.22}\\ \hline
Forktailed-tri&\raisebox{-.5\totalheight}{\includegraphics[height=1cm,width=1cm]{5_Forktailed-tri.png}}&\shortstack{94637.01\\$\pm$19908.02}&\shortstack{4439.74\\$\pm$1976.44}&\shortstack{260.7\\$\pm$178.46}&\shortstack{16807.84\\$\pm$3030.36}&\shortstack{931.53\\$\pm$498.16}&\shortstack{2201.95\\$\pm$606.88}&\shortstack{1220332.84\\$\pm$132808.83}&\shortstack{90733.93\\$\pm$17032.67}\\ \hline
Longtailed-tri&\raisebox{-.5\totalheight}{\includegraphics[height=1cm,width=1cm]{5_Lontailed-tri.png}}&\shortstack{96324.79\\$\pm$15433.57}&\shortstack{2824.19\\$\pm$927.22}&\shortstack{189.65\\$\pm$112.93}&\shortstack{25650.21\\$\pm$3987.42}&\shortstack{654.01\\$\pm$303.60}&\shortstack{3439.81\\$\pm$812.89}&\shortstack{1463389.81\\$\pm$101243.85}&\shortstack{116302.16\\$\pm$14645.97}\\ \hline
Doubletailed-tri&\raisebox{-.5\totalheight}{\includegraphics[height=1cm,width=1cm]{5_Doubledtailed-tri.png}}&\shortstack{134849.47\\$\pm$26559.39}&\shortstack{4584.01\\$\pm$2156.15}&\shortstack{310.59\\$\pm$224.46}&\shortstack{29919.87\\$\pm$5222.24}&\shortstack{1093.84\\$\pm$620.48}&\shortstack{3913.59\\$\pm$1029.82}&\shortstack{1802883.01\\$\pm$161417.33}&\shortstack{138182.49\\$\pm$21515.90}\\ \hline
Tailed-4-cycle&\raisebox{-.5\totalheight}{\includegraphics[height=1cm,width=1cm]{5_Tailed-4-cycle.png}}&\shortstack{94376.22\\$\pm$13509.84}&\shortstack{2920.34\\$\pm$835.51}&\shortstack{199.95\\$\pm$108.70}&\shortstack{25477.97\\$\pm$3251.30}&\shortstack{809.83\\$\pm$300.99}&\shortstack{3330.35\\$\pm$599.83}&\shortstack{1436307.44\\$\pm$96152.48}&\shortstack{118105.54\\$\pm$14268.97}\\ \hline
5-cycle&\raisebox{-.5\totalheight}{\includegraphics[height=1cm,width=1cm]{5_5-cycle.png}}&\shortstack{13146.44\\$\pm$1464.35}&\shortstack{322.61\\$\pm$65.57}&\shortstack{24.13\\$\pm$10.59}&\shortstack{4231.93\\$\pm$455.77}&\shortstack{94.91\\$\pm$30.36}&\shortstack{566.20\\$\pm$85.22}&\shortstack{228900.30\\$\pm$11380.93}&\shortstack{19606.57\\$\pm$1809.66}\\ \hline
Hourglass&\raisebox{-.5\totalheight}{\includegraphics[height=1cm,width=1cm]{5_Hourglass.png}}&\shortstack{2481.25\\$\pm$754.95}&\shortstack{47.43\\$\pm$32.70}&\shortstack{3.31\\$\pm$4.10}&\shortstack{441.65\\$\pm$130.11}&\shortstack{11.28\\$\pm$10.39}&\shortstack{44.57\\$\pm$21.25}&\shortstack{45064.33\\$\pm$6104.16}&\shortstack{6288.93\\$\pm$1438.60}\\ \hline
Cobra&\raisebox{-.5\totalheight}{\includegraphics[height=1cm,width=1cm]{5_Cobra.png}}&\shortstack{10476.32\\$\pm$3654.94}&\shortstack{175.32\\$\pm$141.36}&\shortstack{12.48\\$\pm$20.12}&\shortstack{1915.22\\$\pm$647.21}&\shortstack{40.53\\$\pm$44.67}&\shortstack{191.14\\$\pm$102.21}&\shortstack{169096.06\\$\pm$23792.62}&\shortstack{23100.66\\$\pm$5313.78}\\ \hline
Stingray&\raisebox{-.5\totalheight}{\includegraphics[height=1cm,width=1cm]{5_Stingray.png}}&\shortstack{14745.66\\$\pm$5367.17}&\shortstack{399.48\\$\pm$405.05}&\shortstack{23.57\\$\pm$35.30}&\shortstack{2135.34\\$\pm$739.42}&\shortstack{78.60\\$\pm$83.97}&\shortstack{210.99\\$\pm$117.21}&\shortstack{224529.11\\$\pm$36993.41}&\shortstack{29565.64\\$\pm$7724.89}\\ \hline
Hatted-4-cycle&\raisebox{-.5\totalheight}{\includegraphics[height=1cm,width=1cm]{5_Hatted-4-cycle.png}}&\shortstack{7186.27\\$\pm$1917.30}&\shortstack{105.49\\$\pm$56.69}&\shortstack{8.48\\$\pm$8.71}&\shortstack{1572.43\\$\pm$372.13}&\shortstack{32.27\\$\pm$23.48}&\shortstack{157.39\\$\pm$54.01}&\shortstack{135311.30\\$\pm$14879.55}&\shortstack{20076.45\\$\pm$3731.70}\\ \hline
3-wedge-col&\raisebox{-.5\totalheight}{\includegraphics[height=1cm,width=1cm]{5_3-wedgeCol.png}}&\shortstack{1259.95\\$\pm$333.82}&\shortstack{28.54\\$\pm$22.39}&\shortstack{1.73\\$\pm$2.49}&\shortstack{273.71\\$\pm$66.85}&\shortstack{7.82\\$\pm$6.26}&\shortstack{26.58\\$\pm$10.66}&\shortstack{22963.38\\$\pm$2541.71}&\shortstack{3489.66\\$\pm$659.74}\\ \hline
3-tri-collision&\raisebox{-.5\totalheight}{\includegraphics[height=1cm,width=1cm]{5_3-TriCollison.png}}&\shortstack{277.13\\$\pm$164.60}&\shortstack{7.32\\$\pm$14.56}&\shortstack{0.31\\$\pm$1.11}&\shortstack{26.26\\$\pm$17.19}&\shortstack{0.91\\$\pm$2.18}&\shortstack{1.97\\$\pm$2.44}&\shortstack{4566.32\\$\pm$1237.44}&\shortstack{989.83\\$\pm$381.89}\\ \hline
Tailed-4-clique&\raisebox{-.5\totalheight}{\includegraphics[height=1cm,width=1cm]{5_Tailed-4-clique.png}}&\shortstack{544.64\\$\pm$411.29}&\shortstack{6.28\\$\pm$18.16}&\shortstack{0.34\\$\pm$2.85}&\shortstack{57.70\\$\pm$64.05}&\shortstack{0.80\\$\pm$5.06}&\shortstack{4.16\\$\pm$10.62}&\shortstack{8403.58\\$\pm$2371.79}&\shortstack{1757.76\\$\pm$724.94}\\ \hline
Triangle-strip&\raisebox{-.5\totalheight}{\includegraphics[height=1cm,width=1cm]{5_TriangleStrip.png}}&\shortstack{1110.45\\$\pm$612.68}&\shortstack{11.75\\$\pm$16.04}&\shortstack{0.73\\$\pm$2.03}&\shortstack{137.45\\$\pm$78.48}&\shortstack{2.34\\$\pm$4.24}&\shortstack{10.02\\$\pm$9.72}&\shortstack{20146.32\\$\pm$4529.78}&\shortstack{4651.54\\$\pm$1580.99}\\ \hline
Diamond-wed-col&\raisebox{-.5\totalheight}{\includegraphics[height=1cm,width=1cm]{5_Diamond-wedCol.png}}&\shortstack{408.99\\$\pm$205.74}&\shortstack{3.36\\$\pm$4.27}&\shortstack{0.26\\$\pm$0.90}&\shortstack{61.97\\$\pm$30.54}&\shortstack{0.93\\$\pm$1.82}&\shortstack{4.59\\$\pm$3.94}&\shortstack{7753.63\\$\pm$1463.12}&\shortstack{1915.81\\$\pm$559.37}\\ \hline
4-wheel&\raisebox{-.5\totalheight}{\includegraphics[height=1cm,width=1cm]{5_4-Wheel.png}}&\shortstack{31.73\\$\pm$28.81}&\shortstack{0.13\\$\pm$0.55}&\shortstack{0.01\\$\pm$0.14}&\shortstack{2.90\\$\pm$3.47}&\shortstack{0.03\\$\pm$0.21}&\shortstack{0.14\\$\pm$0.44}&\shortstack{555.67\\$\pm$184.19}&\shortstack{207.22\\$\pm$97.78}\\ \hline
Hatted-4-clique&\raisebox{-.5\totalheight}{\includegraphics[height=1cm,width=1cm]{5_Hatted-4-clique.png}}&\shortstack{87.04\\$\pm$90.16}&\shortstack{0.57\\$\pm$2.12}&\shortstack{0.03\\$\pm$0.33}&\shortstack{6.59\\$\pm$10.31}&\shortstack{0.05\\$\pm$0.42}&\shortstack{0.35\\$\pm$1.13}&\shortstack{1451.08\\$\pm$570.85}&\shortstack{495.47\\$\pm$267.53}\\ \hline
Almost-5-clique&\raisebox{-.5\totalheight}{\includegraphics[height=1cm,width=1cm]{5_Almost-5-clique.png}}&\shortstack{4.50\\$\pm$8.17}&\shortstack{0.01\\$\pm$0.13}&\shortstack{0.00\\$\pm$0.03}&\shortstack{0.23\\$\pm$0.90}&\shortstack{0.00\\$\pm$0.00}&\shortstack{0.00\\$\pm$0.06}&\shortstack{66.77\\$\pm$42.08}&\shortstack{32.81\\$\pm$27.16}\\ \hline
5-clique&\raisebox{-.5\totalheight}{\includegraphics[height=1cm,width=1cm]{5_5-clique.png}}&\shortstack{0.10\\$\pm$0.41}&\shortstack{0.00\\$\pm$0.00}&\shortstack{0.00\\$\pm$0.00}&\shortstack{0.00\\$\pm$0.05}&\shortstack{0.00\\$\pm$0.00}&\shortstack{0.00\\$\pm$0.00}&\shortstack{1.15\\$\pm$1.54}&\shortstack{0.72\\$\pm$1.22}\\ 
\end{longtable}
\end{ThreePartTable}
\end{landscape}
}
\small{
\begin{landscape}
\begin{ThreePartTable}
    \label{Tab:5nodemotif_gnm_avg_nonind} 
\begin{longtable}{llllllllll}
    \caption{Average of 5-node non-induced motifs in \textit{C.~elegans} multi-layer network using the gnm random network generation method.}\\ 
\toprule
& \textbf{Motif} & \textbf{ACh}& \textbf{Electrical}& \textbf{GABA} & \textbf{Glu}& \textbf{MA}& \textbf{Peptides}& \textbf{All}& \textbf{LC}\\
\midrule
\endhead 
\hline
\multicolumn{10}{l}{Continued}\\   \bottomrule
\endfoot
\bottomrule
\endlastfoot
Ind set&\raisebox{-.5\totalheight}{\includegraphics[height=1cm,width=1cm]{5_Indset.png}}&\shortstack{ 9161897856\\$\pm$0.00}&\shortstack{8301429675\\$\pm$0.00}&\shortstack{83291670\\$\pm$0.00}&\shortstack{2349279569\\$\pm$0.00}&\shortstack{297602656\\$\pm$0.00}&\shortstack{873642672\\$\pm$0.00}&\shortstack{13589056305\\$\pm$0.00}&\shortstack{29034396\\$\pm$0.00}\\ \hline
Only link&\raisebox{-.5\totalheight}{\includegraphics[height=1cm,width=1cm]{5_Onlyedge.png}}&\shortstack{2984601600\\$\pm$0.00}&\shortstack{1338520250\\$\pm$0.00}&\shortstack{21021000\\$\pm$0.00}&\shortstack{861540420\\$\pm$0.00}&\shortstack{70949312\\$\pm$0.00}&\shortstack{268637920\\$\pm$0.00}&\shortstack{8013762350\\$\pm$0.00}&\shortstack{37114200\\$\pm$0.00}\\ \hline
Matching&\raisebox{-.5\totalheight}{\includegraphics[height=1cm,width=1cm]{5_Matching.png}}&\shortstack{145710202.36\\$\pm$22769.54}&\shortstack{32311453.38\\$\pm$10951.11}&\shortstack{789813.85\\$\pm$1624.41}&\shortstack{47327808.94\\$\pm$13156.39}&\shortstack{2525113.82\\$\pm$2931.74}&\shortstack{12360482.36\\$\pm$6502.41}&\shortstack{708594073\\$\pm$46358.68}&\shortstack{7102107.98\\$\pm$4646.53}\\ \hline

Wedge+link&\raisebox{-.5\totalheight}{\includegraphics[height=1cm,width=1cm]{5_wedge+edge.png}}&\shortstack{9475531.58\\$\pm$91427.32}&\shortstack{1037924.06\\$\pm$21282.06}&\shortstack{39275.31\\$\pm$1839.89}&\shortstack{3461860.08\\$\pm$44767.78}&\shortstack{119039.99\\$\pm$4154.95}&\shortstack{756857.87\\$\pm$15010.87}&\shortstack{83487911.29\\$\pm$366089.38}&\shortstack{1808384.13\\$\pm$21324.8}\\ \hline
Triangle+link&\raisebox{-.5\totalheight}{\includegraphics[height=1cm,width=1cm]{5_Triangle+edge.png}}&\shortstack{102085.44\\$\pm$10265.62}&\shortstack{5546.23\\$\pm$1650.46}&\shortstack{322.82\\$\pm$194.53}&\shortstack{42162.76\\$\pm$5452.00}&\shortstack{949.76\\$\pm$407.88}&\shortstack{7665.21\\$\pm$1738.95}&\shortstack{1639650.23\\$\pm$60660.05}&\shortstack{76442.95\\$\pm$5446.38}\\ \hline
4-star&\raisebox{-.5\totalheight}{\includegraphics[height=1cm,width=1cm]{5_4-star.png}}&\shortstack{51277.76\\$\pm$3356.75}&\shortstack{2772.37\\$\pm$403.34}&\shortstack{160.14\\$\pm$62.21}&\shortstack{21085.27\\$\pm$1837.90}&\shortstack{463.33\\$\pm$129.39}&\shortstack{3848.88\\$\pm$565.75}&\shortstack{818595.54\\$\pm$23125.48}&\shortstack{38288.61\\$\pm$3138.37}\\ \hline
Prong&\raisebox{-.5\totalheight}{\includegraphics[height=1cm,width=1cm]{5_Prong.png}}&\shortstack{615414.90\\$\pm$25108.56}&\shortstack{33315.87\\$\pm$2969.88}&\shortstack{1934.67\\$\pm$425.48}&\shortstack{252902.71\\$\pm$13808.28}&\shortstack{5562.23\\$\pm$883.96}&\shortstack{46229.76\\$\pm$4052.25}&\shortstack{9827894.43\\$\pm$178358.34}&\shortstack{459452.01\\$\pm$23694.60}\\ \hline
4-path&\raisebox{-.5\totalheight}{\includegraphics[height=1cm,width=1cm]{5_4-path.png}}&\shortstack{615473.90\\$\pm$18626.54}&\shortstack{33325.20\\$\pm$2224.89}&\shortstack{1936.27\\$\pm$310.22}&\shortstack{252857.96\\$\pm$10245.80}&\shortstack{5569.91\\$\pm$657.36}&\shortstack{46234.96\\$\pm$2978.57}&\shortstack{9830537.29\\$\pm$132926.08}&\shortstack{459497.04\\$\pm$17387.90}\\ \hline
Forktailed-tri&\raisebox{-.5\totalheight}{\includegraphics[height=1cm,width=1cm]{5_Forktailed-tri.png}}&\shortstack{9917.03\\$\pm$1323.84}&\shortstack{267.62\\$\pm$100.21}&\shortstack{23.12\\$\pm$20.43}&\shortstack{4603.84\\$\pm$825.72}&\shortstack{66.22\\$\pm$40.93}&\shortstack{705.23\\$\pm$227.38}&\shortstack{289247.67\\$\pm$15027.1}&\shortstack{29049.18\\$\pm$3630.40}\\ \hline
Longtailed-tri&\raisebox{-.5\totalheight}{\includegraphics[height=1cm,width=1cm]{5_Lontailed-tri.png}}&\shortstack{19849.53\\$\pm$2283.71}&\shortstack{535.8\\$\pm$175.47}&\shortstack{46.83\\$\pm$32.98}&\shortstack{9216.53\\$\pm$1409.61}&\shortstack{132.53\\$\pm$67.04}&\shortstack{1402.72\\$\pm$372.99}&\shortstack{578776.7\\$\pm$25733.03}&\shortstack{58134.38\\$\pm$5776.44}\\ \hline
Doubletailed-tri&\raisebox{-.5\totalheight}{\includegraphics[height=1cm,width=1cm]{5_Doubledtailed-tri.png}}&\shortstack{19827.12\\$\pm$2495.8}&\shortstack{535.85\\$\pm$196.01}&\shortstack{46.81\\$\pm$38.66}&\shortstack{9205.79\\$\pm$1553.78}&\shortstack{132.06\\$\pm$77.27}&\shortstack{1409.44\\$\pm$428.25}&\shortstack{578635.28\\$\pm$28050.75}&\shortstack{58108.52\\$\pm$6529.90}\\ \hline
Tailed-4-cycle&\raisebox{-.5\totalheight}{\includegraphics[height=1cm,width=1cm]{5_Tailed-4-cycle.png}}&\shortstack{19971.53\\$\pm$1470.95}&\shortstack{533.79\\$\pm$118.71}&\shortstack{48.64\\$\pm$27.27}&\shortstack{9233.40\\$\pm$884.86}&\shortstack{129.83\\$\pm$52.02}&\shortstack{1403.36\\$\pm$252.10}&\shortstack{578372.27\\$\pm$16768.67}&\shortstack{58131.02\\$\pm$4723.06}\\ \hline
5-cycle&\raisebox{-.5\totalheight}{\includegraphics[height=1cm,width=1cm]{5_5-cycle.png}}&\shortstack{3993.27\\$\pm$215.91}&\shortstack{107.23\\$\pm$16.60}&\shortstack{9.36\\$\pm$4.18}&\shortstack{1844.03\\$\pm$133.12}&\shortstack{25.58\\$\pm$7.30}&\shortstack{281.22\\$\pm$34.47}&\shortstack{115702.60\\$\pm$2660.32}&\shortstack{11654.32\\$\pm$755.20}\\ \hline
Hourglass&\raisebox{-.5\totalheight}{\includegraphics[height=1cm,width=1cm]{5_Hourglass.png}}&\shortstack{159.78\\$\pm$38.42}&\shortstack{2.16\\$\pm$2.07}&\shortstack{0.27\\$\pm$0.66}&\shortstack{83.86\\$\pm$27.49}&\shortstack{0.80\\$\pm$1.18}&\shortstack{10.56\\$\pm$6.83}&\shortstack{8516.31\\$\pm$737.66}&\shortstack{1833.40\\$\pm$338.19}\\ \hline
Cobra&\raisebox{-.5\totalheight}{\includegraphics[height=1cm,width=1cm]{5_Cobra.png}}&\shortstack{642.71\\$\pm$178.78}&\shortstack{8.60\\$\pm$10.56}&\shortstack{1.19\\$\pm$3.01}&\shortstack{336.27\\$\pm$130.98}&\shortstack{2.84\\$\pm$5.34}&\shortstack{42.88\\$\pm$33.15}&\shortstack{34026.44\\$\pm$3107.27}&\shortstack{7321.16\\$\pm$1407.80}\\ \hline
Stingray&\raisebox{-.5\totalheight}{\includegraphics[height=1cm,width=1cm]{5_Stingray.png}}&\shortstack{642.22\\$\pm$185.27}&\shortstack{8.46\\$\pm$10.61}&\shortstack{1.18\\$\pm$2.88}&\shortstack{336.74\\$\pm$138.12}&\shortstack{2.80\\$\pm$5.42}&\shortstack{43.27\\$\pm$34.96}&\shortstack{34018.77\\$\pm$3243.51}&\shortstack{7314.75\\$\pm$1501.01}\\ \hline
Hatted-4-cycle&\raisebox{-.5\totalheight}{\includegraphics[height=1cm,width=1cm]{5_Hatted-4-cycle.png}}&\shortstack{642.84\\$\pm$97.48}&\shortstack{8.61\\$\pm$4.92}&\shortstack{1.16\\$\pm$1.66}&\shortstack{336.03\\$\pm$68.05}&\shortstack{2.89\\$\pm$2.58}&\shortstack{41.99\\$\pm$15.92}&\shortstack{34034.34\\$\pm$1925.82}&\shortstack{7340.42\\$\pm$1011.43}\\ \hline
3-wedge-col&\raisebox{-.5\totalheight}{\includegraphics[height=1cm,width=1cm]{5_3-wedgeCol.png}}&\shortstack{107.79\\$\pm$19.55}&\shortstack{1.42\\$\pm$1.39}&\shortstack{0.17\\$\pm$0.45}&\shortstack{56.17\\$\pm$13.09}&\shortstack{0.48\\$\pm$0.82}&\shortstack{7.06\\$\pm$3.83}&\shortstack{5666.54\\$\pm$295.45}&\shortstack{1221.05\\$\pm$167.99}\\ \hline
3-tri-collision&\raisebox{-.5\totalheight}{\includegraphics[height=1cm,width=1cm]{5_3-TriCollison.png}}&\shortstack{3.47\\$\pm$2.85}&\shortstack{0.02\\$\pm$0.13}&\shortstack{0.00\\$\pm$0.03}&\shortstack{2.07\\$\pm$2.31}&\shortstack{0.01\\$\pm$0.15}&\shortstack{0.23\\$\pm$0.60}&\shortstack{333.15\\$\pm$57.46}&\shortstack{152.86\\$\pm$51.00}\\ \hline
Tailed-4-clique&\raisebox{-.5\totalheight}{\includegraphics[height=1cm,width=1cm]{5_Tailed-4-clique.png}}&\shortstack{7.01\\$\pm$14.95}&\shortstack{0.04\\$\pm$0.81}&\shortstack{0.00\\$\pm$0.00}&\shortstack{4.09\\$\pm$12.11}&\shortstack{0.01\\$\pm$0.22}&\shortstack{0.64\\$\pm$3.67}&\shortstack{667.24\\$\pm$226.66}&\shortstack{306.9\\$\pm$146.56}\\ \hline
Triangle-strip&\raisebox{-.5\totalheight}{\includegraphics[height=1cm,width=1cm]{5_TriangleStrip.png}}&\shortstack{20.81\\$\pm$11.29}&\shortstack{0.14\\$\pm$0.49}&\shortstack{0.03\\$\pm$0.18}&\shortstack{12.38\\$\pm$9.23}&\shortstack{0.06\\$\pm$0.30}&\shortstack{1.28\\$\pm$2.21}&\shortstack{2001.56\\$\pm$292.29}&\shortstack{919.73\\$\pm$271.41}\\ \hline
Diamond-wed-col&\raisebox{-.5\totalheight}{\includegraphics[height=1cm,width=1cm]{5_Diamond-wedCol.png}}&\shortstack{10.44\\$\pm$4.88}&\shortstack{0.07\\$\pm$0.29}&\shortstack{0.01\\$\pm$0.12}&\shortstack{6.17\\$\pm$3.96}&\shortstack{0.03\\$\pm$0.19}&\shortstack{0.58\\$\pm$0.94}&\shortstack{999.91\\$\pm$111.44}&\shortstack{460.89\\$\pm$111.01}\\ \hline
4-wheel&\raisebox{-.5\totalheight}{\includegraphics[height=1cm,width=1cm]{5_4-Wheel.png}}&\shortstack{0.17\\$\pm$0.44}&\shortstack{0.00\\$\pm$0.00}&\shortstack{0.00\\$\pm$0.00}&\shortstack{0.12\\$\pm$0.43}&\shortstack{0.00\\$\pm$0.00}&\shortstack{0.01\\$\pm$0.08}&\shortstack{29.39\\$\pm$8.77}&\shortstack{28.72\\$\pm$13.88}\\ \hline
Hatted-4-clique&\raisebox{-.5\totalheight}{\includegraphics[height=1cm,width=1cm]{5_Hatted-4-clique.png}}&\shortstack{0.34\\$\pm$1.09}&\shortstack{0.00\\$\pm$0.00}&\shortstack{0.00\\$\pm$0.00}&\shortstack{0.22\\$\pm$0.98}&\shortstack{0.00\\$\pm$0.03}&\shortstack{0.03\\$\pm$0.26}&\shortstack{59.15\\$\pm$25.08}&\shortstack{57.64\\$\pm$36.04}\\ \hline
Almost-5-clique&\raisebox{-.5\totalheight}{\includegraphics[height=1cm,width=1cm]{5_Almost-5-clique.png}}&\shortstack{0.01\\$\pm$0.07}&\shortstack{0.00\\$\pm$0.00}&\shortstack{0.00\\$\pm$0.00}&\shortstack{0.00\\$\pm$0.07}&\shortstack{0.00\\$\pm$0.00}&\shortstack{0.00\\$\pm$0.00}&\shortstack{1.12\\$\pm$1.53}&\shortstack{2.41\\$\pm$3.29}\\ \hline
5-clique&\raisebox{-.5\totalheight}{\includegraphics[height=1cm,width=1cm]{5_5-clique.png}}&\shortstack{0.00\\$\pm$0.00}&\shortstack{0.00\\$\pm$0.00}&\shortstack{0.00\\$\pm$0.00}&\shortstack{0.00\\$\pm$0.00}&\shortstack{0.00\\$\pm$0.00}&\shortstack{0.00\\$\pm$0.00}&\shortstack{0.01\\$\pm$0.08}&\shortstack{0.03\\$\pm$0.19}\\ 
\end{longtable}
\end{ThreePartTable}
\end{landscape}
}

\clearpage
\section{Results for induced 5-node motifs in the original network}
\label{app:5induced}

\begin{landscape}
\begin{ThreePartTable}
    \label{Tab:5nodemotif_induced} 
\begin{longtable}{llllllllll}
    \caption{5-node induced motifs in the \textit{C.~elegans} multi-layer network. The values in bold indicate that the discovered motif count is not significant.} \\
\toprule
 & \textbf{Motif} & \textbf{ACh}& \textbf{Electrical}& \textbf{GABA} & \textbf{Glu}& \textbf{MA}& \textbf{Peptides}& \textbf{All}& \textbf{LC}\\
\midrule
\endhead 
\hline
\multicolumn{10}{l}{Continued}\\   \bottomrule
\endfoot
\bottomrule
\endlastfoot
Ind set&\raisebox{-.5\totalheight}{\includegraphics[height=1cm,width=1cm]{5_Indset.png}}&6740603526&7101053698&65779414&1653421431&239778812&648734427&7765614959&9616064\\ \hline

Only link&\raisebox{-.5\totalheight}{\includegraphics[height=1cm,width=1cm]{5_Onlyedge.png}}&1961317251&1077646814&14473518&557053610&46487282&186951607&4165138807&9254750\\ \hline

Matching&\raisebox{-.5\totalheight}{\includegraphics[height=1cm,width=1cm]{5_Matching.png}}&81968440&24630869&489193&27068856&1412560&8050245&334089336&1357434\\ \hline
Wedge+link&\raisebox{-.5\totalheight}{\includegraphics[height=1cm,width=1cm]{5_wedge+edge.png}} &7760852&1281903&47297&\textbf{2789744$^{\star}$}&167551&715979&48067402&390866\\ \hline
Triangle+link&\raisebox{-.5\totalheight}{\includegraphics[height=1cm,width=1cm]{5_Triangle+edge.png}}&575040&62720&1260&146406&3469&34252&4396960&55008\\ \hline
4-star&\raisebox{-.5\totalheight}{\includegraphics[height=1cm,width=1cm]{5_4-star.png}}&688500&129768&5712&78525&27712&15831&5563054&190900\\ \hline
Prong&\raisebox{-.5\totalheight}{\includegraphics[height=1cm,width=1cm]{5_Prong.png}}&1658015&133846&10568&476066&41387&87142&13208481&305438\\ \hline
4-path&\raisebox{-.5\totalheight}{\includegraphics[height=1cm,width=1cm]{5_4-path.png}}&55959&\textbf{53350$^{\star}$}&\textbf{2618$^{\star}$}&353578&14198&61360&6627552&107706\\ \hline
Forktailed-tri&\raisebox{-.5\totalheight}{\includegraphics[height=1cm,width=1cm]{5_Forktailed-tri.png}}&370878&30103&1286&46876&3578&5385&3297305&131420\\ \hline
Longtailed-tri&\raisebox{-.5\totalheight}{\includegraphics[height=1cm,width=1cm]{5_Lontailed-tri.png}}&214157&10032&351&60261&997&8327&2109920&58681\\ \hline
Doubletailed-tri&\raisebox{-.5\totalheight}{\includegraphics[height=1cm,width=1cm]{5_Doubledtailed-tri.png}}&385163&13739&1155&75827&3632&8129&3016956&93356\\ \hline
Tailed-4-cycle&\raisebox{-.5\totalheight}{\includegraphics[height=1cm,width=1cm]{5_Tailed-4-cycle.png}}&113972&3910&\textbf{151$^{\star}$}&67539&3409&7929&972426&\textbf{27066$^{\dagger}$}\\ \hline
5-cycle&\raisebox{-.5\totalheight}{\includegraphics[height=1cm,width=1cm]{5_5-cycle.png}}&8270&\textbf{308$^{\star}$}&\textbf{10$^{\star \dagger}$}&3876&129&594&94414&2518\\ \hline
Hourglass&\raisebox{-.5\totalheight}{\includegraphics[height=1cm,width=1cm]{5_Hourglass.png}}&13048&555&12&2820&37&260&178783&7811\\ \hline
Cobra&\raisebox{-.5\totalheight}{\includegraphics[height=1cm,width=1cm]{5_Cobra.png}}&94517&2893&155&15651&\textbf{95$^{\star}$}&1664&705810&33169\\ \hline
Stingray&\raisebox{-.5\totalheight}{\includegraphics[height=1cm,width=1cm]{5_Stingray.png}}&187957&11045&432&17861&870&1678&1385604&71963\\ \hline
Hatted-4-cycle&\raisebox{-.5\totalheight}{\includegraphics[height=1cm,width=1cm]{5_Hatted-4-cycle.png}}&19558&330&\textbf{13$^{\star}$}&7617&162&809&204890&\textbf{8138$^{\star}$}\\ \hline
3-wedge-col&\raisebox{-.5\totalheight}{\includegraphics[height=1cm,width=1cm]{5_3-wedgeCol.png}}&4082&\textbf{35$^{\star}$}&\textbf{4$^{\star}$}&4525&214&308&22070&\textbf{612$^{\dagger}$}\\ \hline
3-tri-collision&\raisebox{-.5\totalheight}{\includegraphics[height=1cm,width=1cm]{5_3-TriCollison.png}}&23861&2961&58&1085&32&141&150693&15981\\ \hline
Tailed-4-clique&\raisebox{-.5\totalheight}{\includegraphics[height=1cm,width=1cm]{5_Tailed-4-clique.png}}&30636&541&\textbf{0$^{\star \dagger}$}&1859&\textbf{9$^{\star}$}&139&195643&10625\\ \hline
Triangle-strip&\raisebox{-.5\totalheight}{\includegraphics[height=1cm,width=1cm]{5_TriangleStrip.png}}&24651&633&14&2680&\textbf{7$^{\star}$}&261&204396&13383\\ \hline
Diamond-wed-col&\raisebox{-.5\totalheight}{\includegraphics[height=1cm,width=1cm]{5_Diamond-wedCol.png}}&4560&48&4&1946&13&170&33092&1614\\ \hline
4-wheel&\raisebox{-.5\totalheight}{\includegraphics[height=1cm,width=1cm]{5_4-Wheel.png}}&1766&23&1&339&\textbf{0$^{\star \dagger}$}&24&10905&682\\ \hline
Hatted-4-clique&\raisebox{-.5\totalheight}{\includegraphics[height=1cm,width=1cm]{5_Hatted-4-clique.png}}&14530&420&\textbf{0$^{\star \dagger}$}&512&4&72&91445&8786\\ \hline
Almost-5-clique&\raisebox{-.5\totalheight}{\includegraphics[height=1cm,width=1cm]{5_Almost-5-clique.png}}&3069&20&\textbf{0$^{\star \dagger}$}&72&1&2&15869&1938\\ \hline
5-clique&\raisebox{-.5\totalheight}{\includegraphics[height=1cm,width=1cm]{5_5-clique.png}}&386&\textbf{0$^{\star \dagger}$}&\textbf{0$^{\star \dagger}$}&1&\textbf{0$^{\star \dagger}$}&\textbf{0$^{\star \dagger}$}&1670&193\\ 
\hline
\multicolumn{10}{l}{\textbf{$\star$ indicates link-swapping random network generation method}} \\
\multicolumn{10}{l}{\textbf{$\dagger$ indicates gnm random network generation method}}\\
\end{longtable}
\end{ThreePartTable}
\end{landscape}

\begin{landscape}
\begin{ThreePartTable}
    \label{Tab:5node_indVsnonind_motif} 
\begin{longtable}{llllllllll}
    \caption{Ratio of 5-node induced vs non-induced motifs in the \textit{C.~elegans} multi-layer network.} \\
\toprule
& \textbf{Motif} & \textbf{ACh}& \textbf{Electrical}& \textbf{GABA} & \textbf{Glu}& \textbf{MA}& \textbf{Peptides}& \textbf{All}& \textbf{LC}\\
\midrule
\endhead 
\hline
\multicolumn{10}{l}{Continued}\\   \bottomrule
\endfoot
\bottomrule
\endlastfoot
Ind set&\raisebox{-.5\totalheight}{\includegraphics[height=1cm,width=1cm]{5_Indset.png}}&0.74&0.86&0.79&0.70&0.81&0.74&0.57&0.33\\ \hline
Only link&\raisebox{-.5\totalheight}{\includegraphics[height=1cm,width=1cm]{5_Onlyedge.png}}&0.66&0.81&0.69&0.65&0.66&0.70&0.52&0.25\\ \hline
Matching&\raisebox{-.5\totalheight}{\includegraphics[height=1cm,width=1cm]{5_Matching.png}}&0.57&0.77&0.65&0.58&0.59&0.66&0.48&0.2\\ \hline

Wedge+link&\raisebox{-.5\totalheight}{\includegraphics[height=1cm,width=1cm]{5_wedge+edge.png}}&0.45&0.67&0.63&0.52&0.6&0.63&0.38&0.15\\ \hline
Triangle+link&\raisebox{-.5\totalheight}{\includegraphics[height=1cm,width=1cm]{5_Triangle+edge.png}}&0.53&0.78&0.68&0.59&0.71&0.73&0.50&0.22\\ \hline
4-star&\raisebox{-.5\totalheight}{\includegraphics[height=1cm,width=1cm]{5_4-star.png}}&0.49&0.72&0.75&0.51&0.86&0.66&0.49&0.40\\ \hline
Prong&\raisebox{-.5\totalheight}{\includegraphics[height=1cm,width=1cm]{5_Prong.png}}&0.26&0.36&0.53&0.39&0.58&0.52&0.26&0.14\\ \hline
4-path&\raisebox{-.5\totalheight}{\includegraphics[height=1cm,width=1cm]{5_4-path.png}}&0.19&0.31&0.37&0.36&0.44&0.46&0.21&0.08\\ \hline
Forktailed-tri&\raisebox{-.5\totalheight}{\includegraphics[height=1cm,width=1cm]{5_Forktailed-tri.png}}&0.29&0.38&0.50&0.41&0.63&0.45&0.34&0.23\\ \hline
Longtailed-tri&\raisebox{-.5\totalheight}{\includegraphics[height=1cm,width=1cm]{5_Lontailed-tri.png}}&0.20&0.33&0.41&0.36&0.53&0.45&0.25&0.12\\ \hline
Doubletailed-tri&\raisebox{-.5\totalheight}{\includegraphics[height=1cm,width=1cm]{5_Doubledtailed-tri.png}}&0.22&0.19&0.41&0.37&0.59&0.39&0.24&0.13\\ \hline
Tailed-4-cycle&\raisebox{-.5\totalheight}{\includegraphics[height=1cm,width=1cm]{5_Tailed-4-cycle.png}}&0.10&0.08&0.12&0.35&0.53&0.40&0.12&0.05\\ \hline
5-cycle&\raisebox{-.5\totalheight}{\includegraphics[height=1cm,width=1cm]{5_5-cycle.png}}&0.07&0.13&0.20&0.19&0.38&0.26&0.10&0.04\\ \hline
Hourglass&\raisebox{-.5\totalheight}{\includegraphics[height=1cm,width=1cm]{5_Hourglass.png}}&0.14&0.25&0.43&0.37&0.64&0.36&0.25&0.14\\ \hline
Cobra&\raisebox{-.5\totalheight}{\includegraphics[height=1cm,width=1cm]{5_Cobra.png}}&0.2&0.29&0.78&0.40&0.44&0.44&0.23&0.14\\ \hline
Stingray&\raisebox{-.5\totalheight}{\includegraphics[height=1cm,width=1cm]{5_Stingray.png}}&0.26&0.3&0.53&0.41&0.75&0.39&0.29&0.19\\ \hline
Hatted-4-cycle&\raisebox{-.5\totalheight}{\includegraphics[height=1cm,width=1cm]{5_Hatted-4-cycle.png}}&0.07&0.07&0.19&0.26&0.59&0.29&0.11&0.05\\ \hline
3-wedge-col&\raisebox{-.5\totalheight}{\includegraphics[height=1cm,width=1cm]{5_3-wedgeCol.png}}&0.06&0.01&0.06&0.5&0.8&0.41&0.06&0.02\\ \hline
3-tri-collision&\raisebox{-.5\totalheight}{\includegraphics[height=1cm,width=1cm]{5_3-TriCollison.png}}&0.46&0.86&1.00&0.60&0.82&0.64&0.49&0.49\\ \hline
Tailed-4-clique&\raisebox{-.5\totalheight}{\includegraphics[height=1cm,width=1cm]{5_Tailed-4-clique.png}}&0.36&0.36&-&0.56&0.39&0.47&0.39&0.24\\ \hline
Triangle-strip&\raisebox{-.5\totalheight}{\includegraphics[height=1cm,width=1cm]{5_TriangleStrip.png}}&0.15&0.23&0.78&0.36&0.17&0.38&0.2&0.14\\ \hline
Diamond-wed-col&\raisebox{-.5\totalheight}{\includegraphics[height=1cm,width=1cm]{5_Diamond-wedCol.png}}&0.07&0.06&0.50&0.43&0.5&0.48&0.09&0.04\\ \hline
4-wheel&\raisebox{-.5\totalheight}{\includegraphics[height=1cm,width=1cm]{5_4-Wheel.png}}&0.11&0.28&1.00&0.59&0.00&0.80&0.13&0.07\\ \hline
Hatted-4-clique&\raisebox{-.5\totalheight}{\includegraphics[height=1cm,width=1cm]{5_Hatted-4-clique.png}}&0.33&0.78&1.00&0.53&0.40&0.86&0.39&0.34\\ \hline
Almost-5-clique&\raisebox{-.5\totalheight}{\includegraphics[height=1cm,width=1cm]{5_Almost-5-clique.png}}&0.44&1.00&-&0.88&1.00&1.00&0.49&0.50\\
\end{longtable}
\end{ThreePartTable}
\end{landscape}

\small{
\begin{landscape}
\begin{ThreePartTable}
    \label{Tab:Percentage_5node_noninduced_motif} 
\begin{longtable}{lccccccc}
    \caption{Percentage of 5-node non-induced motifs in the \textit{C.~elegans} multi-layer network relative to ALL} 
\endhead 
\toprule
& \textbf{Motif}& \textbf{ACh}& \textbf{Electrical}& \textbf{GABA} & \textbf{Glu}& \textbf{MA}& \textbf{Peptides}\\
\midrule
Ind set&\raisebox{-.5\totalheight}{\includegraphics[height=1cm,width=1cm]{5_Indset.png}}&67.42&61.09&0.61&17.29&2.19&6.43\\ \hline
Only link&\raisebox{-.5\totalheight}{\includegraphics[height=1cm,width=1cm]{5_Onlyedge.png}}&37.24&16.70&0.26&10.75&0.89&3.35\\ \hline
Matching&\raisebox{-.5\totalheight}{\includegraphics[height=1cm,width=1cm]{5_Matching.png}}&20.44&4.53&0.11&6.65&0.34&1.73\\ \hline
Wedge+link&\raisebox{-.5\totalheight}{\includegraphics[height=1cm,width=1cm]{5_wedge+edge.png}}&13.74&1.52&0.06&4.30&0.22&0.91\\ \hline
Triangle+link&\raisebox{-.5\totalheight}{\includegraphics[height=1cm,width=1cm]{5_Triangle+edge.png}}&12.30&0.91&0.02&2.82&0.05&0.53\\ \hline
4-star&\raisebox{-.5\totalheight}{\includegraphics[height=1cm,width=1cm]{5_4-star.png}}&12.35&1.58&0.07&1.36&0.28&0.21\\ \hline
Prong&\raisebox{-.5\totalheight}{\includegraphics[height=1cm,width=1cm]{5_Prong.png}}&12.80&0.73&0.04&2.47&0.14&0.34\\ \hline
4-path&\raisebox{-.5\totalheight}{\includegraphics[height=1cm,width=1cm]{5_4-path.png}}&12.39&0.55&0.02&3.15&0.10&0.42\\ \hline
Forktailed-tri&\raisebox{-.5\totalheight}{\includegraphics[height=1cm,width=1cm]{5_Forktailed-tri.png}}&13.16&0.82&0.03&1.20&0.06&0.13\\ \hline
Longtailed-tri&\raisebox{-.5\totalheight}{\includegraphics[height=1cm,width=1cm]{5_Lontailed-tri.png}}&12.69&0.36&0.01&1.94&0.02&0.22\\ \hline
Doubletailed-tri&\raisebox{-.5\totalheight}{\includegraphics[height=1cm,width=1cm]{5_Doubledtailed-tri.png}}&14.08&0.58&0.02&1.61&0.05&0.16\\ \hline
Tailed-4-cycle&\raisebox{-.5\totalheight}{\includegraphics[height=1cm,width=1cm]{5_Tailed-4-cycle.png}}&14.30&0.59&0.02&2.44&0.08&0.25\\ \hline
5-cycle&\raisebox{-.5\totalheight}{\includegraphics[height=1cm,width=1cm]{5_5-cycle.png}}&13.25&0.27&0.01&2.29&0.04&0.25\\ \hline
Hourglass&\raisebox{-.5\totalheight}{\includegraphics[height=1cm,width=1cm]{5_Hourglass.png}}&13.34&0.31&0.00&1.08&0.01&0.10\\ \hline
Cobra&\raisebox{-.5\totalheight}{\includegraphics[height=1cm,width=1cm]{5_Cobra.png}}&15.39&0.32&0.01&1.27&0.01&0.12\\ \hline
Stingray&\raisebox{-.5\totalheight}{\includegraphics[height=1cm,width=1cm]{5_Stingray.png}}&15.27&0.77&0.02&0.92&0.02&0.09\\ \hline
Hatted-4-cycle&\raisebox{-.5\totalheight}{\includegraphics[height=1cm,width=1cm]{5_Hatted-4-cycle.png}}&15.33&0.27&0.00&1.56&0.01&0.15\\ \hline
3-wedge-col&\raisebox{-.5\totalheight}{\includegraphics[height=1cm,width=1cm]{5_3-wedgeCol.png}}&16.71&0.90&0.02&2.27&0.07&0.19\\ \hline
3-tri-collision&\raisebox{-.5\totalheight}{\includegraphics[height=1cm,width=1cm]{5_3-TriCollison.png}}&16.79&1.12&0.02&0.59&0.01&0.07\\ \hline
Tailed-4-clique&\raisebox{-.5\totalheight}{\includegraphics[height=1cm,width=1cm]{5_Tailed-4-clique.png}}&16.92&0.30&0.00&0.66&0.00&0.06\\ \hline
Triangle-strip&\raisebox{-.5\totalheight}{\includegraphics[height=1cm,width=1cm]{5_TriangleStrip.png}}&16.83&0.28&0.00&0.74&0.00&0.07\\ \hline
Diamond-wed-col&\raisebox{-.5\totalheight}{\includegraphics[height=1cm,width=1cm]{5_Diamond-wedCol.png}}&18.10&0.20&0.00&1.24&0.01&0.10\\ \hline
4-wheel&\raisebox{-.5\totalheight}{\includegraphics[height=1cm,width=1cm]{5_4-Wheel.png}}&20.06&0.10&0.00&0.68&0.00&0.04\\ \hline
Hatted-4-clique&\raisebox{-.5\totalheight}{\includegraphics[height=1cm,width=1cm]{5_Hatted-4-clique.png}}&18.81&0.23&0.00&0.41&0.00&0.04\\ \hline
Almost-5-clique&\raisebox{-.5\totalheight}{\includegraphics[height=1cm,width=1cm]{5_Almost-5-clique.png}}&21.27&0.06&0.00&0.25&0.00&0.01\\ \hline
5-clique&\raisebox{-.5\totalheight}{\includegraphics[height=1cm,width=1cm]{5_5-clique.png}}&23.11&0.00&0.00&0.06&0.00&0.00\\ 
\hline
\end{longtable}
\end{ThreePartTable}
\end{landscape}
}

\clearpage
\section{Results for induced 5-node motifs in the random reference networks}
\label{app:5induced_random}

\small{
\begin{landscape}
\begin{ThreePartTable}
    \label{Tab:5nodemotif_es_avg_ind} 
\begin{longtable}{p{1.4cm}lllllllll}
    \caption{Average of 5-node induced motifs in the \textit{C.~elegans} multi-layer network using link-swapping random network generation method} \\
\toprule
& \textbf{Motif} & \textbf{ACh}& \textbf{Electrical}& \textbf{GABA} & \textbf{Glu}& \textbf{MA}& \textbf{Peptides}& \textbf{All}& \textbf{LC}\\
\midrule
\endhead 
\hline
\multicolumn{10}{l}{Continued}\\   \bottomrule
\endfoot
\bottomrule
\endlastfoot
Ind set&\raisebox{-.5\totalheight}{\includegraphics[height=1cm,width=1cm]{5_Indset.png}}&\shortstack{6638371083.85\\$\pm$5229480.35}&\shortstack{7072844815.9\\$\pm$2691785.78}&\shortstack{64932551.47\\$\pm$120832.81}&\shortstack{1629462881.71\\$\pm$1525880.11}&\shortstack{235850729.88\\$\pm$351790.57}&\shortstack{642555395.89\\$\pm$604903.08}&\shortstack{7531898979.22\\$\pm$9971651.86}&\shortstack{8035600.37\\$\pm$114950.68}\\ \hline
Only link&\raisebox{-.5\totalheight}{\includegraphics[height=1cm,width=1cm]{5_Onlyedge.png}}&\shortstack{2115708236.85\\$\pm$8629544.32}&\shortstack{1125545846.85\\$\pm$4890598.66}&\shortstack{15943319.11\\$\pm$211424.78}&\shortstack{594490370.04\\$\pm$2534900.87}&\shortstack{53437636.13\\$\pm$618288.58}&\shortstack{197138635.66\\$\pm$1046443.55}&\shortstack{4447858263.12\\$\pm$14257168.69}&\shortstack{10330455.58\\$\pm$99746.4}\\ \hline
Matching&\raisebox{-.5\totalheight}{\includegraphics[height=1cm,width=1cm]{5_Matching.png}}&\shortstack{100017439.93\\$\pm$1047257.75}&\shortstack{26716330.47\\$\pm$273614.95}&\shortstack{580175.54\\$\pm$19798.00}&\shortstack{32229221.98\\$\pm$361024.55}&\shortstack{1802666.56\\$\pm$53253.94}&\shortstack{9012909.30\\$\pm$121488.06}&\shortstack{392532167.53\\$\pm$3198412.76}&\shortstack{1992711.10\\$\pm$66893.27}\\ \hline
Wedge +link&\raisebox{-.5\totalheight}{\includegraphics[height=1cm,width=1cm]{5_wedge+edge.png}}&\shortstack{8051848.84\\$\pm$69827.44}&\shortstack{1093303.71\\$\pm$26037.61}&\shortstack{38261.59\\$\pm$1615.41}&\shortstack{2796428.48\\$\pm$21885.89}&\shortstack{126558.81\\$\pm$4545.63}&\shortstack{654133.57\\$\pm$8796.75}&\shortstack{53666818.69\\$\pm$265746.32}&\shortstack{600960.88\\$\pm$19499.07}\\ \hline
Triangle +link&\raisebox{-.5\totalheight}{\includegraphics[height=1cm,width=1cm]{5_Triangle+edge.png}}&\shortstack{149192.44\\$\pm$11679.31}&\shortstack{10432.00\\$\pm$2169.41}&\shortstack{455.50\\$\pm$193.36}&\shortstack{50607.89\\$\pm$5194.90}&\shortstack{1462.57\\$\pm$454.91}&\shortstack{9714.46\\$\pm$1709.10}&\shortstack{1487234.60\\$\pm$44341.91}&\shortstack{34007.58\\$\pm$2419.85}\\ \hline
4-star&\raisebox{-.5\totalheight}{\includegraphics[height=1cm,width=1cm]{5_4-star.png}}&\shortstack{221680.24\\$\pm$34552.80}&\shortstack{29055.27\\$\pm$8140.29}&\shortstack{1248.76\\$\pm$538.94}&\shortstack{39397.92\\$\pm$4247.40}&\shortstack{5326.67\\$\pm$1607.10}&\shortstack{7308.74\\$\pm$1170.24}&\shortstack{1782894.93\\$\pm$177780.68}&\shortstack{47027.00\\$\pm$8611.91}\\ \hline
Prong&\raisebox{-.5\totalheight}{\includegraphics[height=1cm,width=1cm]{5_Prong.png}}&\shortstack{1156830.80\\$\pm$75903.16}&\shortstack{89597.09\\$\pm$12700.75}&\shortstack{4712.99\\$\pm$1210.39}&\shortstack{340940.65\\$\pm$20760.49}&\shortstack{17564.76\\$\pm$3279.17}&\shortstack{62716.83\\$\pm$5783.52}&\shortstack{10745479.93\\$\pm$365075.66}&\shortstack{263514.84\\$\pm$13608.48}\\ \hline
4-path&\raisebox{-.5\totalheight}{\includegraphics[height=1cm,width=1cm]{5_4-path.png}}&\shortstack{794020.37\\$\pm$26731.77}&\shortstack{51443.45\\$\pm$4207.49}&\shortstack{2680.49\\$\pm$428.57}&\shortstack{286470.67\\$\pm$11876.15}&\shortstack{9720.22\\$\pm$1153.87}&\shortstack{53356.91\\$\pm$3475.44}&\shortstack{8248965.73\\$\pm$117470.96}&\shortstack{198303.27\\$\pm$5846.76}\\ \hline
Forktailed-tri&\raisebox{-.5\totalheight}{\includegraphics[height=1cm,width=1cm]{5_Forktailed-tri.png}}&\shortstack{66052.51\\$\pm$11558.34}&\shortstack{3638.95\\$\pm$1361.61}&\shortstack{211.79\\$\pm$128.35}&\shortstack{12335.98\\$\pm$1813.85}&\shortstack{766.09\\$\pm$362.49}&\shortstack{1741.92\\$\pm$414.34}&\shortstack{781329.24\\$\pm$74126.36}&\shortstack{39869.65\\$\pm$6264.29}\\ \hline
Longtailed-tri&\raisebox{-.5\totalheight}{\includegraphics[height=1cm,width=1cm]{5_Lontailed-tri.png}}&\shortstack{60402.07\\$\pm$6411.55}&\shortstack{2173.67\\$\pm$568.71}&\shortstack{141\\$\pm$67.56}&\shortstack{18159.26\\$\pm$2199.29}&\shortstack{484.18\\$\pm$182.37}&\shortstack{2658.37\\$\pm$531.72}&\shortstack{842558.45\\$\pm$35954.10}&\shortstack{41149.65\\$\pm$2117.00}\\ \hline
Doubletailed -tri&\raisebox{-.5\totalheight}{\includegraphics[height=1cm,width=1cm]{5_Doubledtailed-tri.png}}&\shortstack{87867.79\\$\pm$12821.39}&\shortstack{3473.41\\$\pm$1317.44}&\shortstack{238.14\\$\pm$146.90}&\shortstack{21564.99\\$\pm$2999.27}&\shortstack{847.84\\$\pm$419.66}&\shortstack{3050.80\\$\pm$688.72}&\shortstack{1064837.67\\$\pm$67829.72}&\shortstack{51697.92\\$\pm$4490.93}\\ \hline
Tailed-4-cycle&\raisebox{-.5\totalheight}{\includegraphics[height=1cm,width=1cm]{5_Tailed-4-cycle.png}}&\shortstack{56809.34\\$\pm$5518.77}&\shortstack{2091.28\\$\pm$535.28}&\shortstack{144.01\\$\pm$73.06}&\shortstack{17897.89\\$\pm$1923.37}&\shortstack{602.73\\$\pm$201.41}&\shortstack{2548.79\\$\pm$421.43}&\shortstack{805132.64\\$\pm$37359.72}&\shortstack{41173.32\\$\pm$3381.20}\\ \hline
5-cycle&\raisebox{-.5\totalheight}{\includegraphics[height=1cm,width=1cm]{5_5-cycle.png}}&\shortstack{7613.42\\$\pm$507.89}&\shortstack{233.98\\$\pm$41.32}&\shortstack{16.81\\$\pm$6.97}&\shortstack{2897.47\\$\pm$247.84}&\shortstack{66.60\\$\pm$18.35}&\shortstack{426.79\\$\pm$58.11}&\shortstack{124504.51\\$\pm$3676.76}&\shortstack{6381.69\\$\pm$395.93}\\ \hline
Hourglass&\raisebox{-.5\totalheight}{\includegraphics[height=1cm,width=1cm]{5_Hourglass.png}}&\shortstack{1582.81\\$\pm$365.10}&\shortstack{37.03\\$\pm$22.69}&\shortstack{2.65\\$\pm$3.16}&\shortstack{321.83\\$\pm$79.85}&\shortstack{9.10\\$\pm$7.98}&\shortstack{35.50\\$\pm$15.56}&\shortstack{28548.19\\$\pm$3033.45}&\shortstack{2856.68\\$\pm$480.89}\\ \hline
Cobra&\raisebox{-.5\totalheight}{\includegraphics[height=1cm,width=1cm]{5_Cobra.png}}&\shortstack{6651.58\\$\pm$1695.95}&\shortstack{131.64\\$\pm$89.57}&\shortstack{9.76\\$\pm$13.28}&\shortstack{1413.78\\$\pm$382.71}&\shortstack{32.26\\$\pm$31.79}&\shortstack{153.21\\$\pm$71.57}&\shortstack{102606.15\\$\pm$10415.27}&\shortstack{9569.84\\$\pm$1336.23}\\ \hline
Stingray&\raisebox{-.5\totalheight}{\includegraphics[height=1cm,width=1cm]{5_Stingray.png}}&\shortstack{10123.29\\$\pm$2958.17}&\shortstack{319.24\\$\pm$287.75}&\shortstack{19.51\\$\pm$25.59}&\shortstack{1601.83\\$\pm$444.54}&\shortstack{66.72\\$\pm$64.70}&\shortstack{170.56\\$\pm$81.64}&\shortstack{146828.00\\$\pm$19845.37}&\shortstack{14089.51\\$\pm$2798.39}\\ \hline
Hatted-4-cycle&\raisebox{-.5\totalheight}{\includegraphics[height=1cm,width=1cm]{5_Hatted-4-cycle.png}}&\shortstack{4130.30\\$\pm$656.61}&\shortstack{73.32\\$\pm$28.86}&\shortstack{6.26\\$\pm$5.21}&\shortstack{1118.58\\$\pm$196.28}&\shortstack{24.58\\$\pm$15.02}&\shortstack{122.63\\$\pm$34.27}&\shortstack{77845.4\\$\pm$5284.27}&\shortstack{7825.24\\$\pm$717.04}\\ \hline
3-wedge-col&\raisebox{-.5\totalheight}{\includegraphics[height=1cm,width=1cm]{5_3-wedgeCol.png}}&\shortstack{707.31\\$\pm$155.37}&\shortstack{18.66\\$\pm$17.46}&\shortstack{1.20\\$\pm$1.89}&\shortstack{196.96\\$\pm$46.36}&\shortstack{6.10\\$\pm$5.16}&\shortstack{20.63\\$\pm$8.87}&\shortstack{12950.30\\$\pm$1176.47}&\shortstack{1369.87\\$\pm$255.21}\\ \hline
3-tri-collision&\raisebox{-.5\totalheight}{\includegraphics[height=1cm,width=1cm]{5_3-TriCollison.png}}&\shortstack{202.61\\$\pm$106.78}&\shortstack{6.78\\$\pm$13.94}&\shortstack{0.29\\$\pm$1.07}&\shortstack{20.35\\$\pm$11.33}&\shortstack{0.86\\$\pm$2.09}&\shortstack{1.64\\$\pm$2.01}&\shortstack{3304.02\\$\pm$822.07}&\shortstack{585.61\\$\pm$206.18}\\ \hline
Tailed-4-clique&\raisebox{-.5\totalheight}{\includegraphics[height=1cm,width=1cm]{5_Tailed-4-clique.png}}&\shortstack{395.60\\$\pm$274.55}&\shortstack{5.20\\$\pm$14.71}&\shortstack{0.29\\$\pm$2.41}&\shortstack{45.87\\$\pm$48.57}&\shortstack{0.70\\$\pm$4.38}&\shortstack{3.50\\$\pm$8.89}&\shortstack{5878.97\\$\pm$1465.53}&\shortstack{949.32\\$\pm$333.92}\\ \hline
Triangle-strip&\raisebox{-.5\totalheight}{\includegraphics[height=1cm,width=1cm]{5_TriangleStrip.png}}&\shortstack{710.51\\$\pm$287.91}&\shortstack{9.11\\$\pm$11.03}&\shortstack{0.59\\$\pm$1.47}&\shortstack{103.51\\$\pm$45.94}&\shortstack{2.01\\$\pm$3.34}&\shortstack{8.16\\$\pm$6.93}&\shortstack{13251.97\\$\pm$2315.85}&\shortstack{2388.28\\$\pm$590.92}\\ \hline
Diamond-wed-col&\raisebox{-.5\totalheight}{\includegraphics[height=1cm,width=1cm]{5_Diamond-wedCol.png}}&\shortstack{232.60\\$\pm$74.49}&\shortstack{2.36\\$\pm$2.51}&\shortstack{0.19\\$\pm$0.59}&\shortstack{45.79\\$\pm$16.94}&\shortstack{0.75\\$\pm$1.32}&\shortstack{3.73\\$\pm$2.78}&\shortstack{4646.19\\$\pm$569.31}&\shortstack{865.20\\$\pm$148.16}\\ \hline
4-wheel&\raisebox{-.5\totalheight}{\includegraphics[height=1cm,width=1cm]{5_4-Wheel.png}}&\shortstack{19.69\\$\pm$12.48}&\shortstack{0.10\\$\pm$0.40}&\shortstack{0.01\\$\pm$0.10}&\shortstack{2.25\\$\pm$2.29}&\shortstack{0.03\\$\pm$0.21}&\shortstack{0.13\\$\pm$0.37}&\shortstack{372.66\\$\pm$91.97}&\shortstack{119.56\\$\pm$41.46}\\ \hline
Hatted-4-clique&\raisebox{-.5\totalheight}{\includegraphics[height=1cm,width=1cm]{5_Hatted-4-clique.png}}&\shortstack{62.96\\$\pm$53.53}&\shortstack{0.51\\$\pm$1.69}&\shortstack{0.02\\$\pm$0.22}&\shortstack{5.28\\$\pm$6.82}&\shortstack{0.05\\$\pm$0.42}&\shortstack{0.32\\$\pm$0.97}&\shortstack{1085.06\\$\pm$365.50}&\shortstack{320.15\\$\pm$141.48}\\ \hline
Almost-5-clique&\raisebox{-.5\totalheight}{\includegraphics[height=1cm,width=1cm]{5_Almost-5-clique.png}}&\shortstack{3.53\\$\pm$5.19}&\shortstack{0.01\\$\pm$0.13}&\shortstack{0.00\\$\pm$0.03}&\shortstack{0.20\\$\pm$0.64}&\shortstack{0.00\\$\pm$0.00}&\shortstack{0.00\\$\pm$0.06}&\shortstack{55.24\\$\pm$30.17}&\shortstack{25.63\\$\pm$17.73}\\ \hline
5-clique&\raisebox{-.5\totalheight}{\includegraphics[height=1cm,width=1cm]{5_5-clique.png}}&\shortstack{0.10\\$\pm$0.41}&\shortstack{0.00\\$\pm$0.00}&\shortstack{0.00\\$\pm$0.00}&\shortstack{0.00\\$\pm$0.05}&\shortstack{0.00\\$\pm$0.00}&\shortstack{0.00\\$\pm$0.00}&\shortstack{1.15\\$\pm$1.54}&\shortstack{0.72\\$\pm$1.22}\\
\end{longtable}
\end{ThreePartTable}
\end{landscape}
}


\small{
\begin{landscape}
\begin{ThreePartTable}
    \label{Tab:5nodemotif_gnm_avg_ind} 
\begin{longtable}{p{1.4cm}lllllllll}
    \caption{Average of 5-node induced motifs in the \textit{C.~elegans} multi-layer network using the gnm random network generation method} \\
\toprule
& \textbf{Motif} & \textbf{ACh}& \textbf{Electrical}& \textbf{GABA} & \textbf{Glu}& \textbf{MA}& \textbf{Peptides}& \textbf{All}& \textbf{LC}\\
\midrule
\endhead 
\hline
\multicolumn{10}{l}{Continued}\\   \bottomrule
\endfoot
\bottomrule
\endlastfoot
Ind set&\raisebox{-.5\totalheight}{\includegraphics[height=1cm,width=1cm]{5_Indset.png}}&\shortstack{6578587162.18\\$\pm$2218300.93}&\shortstack{7055801840.73\\$\pm$1201243.91}&\shortstack{64489879.45\\$\pm$64937.65}&\shortstack{1616723295.44\\$\pm$943091.19}&\shortstack{233762601.6\\$\pm$153206.55}&\shortstack{639227916.13\\$\pm$398518.73}&\shortstack{7398956868.03\\$\pm$3962967.47}&\shortstack{7380806.6\\$\pm$62411.58}\\ \hline
Only link&\raisebox{-.5\totalheight}{\includegraphics[height=1cm,width=1cm]{5_Onlyedge.png}}&\shortstack{2215842909.09\\$\pm$3869772.26}&\shortstack{1156661639.44\\$\pm$2259752.04}&\shortstack{16726650.94\\$\pm$117753.49}&\shortstack{615761847.79\\$\pm$1616945}&\shortstack{57169207.7\\$\pm$278645.58}&\shortstack{202912466.95\\$\pm$701227.58}&\shortstack{4638352742.93\\$\pm$6095816.34}&\shortstack{10849646.9\\$\pm$58400.44}\\ \hline
Matching&\raisebox{-.5\totalheight}{\includegraphics[height=1cm,width=1cm]{5_Matching.png}}&\shortstack{111846848.16\\$\pm$448472.92}&\shortstack{28384329.33\\$\pm$127044.13}&\shortstack{645648.57\\$\pm$11163.6}&\shortstack{35127451.28\\$\pm$214967.79}&\shortstack{2086901.45\\$\pm$24398.46}&\shortstack{9636868.56\\$\pm$78450}&\shortstack{435961597.06\\$\pm$1405103.04}&\shortstack{2385766.75\\$\pm$46589.13}\\ \hline

Wedge+ link&\raisebox{-.5\totalheight}{\includegraphics[height=1cm,width=1cm]{5_wedge+edge.png}}&\shortstack{7520875.51\\$\pm$39137.84}&\shortstack{926791.41\\$\pm$14845.66}&\shortstack{32976.59\\$\pm$1057.02}&\shortstack{2667882.72\\$\pm$17456.78}&\shortstack{100817.06\\$\pm$2389.44}&\shortstack{609192.44\\$\pm$6937.41}&\shortstack{54593519.96\\$\pm$82503.95}&\shortstack{697764.59\\$\pm$10994.98}\\ \hline
Triangle+ link&\raisebox{-.5\totalheight}{\includegraphics[height=1cm,width=1cm]{5_Triangle+edge.png}}&\shortstack{83773.15\\$\pm$8387.30}&\shortstack{5031.49\\$\pm$1498.74}&\shortstack{278.79\\$\pm$168.74}&\shortstack{33746.16\\$\pm$4322.97}&\shortstack{824.37\\$\pm$353.89}&\shortstack{6364.24\\$\pm$1435.80}&\shortstack{1139585.12\\$\pm$41860.29}&\shortstack{33842.51\\$\pm$2459.73}\\ \hline
4-star&\raisebox{-.5\totalheight}{\includegraphics[height=1cm,width=1cm]{5_4-star.png}}&\shortstack{42128.79\\$\pm$2731.88}&\shortstack{2515.16\\$\pm$370.33}&\shortstack{138.44\\$\pm$54.07}&\shortstack{16881.96\\$\pm$1459.24}&\shortstack{400.62\\$\pm$113.49}&\shortstack{3195.18\\$\pm$465.62}&\shortstack{568692.20\\$\pm$16190.52}&\shortstack{16992.19\\$\pm$1442.04}\\ \hline
Prong&\raisebox{-.5\totalheight}{\includegraphics[height=1cm,width=1cm]{5_Prong.png}}&\shortstack{505365.82\\$\pm$18901.37}&\shortstack{30231.17\\$\pm$2610.39}&\shortstack{1667.67\\$\pm$346.74}&\shortstack{202404.97\\$\pm$10060.57}&\shortstack{4815.31\\$\pm$715.85}&\shortstack{38405.16\\$\pm$3082.33}&\shortstack{6828721.93\\$\pm$109047.13}&\shortstack{203860.10\\$\pm$7897.02}\\ \hline
4-path&\raisebox{-.5\totalheight}{\includegraphics[height=1cm,width=1cm]{5_4-path.png}}&\shortstack{505272.07\\$\pm$12998.57}&\shortstack{30240.02\\$\pm$1886.57}&\shortstack{1668.56\\$\pm$239.98}&\shortstack{202334.90\\$\pm$6886.17}&\shortstack{4827.29\\$\pm$506.33}&\shortstack{38418.52\\$\pm$2124.19}&\shortstack{6831252.69\\$\pm$70495.45}&\shortstack{203776.88\\$\pm$4152.66}\\ \hline
Forktailed-tri&\raisebox{-.5\totalheight}{\includegraphics[height=1cm,width=1cm]{5_Forktailed-tri.png}}&\shortstack{8414.02\\$\pm$1001.20}&\shortstack{247.02\\$\pm$86.83}&\shortstack{20.31\\$\pm$16.30}&\shortstack{3822.28\\$\pm$590.00}&\shortstack{59.29\\$\pm$33.75}&\shortstack{604.38\\$\pm$173.76}&\shortstack{213608.69\\$\pm$8981.44}&\shortstack{14808.97\\$\pm$1289.28}\\ \hline
Longtailed-tri&\raisebox{-.5\totalheight}{\includegraphics[height=1cm,width=1cm]{5_Lontailed-tri.png}}&\shortstack{16841.26\\$\pm$1752.86}&\shortstack{494.14\\$\pm$152.80}&\shortstack{41.31\\$\pm$26.50}&\shortstack{7655.63\\$\pm$1034.12}&\shortstack{118.44\\$\pm$55.56}&\shortstack{1203.31\\$\pm$289.18}&\shortstack{427453.60\\$\pm$15661.46}&\shortstack{29601.01\\$\pm$1834.75}\\ \hline
Doubletailed -tri&\raisebox{-.5\totalheight}{\includegraphics[height=1cm,width=1cm]{5_Doubledtailed-tri.png}}&\shortstack{16816.51\\$\pm$1905.37}&\shortstack{494.42\\$\pm$170.98}&\shortstack{41.11\\$\pm$31.08}&\shortstack{7642.87\\$\pm$1126.36}&\shortstack{118.43\\$\pm$64.95}&\shortstack{1208.49\\$\pm$332.80}&\shortstack{427372.77\\$\pm$16916.69}&\shortstack{29615.59\\$\pm$2168.43}\\ \hline
Tailed-4-cycle&\raisebox{-.5\totalheight}{\includegraphics[height=1cm,width=1cm]{5_Tailed-4-cycle.png}}&\shortstack{16956.21\\$\pm$1196.69}&\shortstack{492.37\\$\pm$108.03}&\shortstack{43.15\\$\pm$23.55}&\shortstack{7670.42\\$\pm$695.25}&\shortstack{116.14\\$\pm$46.09}&\shortstack{1203.39\\$\pm$206.78}&\shortstack{427118.49\\$\pm$11531.43}&\shortstack{29610.88\\$\pm$2050.83}\\ \hline
5-cycle&\raisebox{-.5\totalheight}{\includegraphics[height=1cm,width=1cm]{5_5-cycle.png}}&\shortstack{3390.80\\$\pm$170.12}&\shortstack{98.91\\$\pm$15.34}&\shortstack{8.26\\$\pm$3.72}&\shortstack{1531.83\\$\pm$104.57}&\shortstack{22.81\\$\pm$6.59}&\shortstack{241.57\\$\pm$28.97}&\shortstack{85440.42\\$\pm$1740.06}&\shortstack{5939.3\\$\pm$305.44}\\ \hline
Hourglass&\raisebox{-.5\totalheight}{\includegraphics[height=1cm,width=1cm]{5_Hourglass.png}}&\shortstack{139.95\\$\pm$32.19}&\shortstack{2.02\\$\pm$1.93}&\shortstack{0.24\\$\pm$0.62}&\shortstack{72.14\\$\pm$22.27}&\shortstack{0.75\\$\pm$1.11}&\shortstack{9.36\\$\pm$5.89}&\shortstack{6685.20\\$\pm$519.42}&\shortstack{1072.47\\$\pm$155.40}\\ \hline
Cobra&\raisebox{-.5\totalheight}{\includegraphics[height=1cm,width=1cm]{5_Cobra.png}}&\shortstack{563.12\\$\pm$146.07}&\shortstack{8.07\\$\pm$9.56}&\shortstack{1.11\\$\pm$2.73}&\shortstack{289.54\\$\pm$101.77}&\shortstack{2.64\\$\pm$4.83}&\shortstack{37.57\\$\pm$26.19}&\shortstack{26703.56\\$\pm$2130.87}&\shortstack{4274.40\\$\pm$621.83}\\ \hline
Stingray&\raisebox{-.5\totalheight}{\includegraphics[height=1cm,width=1cm]{5_Stingray.png}}&\shortstack{562.69\\$\pm$147.36}&\shortstack{7.97\\$\pm$9.46}&\shortstack{1.12\\$\pm$2.65}&\shortstack{289.91\\$\pm$103.92}&\shortstack{2.60\\$\pm$4.72}&\shortstack{37.81\\$\pm$26.97}&\shortstack{26697.51\\$\pm$2170.22}&\shortstack{4273.05\\$\pm$652.33}\\ \hline
Hatted-4-cycle&\raisebox{-.5\totalheight}{\includegraphics[height=1cm,width=1cm]{5_Hatted-4-cycle.png}}&\shortstack{563.39\\$\pm$75.18}&\shortstack{8.03\\$\pm$4.38}&\shortstack{1.05\\$\pm$1.43}&\shortstack{289.27\\$\pm$49.10}&\shortstack{2.65\\$\pm$2.28}&\shortstack{37.4\\$\pm$12.86}&\shortstack{26712.63\\$\pm$1188.73}&\shortstack{4292.15\\$\pm$385.82}\\ \hline
3-wedge-col&\raisebox{-.5\totalheight}{\includegraphics[height=1cm,width=1cm]{5_3-wedgeCol.png}}&\shortstack{94.53\\$\pm$17.85}&\shortstack{1.33\\$\pm$1.35}&\shortstack{0.16\\$\pm$0.42}&\shortstack{48.38\\$\pm$11.49}&\shortstack{0.43\\$\pm$0.77}&\shortstack{6.30\\$\pm$3.53}&\shortstack{4446.98\\$\pm$237.16}&\shortstack{713.09\\$\pm$101.60}\\ \hline
3-tri-collision&\raisebox{-.5\totalheight}{\includegraphics[height=1cm,width=1cm]{5_3-TriCollison.png}}&\shortstack{3.15\\$\pm$2.56}&\shortstack{0.02\\$\pm$0.13}&\shortstack{0.00\\$\pm$0.03}&\shortstack{1.86\\$\pm$2.10}&\shortstack{0.01\\$\pm$0.15}&\shortstack{0.19\\$\pm$0.54}&\shortstack{277.31\\$\pm$43.88}&\shortstack{102.09\\$\pm$29.15}\\ \hline
Tailed-4-clique&\raisebox{-.5\totalheight}{\includegraphics[height=1cm,width=1cm]{5_Tailed-4-clique.png}}&\shortstack{6.36\\$\pm$13.47}&\shortstack{0.04\\$\pm$0.81}&\shortstack{0.00\\$\pm$0.00}&\shortstack{3.67\\$\pm$10.68}&\shortstack{0.01\\$\pm$0.16}&\shortstack{0.57\\$\pm$3.28}&\shortstack{555.56\\$\pm$185.32}&\shortstack{205.36\\$\pm$91.57}\\ \hline
Triangle-strip&\raisebox{-.5\totalheight}{\includegraphics[height=1cm,width=1cm]{5_TriangleStrip.png}}&\shortstack{18.86\\$\pm$9.29}&\shortstack{0.14\\$\pm$0.49}&\shortstack{0.03\\$\pm$0.18}&\shortstack{11.07\\$\pm$7.38}&\shortstack{0.05\\$\pm$0.28}&\shortstack{1.12\\$\pm$1.72}&\shortstack{1667.24\\$\pm$215.09}&\shortstack{615.55\\$\pm$143.84}\\ \hline
Diamond-wed-col&\raisebox{-.5\totalheight}{\includegraphics[height=1cm,width=1cm]{5_Diamond-wedCol.png}}&\shortstack{9.47\\$\pm$4.13}&\shortstack{0.07\\$\pm$0.29}&\shortstack{0.01\\$\pm$0.12}&\shortstack{5.50\\$\pm$3.12}&\shortstack{0.03\\$\pm$0.18}&\shortstack{0.52\\$\pm$0.81}&\shortstack{833.12\\$\pm$77.47}&\shortstack{309.01\\$\pm$53.82}\\ \hline
4-wheel&\raisebox{-.5\totalheight}{\includegraphics[height=1cm,width=1cm]{5_4-Wheel.png}}&\shortstack{0.15\\$\pm$0.40}&\shortstack{0.00\\$\pm$0.00}&\shortstack{0.00\\$\pm$0.00}&\shortstack{0.11\\$\pm$0.37}&\shortstack{0.00\\$\pm$0.00}&\shortstack{0.01\\$\pm$0.08}&\shortstack{26.11\\$\pm$7.16}&\shortstack{22.01\\$\pm$9.20}\\ \hline
Hatted-4-clique&\raisebox{-.5\totalheight}{\includegraphics[height=1cm,width=1cm]{5_Hatted-4-clique.png}}&\shortstack{0.31\\$\pm$0.90}&\shortstack{0.00\\$\pm$0.00}&\shortstack{0.00\\$\pm$0.00}&\shortstack{0.21\\$\pm$0.77}&\shortstack{0.00\\$\pm$0.03}&\shortstack{0.03\\$\pm$0.26}&\shortstack{52.58\\$\pm$20.14}&\shortstack{44.23\\$\pm$22.78}\\ \hline
Almost-5-clique&\raisebox{-.5\totalheight}{\includegraphics[height=1cm,width=1cm]{5_Almost-5-clique.png}}&\shortstack{0.01\\$\pm$0.07}&\shortstack{0.00\\$\pm$0.00}&\shortstack{0.00\\$\pm$0.00}&\shortstack{0.00\\$\pm$0.07}&\shortstack{0.00\\$\pm$0.00}&\shortstack{0.00\\$\pm$0.00}&\shortstack{1.07\\$\pm$1.23}&\shortstack{2.07\\$\pm$2.33}\\ \hline
5-clique&\raisebox{-.5\totalheight}{\includegraphics[height=1cm,width=1cm]{5_5-clique.png}}&\shortstack{0.00\\$\pm$0.00}&\shortstack{0.00\\$\pm$0.00}&\shortstack{0.00\\$\pm$0.00}&\shortstack{0.00\\$\pm$0.00}&\shortstack{0.00\\$\pm$0.00}&\shortstack{0.00\\$\pm$0.00}&\shortstack{0.01\\$\pm$0.08}&\shortstack{0.03\\$\pm$0.19}\\
\end{longtable}
\end{ThreePartTable}
\end{landscape}
}

\end{appendices}
\end{document}